\documentclass[10pt]{article}

\usepackage{setspace}
\usepackage[T1]{fontenc}                % Use 8-bit encoding that has 256 glyphs
\usepackage{microtype} 				    % Slightly tweak font spacing for aesthetics
\usepackage{amssymb,amsmath,amscd}
\usepackage[UKenglish]{babel}
\usepackage{titlesec}
\usepackage{hyperref}
\usepackage{graphicx,xcolor,xparse}
\usepackage{enumitem}
\usepackage{multirow}
\usepackage[absolute]{textpos}
\usepackage[margin=1.5em,labelfont=bf]{caption}
\usepackage{cite}
\usepackage[bbgreekl]{mathbbol}
\usepackage[hang,flushmargin]{footmisc} % Remove footnote indents

\usepackage{tikz}

\usepackage{orcidlink}

\hypersetup{
	colorlinks=true,
	linkcolor=dark-blue,
	citecolor=dark-red,
	urlcolor=dark-green,
	linktoc=all
}

\usepackage[vcentermath]{youngtab}

\usepackage{tikz} 
\usepackage{tikz-cd} 
\usetikzlibrary{matrix,calc} 
\usetikzlibrary{decorations.markings,shapes.geometric,shapes.misc} 
\usetikzlibrary{decorations.pathreplacing,positioning} 
\usetikzlibrary{arrows}
\usepackage{mathrsfs}

\graphicspath{{./figures/}}

\usepackage{geometry} 					% Document margins
\numberwithin{equation}{section} 

\titleformat{\section}[block]{\Large\bfseries\centering}{\thesection}{1em}{} 
\titleformat{\subsection}[block]{\bfseries}{\thesubsection}{1em}{} 
\titlespacing*{\section}{0pt}{1em}{1em}
\titlespacing*{\subsection}{0pt}{0.75em}{0.75em}

\addto\captionsenglish{
  \renewcommand{\contentsname}%
    {Table of Contents}%
}

\definecolor{dark-gray}{gray}{0.20}
\definecolor{gray}{gray}{0.30}
\definecolor{light-gray}{gray}{0.80}
\definecolor{dark-red}{rgb}{0.7,0,0}
\definecolor{dark-green}{rgb}{0.1,0.4,0}
\definecolor{dark-blue}{rgb}{0.3,0.3,0.7}
\definecolor{light-blue}{rgb}{0.8,0.8,1}
\definecolor{cardinal}{rgb}{0.6,0,0}
\definecolor{darkgreen}{rgb}{0,0.5,0}
\definecolor{golden}{rgb}{0.92, 0.7, 0}
\definecolor{midnight}{rgb}{0, 0, 0.5}
\definecolor{darkblue}{rgb}{0.2, 0, 0.8}
\definecolor{forestgreen}{rgb}{0.13, 0.55, 0.13}

\def\mop#1{\mathop{\rm #1}\nolimits}

\def\ii{{\rm i}}

\newcommand{\dd}{{\rm d}}
\newcommand{\e}{\mathrm{e}}
\newcommand{\bz}{\bar{z}}

\newcommand\Tr{\mathrm{Tr}\,}

\newcommand{\vol}{\mathbf{V}}
\newcommand{\sgn}{\mathrm{sgn}}

\newcommand{\z}{\bar{\mathbf{Z}}}
\newcommand{\dotp}{{\dot +}}
\newcommand{\dotm}{{\dot -}}

\newcommand\bbC{\mathbb{C}}
\newcommand\bbR{\mathbb{R}}
\newcommand\bbZ{\mathbb{Z}}

\newcommand\bJ{\mathbf{J}}

\newcommand\bL{\mathbf{L}}

\newcommand\bO{\mathbf{O}}

\newcommand\bQ{\mathbf{Q}}

\newcommand\bS{\mathbf{S}}
\newcommand\bZ{\mathbf{Z}}

\newcommand\cH{\mathcal{H}}
\newcommand\cI{\mathcal{I}}
\newcommand\cJ{\mathcal{J}}
\newcommand\cK{\mathcal{K}}

\newcommand\cM{\mathcal{M}}
\newcommand\cN{\mathcal{N}}
\newcommand\cO{\mathcal{O}}
\newcommand\cP{\mathcal{P}}
\newcommand\cQ{\mathcal{Q}}
\newcommand\cR{\mathcal{R}}
\newcommand\cS{\mathcal{S}}

\newcommand\cY{\mathcal{Y}}

\newcommand\fg{\mathfrak{g}}

\newcommand{\bfbeta}{{\text{\boldmath$\beta$}}}
\newcommand{\bfgamma}{{\text{\boldmath$\gamma$}}}
\newcommand{\bfphi}{{\text{\boldmath$\phi$}}}
\newcommand{\bfpsi}{{\text{\boldmath$\psi$}}}

\newcommand\ald{{\dot{\alpha}}}
\newcommand\bed{{\dot{\beta}}}

\newcommand{\Qbt}{\bQ}

\newcommand{\f}[2]{\frac{#1}{#2}}
\newcommand{\comm}[2]{\left[{#1},{#2}\right]}
\newcommand{\acomm}[2]{\left\{{#1},{#2}\right\}}

\newcommand{\vev}[1]{\left\langle {#1} \right\rangle}
\newcommand{\bra}[1]{\left|{#1}\right\rangle}

\newcommand{\tbrack}[3]{\left[{#1},{#2},{#3}\right]_3}

\newcommand{\nn}{\nonumber}
\newcommand{\wti}[1]{\widetilde{#1}}

\NewDocumentCommand\lbracket{mmg}{%
    \ensuremath{\left\{ {#1}_{\lambda_1}{#2}\IfNoValueTF{#3}{\right\}_2}{_{\lambda_2}{#3}\right\}_3}}%    
}

\newcommand\SO{\mathrm{SO}}

\newcommand\UU{\mathrm{U}}
\newcommand\SU{\mathrm{SU}}

\renewcommand\sl{\mathfrak{sl}}

\newcommand\so{\mathfrak{so}}
\newcommand\su{\mathfrak{su}}

\newcommand\gl{\mathfrak{gl}}

\newcommand\uu{\mathfrak{u}}

\def\overleftrightarrow#1{\vbox{\ialign{##\crcr
	$\leftrightarrow$\crcr\noalign{\kern-0pt\nointerlineskip}
	$\hfil\displaystyle{#1}\hfil$\crcr}}}

\definecolor{darkgreen}{rgb}{0,0.5,0}
\definecolor{darkblue}{rgb}{0,0,0.7}
\definecolor{darkred}{rgb}{0.7,0,0}

\title{~\vspace{10mm}\\
\fontsize{20pt}{23pt}\selectfont\textbf{Unravelling the Holomorphic Twist: \\ Central Charges}\vspace{10mm}}

\author{\Large{Pieter Bomans\textsuperscript{\orcidlink{0000-0002-0907-9830}} and Jingxiang Wu\textsuperscript{\orcidlink{https://orcid.org/0000-0001-6867-1407}}}\\[10mm]
	\large Mathematical Institute, University of Oxford\\
	\normalsize Andrew Wiles Building, Radcliffe Observatory Quarter\\
    \normalsize Woodstock Road, Oxford, OX2 6GG, U.K.\\[5mm]
	\texttt{\normalsize\href{mailto:pieter.bomans@maths.ox.ac.uk}{pieter.bomans@maths.ox.ac.uk}, \href{mailto:jingxiang.wu@maths.ox.ac.uk}{jingxiang.wu@maths.ox.ac.uk}}\\
}

\date{}

\begin{document}

\pagestyle{empty}

\maketitle
\thispagestyle{empty}

\vspace{\stretch{1}}

\begin{abstract}
\noindent The holomorphic twist provides a powerful framework to study minimally protected sectors in supersymmetric quantum field theories. We investigate the algebraic structure underlying the holomorphic twist of $\cN=1$ superconformal field theories in four dimensions. In particular, in holomorphically twisted theories the flavour and conformal symmetry algebras are enhanced to infinite-dimensional higher Kac Moody and higher Virasoro symmetry algebras respectively. We explicitly compute the binary and ternary $\lambda$-brackets and clarify their relation with the underlying infinite-dimensional symmetry algebra. Doing so we show that the central extensions of said symmetry algebras precisely encode the conformal anomalies $a$ and $c$ as well as the flavour central charges of the physical four-dimensional theory. This parallels the familiar story in two dimensions where the conformal anomaly $c$ is encoded in the central extension of the Virasoro algebra.

\end{abstract}

\vspace{\stretch{3}}
 
\newpage

{
	\hypersetup{linkcolor=black}
    \setcounter{tocdepth}{2}
    \setlength{\parskip}{0.4em}
	\tableofcontents
}

\setcounter{page}{0}

\clearpage
\pagestyle{plain}

%%%%%%%%%%%%%%%%%%%%%%%%%%%
%   Summary of notation   %
%%%%%%%%%%%%%%%%%%%%%%%%%%%
%
% To denote full 4d superfields we use mathcal capital letters, while semichiral superfields are denoted with bold capital letters. Similarly the modes of semichiral operators as well as modes of 2d holomorphic operators are denoted by bold letters.
%
% Conserved currents are denoted by J, stress tensors by S or T, a generic superfield by O
% The modes of the stress tensor are denoted by L while the associated vector fields are denoted by l
%     
% By default all z and \bz indices are up and \lambda indices are down. Contraction using $\epsilon_{\alpha\beta}$ is denoted by $(-,-)$ where $\epsilon_{12} = -\epsilon^{12} = 1$. 
% For example, $(\lambda_1,\lambda_2) \equiv \lambda_{1\dotp} \lambda_{2\dotm}-\lambda_{1\dotm} \lambda_{2\dotp}$ and $(\bz,\dd \bz) \equiv \bz^\dotp \dd \bz^\dotm - \z^\dotm \dd \bz^\dotp$   
%     

%%%%%%%%%%%%%%%%%%%%%%%%%%%%%%%%%%%%%%%%%	
\section{Introduction and summary}      %
\label{sec:intro}                       %
%%%%%%%%%%%%%%%%%%%%%%%%%%%%%%%%%%%%%%%%%

The operator product expansion (OPE) lies at the heart of any non-perturbative treatment of quantum field theory. Despite extensive research endeavours, unravelling the detailed structure of the OPE in generic quantum field theories remains an elusive intellectual pursuit. This challenge is highlighted by the scarcity of tools accessible beyond the framework of perturbation theory. 

In order to resolve this impasse it proves fruitful to explore theories with additional symmetries. Indeed, the OPE is best understood in conformal field theories (CFTs) due to the extra constraints imposed by conformal invariance. This becomes most evident in two dimensions, where for a large class of theories, known as rational CFTs, the associativity of the OPE combined with the Virasoro symmetry algebra allows one to obtain exact solutions for the correlation functions of local operators \cite{Belavin:1984vu}. The key factor in achieving this success lies in the enhancement of the two-dimensional (global) conformal algebra to the infinite-dimensional Virasoro algebra. The presence of such infinite-dimensional algebras is tied to the two-dimensional setup, and at first sight, it is not clear how to lift these tools to higher dimensions.

A second situation in which the OPE is under better control arises in the context of supersymmetric quantum field theories (SQFTs). In four-dimensional $\cN=1$ theories, one can employ the power of holomorphy \cite{Seiberg:1994bp} and show that certain observables such as the partition function \cite{Closset:2013vra} or correlation functions in a protected sector of the theory, known as the chiral ring \cite{Lerche:1989uy}, only depend holomorphically on the complex structure moduli. In both examples, the holomorphicity in a protected sector greatly simplifies computations and allows for a precise and detailed understanding of the OPE.

These two developments were neatly combined in 4d $\cN=2$ superconformal field theories (SCFTs), where it was shown that the correlation functions in a subsector of the full theory behave precisely like those of a two-dimensional CFT \cite{Beem:2013sza}. By passing to the cohomology of a supercharge of the schematic form $\Qbt \simeq \cQ+\cS$ one obtains a vertex operator algebra (VOA) encoding the OPE of so-called Schur operators when restricted to a complex plane.\footnote{Alternatively, one can obtain this sector by performing a topological-holomorphic twist \cite{Kapustin:2006hi} followed by a $\Omega$-deformation in the topological directions \cite{Oh:2019bgz,Jeong:2019pzg,Bobev:2020vhe}.} This example demonstrates a first instance in which one can leverage the power of infinite-dimensional symmetry in a higher-dimensional context. 

In this work, our focus lies on four-dimensional $\cN=1$ SCFTs instead, where the supersymmetry algebra does not contain an appropriate supercharge needed to obtain a 2d VOA as its cohomology. However, as demonstrated in \cite{Gwilliam:2018lpo} and further expanded upon in \cite{Saberi:2019fkq,Budzik:2022mpd,Budzik:2023xbr}, one can obtain a similar infinite-dimensional symmetry algebra in four-dimensional $\cN=1$ QFTs.\footnote{Early hints of the presence of such infinite-dimensional symmetry algebras in $\cN=1$ theories were already described in \cite{Johansen:1994aw,Johansen:1994ij,Johansen:1994ud,Losev:1995cr,Losev:1996up,NekrasovThesis} as well as in \cite{Costello:2011np} and subsequent works.} The power of holomorphy can be seen in its full glory by passing to the cohomology of a single supercharge $\Qbt \simeq \cQ$. In the holomorphically twisted sector, the algebraic structure underlying the OPE is a four-dimensional generalisation of a VOA, a higher VOA \cite{faonteHigherKacMoodyAlgebras2019, hennionGelfandFuchsCohomologyAlgebraic2022}. Analogous to 2d CFTs, in the holomorphically twisted theory, the supersymmetry algebra is enhanced to an infinite-dimensional higher Virasoro or higher Kac-Moody algebra.

Operators in the holomorphically twisted theory are annihilated by a single supercharge and for that reason, we call them semi-chiral operators. The semi-chiral sector receives contributions from all chiral (semi-)short multiplets and therefore captures the same Hilbert space as the full superconformal index. In particular, in contrast to the Schur sector of $\cN=2$ SCFTs \cite{Chang:2023ywj}, this cohomology captures states dual to black holes\cite{Chang:2022mjp,Choi:2022caq,Choi:2023znd,Budzik:2023vtr,Chang:2023zqk}. Moreover, the construction of the holomorphic twist does not rely on conformal symmetry and therefore can be extended to any non-conformal $\cN=1$ QFT. Understanding the algebraic structure of this twisted sector, and exploiting this infinite-dimensional symmetry algebra therefore provides us with a window into obtaining a deeper understanding of various aspects of $\cN=1$ SCFTs such as confinement \cite{Budzik:2023xbr}, $a$-theorems, RG flows and the structure of black hole states in said theories.

This work aims to clarify the role of this infinite-dimensional symmetry algebra, and in particular relate observables in the holomorphically twisted theory to observables in the physical 4d $\cN=1$ theory. In particular, we show how the central extensions of the symmetry algebra encode the conformal anomalies and flavour central charges of the original theory and demonstrate this through a series of examples. 

\vspace{-8pt}\paragraph{Warm up: two-dimensional VOAs}~
%%%%%%%%%%%%%%%%%%%

The higher VOAs we encounter after performing the holomorphic twist are very similar to the familiar two-dimensional VOAs encountered in the context of two-dimensional CFTs. Before describing their more exotic four-dimensional cousins, let us briefly review some facts about VOAs in two dimensions. The material below is well-known but phased in a slightly unfamiliar way which generalises easily to higher dimensions. For more details on the two-dimensional setting see for example \cite{2015arXiv151200821K} and references therein.

The algebraic structure underlying a two-dimensional CFT is given by a vertex operator algebra. A VOA consists of the following data:\footnote{Additionally, the OPE has to satisfy appropriate locality axioms, which are crucial to guarantee convergence of the OPE.}
\vspace{-6pt}~
\begin{itemize}
    \item A vector space $V$; the state space.
    \item A linear map $Y$; the state-operator map,
    \begin{equation}
        Y\,\,:\,\, V\rightarrow (\text{End}\,V)[\![z]\!]\,\, : \,\, \bO \mapsto \bO(z) = \sum_{n\in \bZ} \frac{\left\{\bO\,,\bullet \right\}_n}{z^{n+1}}\,. 
    \end{equation}
    \item The translation operator $T$; a derivation satisfying the translation axiom $\left(T\bO\right)(z) = \partial_z \bO(z)$.
    \item The vacuum vector $\bra{0}\in V$; satisfying $\bra{0}(z) = \text{Id}_V$ and $\bO(z)\bra{0} = \bO + (T\bO)z + \cO(z^2)$.
\end{itemize}
A VOA is naturally equipped with an infinite number of products represented by the brackets $\left\{\bullet \,, \bullet \right\}_n$ in terms of which the operator product expansion can be expressed as
\begin{equation}
    \bO_1(z)\bO_2(w) = \sum_{n\in\mathbb{Z}} \frac{\left\{ \bO_1\,, \bO_2 \right\}_n(w)}{ (z-w)^{n+1}}\,.
\end{equation}
Conversely, the brackets $\left\{\bullet\,,\,\bullet\right\}_n$ provide us with a convenient way to unpack the OPE data of the VOA, which can be extracted as
\begin{equation}\label{eq:2d{ab}n}
    \left\{ \bO_1\,, \bO_2 \right\}_n(0) = \oint_{S^1} \frac{\dd z}{2\pi\ii} z^n \bO_1(z)\bO_2(0), \quad n \in \mathbb{Z}\,, 
\end{equation}
This equation has a very transparent physical meaning. Any holomorphic operator $\bO(z)$ can be thought of as a current for a global symmetry with conservation law $\bar{\partial} \bO(z) = 0$. This conservation law continues to hold when we insert any $\bar{\partial}$-closed function $\rho$ to construct another current $\rho\, \bO(z)$. Such functions are elements of the Dolbeault cohomology,
\begin{equation}
    \rho\, \in \, H^{0,0}_{\bar{\partial}} \left(\bbC\backslash \{0\}\right) = \bbC[z,z^{-1}]\,,
\end{equation}
and the collection of all holomorphic currents naturally forms an infinite-dimensional symmetry algebra.

Phrased differently, \eqref{eq:2d{ab}n} is nothing but the action of the topological line operator $\left\{\bO, \bullet\right\}_n$ constructed by integrating the current $z^n \bO(z)$ over the codim-1 manifold $S^1$. For any holomorphic operator, the topological line operators therefore arise in infinite towers labelled by $n\in \mathbb{Z}$.

The non-negative modes ($n\geq 0$) capture the singular part of the OPE. These modes can be conveniently collected in a generating function called the $\lambda$-bracket,
\begin{equation}
    \left\{{\bO_1 \,}_\lambda\, {\bO_2} \right\} \equiv \oint_{S^1} \frac{\dd z}{2\pi\ii} \e^{\lambda z} \bO_1(z)\bO_2(0) = \sum_{n\geq 0} \frac{\lambda^n}{n!}\left\{ \bO_1\,,\, \bO_2\right\}_n\,,
    \label{eq:2dlambdabracket}
\end{equation}
This $\lambda$-bracket naturally generalises to higher dimensional situations and will be a central object in this work. A vector space $V$, equipped with a $\lambda$-bracket and a translation operator $T$ defines a Lie conformal algebra. Any vertex operator algebra automatically defines a Lie conformal algebra simply by forgetting about the regular part of the OPE. To get more familiar with this structure we refer the interested reader to Appendix \ref{app:LieConformal} where we give some more explicit details on the Kac Moody algebra and Virasoro algebra as Lie conformal algebras.

\vspace{-8pt}\paragraph{Higher VOA from higher dimensional holomorphic field theories}~
%%%%%%%%%%%%%%%%%%%

After this two-dimensional intermezzo, we are ready to discuss the case of four-dimensional holomorphically twisted theories. After the twist operators only depend holomorphically on the coordinates of $\bbC^2$ and we can proceed entirely analogous to the two-dimensional case.

For any holomorphic operator $\bO(z_1,z_2, \dots, z_d)$ in $\bbC^d$, $\rho \, \bO$ continues to be holomorphic as long as $\rho$ is $\bar{\partial}$-closed. Analogous to the above we  naturally run into an infinite-dimensional symmetry algebra. Such $\bar{\partial}$-closed $\rho$ are elements of the Dolbeault cohomology,
\begin{equation}
    \rho \, \in\, H^{0,\bullet}_{\bar\partial} \left(\bbC^d\backslash \{0\}\right)\,.
\end{equation}
In higher dimensions, the Dolbeault cohomology $H^{0,\bullet}_{\bar\partial} \left(\bbC^d\backslash \{0\}\right)$ is known to concentrate on degree $(0,0)$ and $(0,d-1)$ and is given as
\begin{equation}
    H^{0,i}_{\bar\partial} \left(\bbC^d\backslash \{0\}\right)=\left\{\begin{array}{ll}
    \bbC\left[z_1, \ldots, z_d\right] & i=0\,, \\
    \bbC\left[\partial_1, \ldots, \partial_d\right]\,\omega_{\rm BM}\qquad& i=d-1\,,\\
    0 & \text{otherwise} \,,\\
\end{array} \right.
\end{equation}
where $\omega_{\rm BM}$ is the Bochner-Martinelli kernel introduced in more detail in Appendix \ref{app:Canalysis}. The degree $(0,0)$ cohomology provides us with a direct generalisation of the positive modes in two dimensions, cf. \eqref{eq:2d{ab}n},
\begin{equation}
    \left\{ \bO_1 \,, \bO_2 \right\}_{n_1,n_2,\dots, n_d}(0) = \oint_{S^{2d-1}} \frac{\dd^{d} z}{(2\pi\ii)^n} \, z_1^{n_1} z_2^{n_2} \cdots z_d^{n_d}\, \bO_1(z)\bO_2(0)\,, \qquad n_i \geq 0\,. \label{eq:4d{ab}np}
\end{equation}
Conversely, the negative modes can be generalised by picking $\rho$'s from the degree $(0,d-1)$ cohomology, 
\begin{equation}
    \left\{ \bO_1 \,, \bO_2 \right\}_{n_1,n_2,\dots, n_d}(0) = \oint_{S^{2d-1}} \frac{\dd^{d} z}{(2\pi\ii)^d}\, \partial_{z_1}^{-n_1-1} \partial_{z_2}^{-n_2-1} \cdots \partial_{z_d}^{-n_d-1} \omega_{\rm BM} \, \bO_1(z)\bO_2(0)\,, \qquad n_i \leq -1 \,. \label{eq:4d(ab)nm}
\end{equation}
Note that in two dimensions this definition precisely reproduces the negative modes in equation \eqref{eq:2d{ab}n}. Following the analogy with two dimensions, one can construct infinitely many codimension-one topological surface operators by integrating $\rho\, \bO_1$ over the $(2d-1)$-sphere, $S^{2d-1}$. For example, we can act on another local operator $\bO_2(0)$ by wrapping the topological surface operator on the $S^{2d-1}$ linking the origin,
\begin{equation}
    \oint_{S^{2d-1}} \frac{\dd^dz}{(2\pi \ii)^d}\   \rho\ \bO_1(z) \bO_2(0)\,.
\end{equation}
Collecting the non-negative modes \eqref{eq:4d{ab}np} we define the $\lambda$-bracket as the following generating function, 
\begin{equation}
    \left\{\bO_1 \,{}_\lambda\, \bO_2\right\} = \oint_{S^{2d-1}}\f{\dd^d z}{(2\pi \ii)^d}\e^{\lambda\cdot z}\bO_1(z)\bO_2(0)\,.
\end{equation}
Unlike in one complex dimension where the brackets generate a Lie conformal algebra and therefore satisfy the Jacobi identity, in higher dimensions, the brackets only satisfy the Jacobi identity homotopically. Higher homotopies are captured by a collection of higher ($\lambda$-)brackets. The definitions of the higher brackets are well-known in $E_d$ algebra using the integration cycles in the configuration space of points \cite{Beem:2018fng}. While we expect the same is true for holomorphic theories, it is impractical to deal with configuration space of more than three points. For this reason we follow an alternative definition of the higher brackets as proposed in \cite{Budzik:2022mpd, Budzik:2023xbr}, which we carefully introduce in Section \ref{subsec:inf-dim-sym}. The collection of all local operators together with the $n$-ary ($\lambda$-)brackets form a holomorphic factorisation algebra, which can be thought of as providing a free field realisation of the higher-VOA on $\bbC^d$.

\vspace{-8pt}\paragraph{Summary of the results}~
%%%%%%%%%%%%%%%%%%%

This higher VOA structure naturally arises in the context of holomorphically twisted four-dimensional $\cN=1$ theories \cite{Gwilliam:2018lpo,Saberi:2019fkq,Budzik:2022mpd,Budzik:2023xbr}. In this case, the holomorphic operators of interest are given by semi-chiral superfields $\bO$, which are in one-to-one correspondence with $\Qbt$-cohomology classes. In particular, whenever the theory preserves some flavour symmetry, the twisted theory contains an associated semi-chiral superfield $\bJ$. Similarly, any (local) unitary $\cN=1$ SCFT contains a stress tensor which gives rise to a semi-chiral superfield $\bS_\ald$ in the twisted theory. 

In two-dimensional VOAs, it is well-known that the modes of a conserved current $\bJ_n = \left\{ \bJ\,, \bullet \right\}_n$ and the modes of the holomorphic stress tensor $\bL_n = \left\{ \bS\,, \bullet \right\}_{n+1}$ generate respectively a Kac-Moody algebra and Virasoro algebra. The central charges of these algebras encode important physical information about the theory. The Virasoro central charge $c$ encodes the conformal anomaly of the 2d CFT, while the central charge of the Kac-Moody algebra equals the flavour central charge $k_G$. In the two-dimensional context, these central charges arise in the most singular terms of respectively the $\bS \bS$ or $\bJ \bJ$ OPE, where $\bS$ and $\bJ$ denote the holomorphic stress tensor and the holomorphic current. Similarly, these central charges can be read off from the $\lambda$-brackets
\begin{equation}
    \left\{ \bS\,{}_\lambda \, \bS \right\} =\lambda^3 \f{c}{12} + \left(  2\lambda + \partial \right) \bS\,, \qquad\qquad \left\{ \bJ^a\,{}_\lambda \, \bJ^b \right\} = f^{ab}{}_c \,\bJ^c +  \delta^{ab}k_G \lambda \,.
\end{equation}
In four-dimensional holomorphically twisted theories the modes of conserved current and stress tensor semi-chiral superfields are given by $\bJ^a_{m,n} = \left\{ \bJ^a\,, \bullet \right\}_{m,n}$ and $\bL^\ald_{m,n} = \left\{ \bS_{\ald}\,, \bullet \right\}_{m+1,n+1}$. Analogous to the two-dimensional setup their modes now generate a higher Kac-Moody and higher Virasoro symmetry algebra. It is then natural to expect that the central charges of these higher VOAs encode physical information about the original untwisted theory. We make this expectation precise by explicitly computing binary and ternary $\lambda$-brackets of conserved current and stress tensor semi-chiral superfields. The binary $\lambda$-brackets are given by
\begin{equation}\label{eq:intlam1}
    \left\{ \bS_\ald \,{}_\lambda \bS_\bed \right\} = \lambda_\ald \bS_\bed + \lambda_\bed\, \bS_\ald + \partial_\ald \bS_\bed  \,, \qquad \qquad \left\{ \bJ^a \,{}_\lambda\, \bJ^b \right\} = f^{ab}{}_c\, \bJ^c \,,
\end{equation}
where $f^{ab}{}_c$ are the anti-symmetric structure constants of the flavour symmetry algebra. Unlike in two dimensions, these expressions do not contain any information about the central charges. However, the main result of this paper lies in the computation of the ternary $\lambda$-brackets. The $\lambda$-bracket for three stress tensor semi-chiral superfields is
\begin{equation}\label{eq:intlam2}
    \left\{ \bS_\ald \,{}_{\lambda_1} \bS_\bed \,{}_{\lambda_2} \bS_{\dot\gamma} \right\} = \f43\, (3c - 2a) \,\Lambda_{\cN=4} + 16 \,(c - a)\, \Lambda_{\rm vector}  \,,
\end{equation}
where $\Lambda_{\cN=4}$ and $\Lambda_{\rm vector}$ are theory-independent kinematic factors which are defined in the main text in \eqref{eq:Lambdaexpression}. The coefficients in this expression are precisely given by the $a$ and $c$ conformal anomalies of the untwisted theory! We compute this $\lambda$-bracket in all generality as an integrated three-point function of (descendants of) stress tensors of the original untwisted theory and check our results in a wide variety of examples. Similarly, we compute the $\lambda$-bracket of the conserved current semi-chiral superfields resulting in the following expression,  
\begin{equation}\label{eq:intlam3}
    \left\{ \bJ^a \,{}_{\lambda_1}\, \bJ^b \,{}_{\lambda_2}\, \bJ^c \right\} = k_G\, d^{abc} (\lambda_1, \lambda_2)\,.
\end{equation}
In this equation $d^{abc}$ are the symmetric structure constants of the flavour symmetry group while $k_G$ encodes the flavour central charge of the untwisted theory. These results therefore present a clear connection between the central charges in the holomorphic theory obtained after performing the holomorphic twist and observables in the original untwisted theory.

\vspace{-8pt}\paragraph{Outlook}~
%%%%%%%%%%%%%%%%%%%

The study of the holomorphic twist is still in its early infancy and many interesting questions remain unanswered. Although mathematical results are steadily emerging (see \cite{Elliott:2020ecf,Saberi:2019ghy,Costello:2016vjw,Costello:2021jvx,Safronov:2021tsf,Gwilliam:2022vja,Garner:2023wrc} for a necessarily incomplete selection) it remains a challenge to connect observables in the twisted theory to physical observables in the untwisted theory such as correlation functions of local operators. Our work presents a first step towards better understanding the physics behind the holomorphic twist and opens a window into making progress on a variety of fronts.

The first direction to make progress lies in better understanding the $L_\infty$ structure of the modes and its representation theory. In this work we compute $\lambda$-brackets but the map to the ternary $L_\infty$ 3-bracket is not immediate \cite{Zeng:2023qqp}.
We plan to come back to this question in the near future. A more challenging question is to understand the $L_\infty$ modules and null vectors. A similar understanding in two dimensions gives rise to the discovery of minimal models \cite{Belavin:1984vu} and allows one to exactly solve the OPE. The discovery of a similar structure in the holomorphic twist would present a major breakthrough in our understanding of supersymmetric quantum field theories.

Although in this work we focus on conformal theories, many aspects of our story continue to hold for non-conformal $\cN=1$ SQFTs. In this case, the semi-chiral superfield corresponding to the stress tensor is no longer given by $\bS_\ald$ but rather by $\wti\bS=\partial^\ald \bS_\ald$. The modes of this semi-chiral superfield generate an infinite dimensional symmetry algebra distinct from the higher Virasoro algebra. It would be interesting to better understand the structure of this algebra and find out in which shape a similar relation between conformal anomalies and central charges persists in this setup. Understanding this mechanism would provide us with an elegant framework to prove for example the $a$-theorem \cite{Komargodski:2011vj} in a similar fashion as the original $c$-theorem in two-dimensions \cite{Zamolodchikov:1986gt} and provide us with a deeper understanding of RG flows in $\cN=1$ SCFTs.

From the structure of the $\lambda$-brackets in \eqref{eq:intlam1}-\eqref{eq:intlam3} we observe a striking similarity with the supersymmetric Casimir energy obtained by integrating the anomaly polynomial \cite{Bobev:2015kza}. It would be interesting to better understand the origin of this similarity and relate the supersymmetric Casimir energy to the symmetry algebras discussed in this work. In addition to the flavour central charges discussed in this work the supersymmetric Casimir energy also contains the $k_{FFR}$ and $k_{FRR}$ anomalies, which can be obtained by computing the three-point functions of resp. two/one flavour current and one/two stress tensors analogous to the computations performed in this work. It would be interesting to explicitly compute these from the holomorphically twisted theories study the combined structure formed by the higher Virasoro and Kac-Moody symmetry algebras. Another observation is that the term in front of $\Lambda_{\cN=4}$ is precisely the combination that is manifestly positive in unitary theories through the Hofman-Maldacena bound \cite{Hofman:2008ar}. This coincidence hints that it might be possible to re-prove this bound directly in the holomorphically twisted theory and could shed more light on the structure of unitary representations of said symmetry algebras.

Finally, holomorphic theories arise in a variety of other examples. For example, minimal twists of six-dimensional theories with at least $\cN = (1,0)$ supersymmetry give rise to six-dimensional higher VOAs. Increasing the supersymmetry to six-dimensional $\cN=(2,0)$ SCFTs, there is an alternative twist analogous to $\cQ+\cS$ twist in four dimensions \cite{Beem:2013sza,Beem:2014kka,Beem:2014rza} giving rise to a two-dimensional VOA. On the other hand, for six-dimensional $\cN=(1,1)$ SQFTs there is holomorphic-topological twist \cite{Elliott:2020ecf} giving rise to an associated four-dimensional higher VOA. It would be very interesting to better understand this unexplored protected sector and the physical observables it encodes.

\vspace{-8pt}\paragraph{Structure of this paper}~
%%%%%%%%%%%%%%%%%%%

The remainder of this paper is organised as follows. In Section \ref{sec:holotwist}, we present the holomorphic twist of four-dimensional $\cN=1$ QFTs, discuss the resulting infinite-dimensional symmetry algebras and carefully define the $\lambda$-brackets. Subsequently, in section \ref{sec:lbracket}, we compute the binary and ternary $\lambda$-bracket of stress tensors and conserved currents and relate them to the conformal anomalies and flavour central charges of the original theory. In section \ref{sec:examples}, we conclude with a series of examples demonstrating the general formulae derived in the previous sections. Appendix \ref{app:4dN1SCA} summarises our conventions and notation and introduces the $\cN=1$ superconformal algebra. Appendices \ref{app:Canalysis}-\ref{app:LieConformal} contain various details omitted in the main text.

%%%%%%%%%%%%%%%%%%%%%%%%%%%%%%%%%%%%%%%%%%%%%%%%%%%%%%%%%%%%
\section{Infinite-dimensional symmetry in four dimensions} %
\label{sec:holotwist}                                      %
%%%%%%%%%%%%%%%%%%%%%%%%%%%%%%%%%%%%%%%%%%%%%%%%%%%%%%%%%%%%

Consider a four-dimensional $\cN=1$ superconformal theory in Euclidean space.\footnote{For most of this work we specialise to superconformal theories but the twisting procedure does not demand superconformal symmetry and many of our results continue to hold in a non-conformal setup with minimal modifications.} The superconformal algebra contains (at least) two spinor supercharges $\cQ_\alpha$ and $\wti\cQ_{\dot\alpha}$, subject to the following commutation relations,\footnote{For the complete set of commutation relations we refer the reader to Appendix \ref{app:4dN1SCA}.}
\begin{equation}
    \acomm{\cQ_\alpha}{\wti \cQ_{\dot\alpha}} = \cP_{\alpha\dot\alpha}\,,
\end{equation}
where $\cP_{\alpha\ald}$ are the translation generators and $\alpha,\dot\alpha = \pm$ are fundamental $\SU(2)_L\times \SU(2)_R \simeq \SO(4)$ indices. In this work we are interested in the holomorphic twist\footnote{We take a modern point of view where twisting is defined by passing to the cohomology of a nilpotent supercharge. See Appendix F of \cite{Budzik:2023xbr} for a disambiguation of the terminology.} with respect to a nilpotent supercharge $\Qbt$, obtained by passing to the $\Qbt$-cohomology. Without loss of generality, we pick $\Qbt = \cQ_-$ as our twisting supercharge. This choice selects a complex structure such that $\left\{z^{\dot\alpha} , \bz^{\dot\alpha}\right\} = \left\{x^{+\dot\alpha},x^{-\dot\alpha}\right\}$ are (anti-)holomorphic coordinates on $\bbC^2$.\footnote{In contrast to \cite{Budzik:2023xbr}, we follow standard conventions and refrain from interchanging dotted and dotless indices.} Moreover, we immediately notice that 
\begin{equation}
    \acomm{\Qbt}{\wti\cQ_{\dot\alpha}} = \partial_{\bz^{\dot\alpha}}\,,
\end{equation}
and therefore the twisted theory is holomorphic in the cohomological sense. As already alluded to in the introduction, this holomorphic theory behaves in many ways similar to a two-dimensional vertex operator algebra. We follow this analogy and describe how an infinite-dimensional symmetry algebra emerges in this protected sector and define infinitely many $n$-ary operations encoding the symmetry algebra. 

The underlying vector space of this higher VOA is defined as the $\Qbt$-cohomology of local operators. The inclusion of non-local operators in this framework presents an interesting extension we hope to address in future work. Since operators in the cohomology are annihilated by one chiral supercharge we call them semi-chiral operators. The harmonic representatives of the cohomology are characterised by the following condition on the quantum numbers 
\begin{equation}
    \acomm{\Qbt}{\Qbt^\dagger} = \Delta + \f32 r + \f12 j_L = 0\,,
\end{equation}
where $\Delta$ is the conformal dimension, $r$ is the $r$-charge and $j_L$ denotes the eigenvalue of the Cartan generator of $\SU(2)_L$. In superconformal theories, the operators are grouped in superconformal multiplets where short multiplets satisfy certain shortening conditions.\footnote{We follow the conventions of \cite{Cordova:2016emh} to label superconformal multiplets.} One can easily see that cohomology classes are in one-to-one correspondence with multiplets with a chiral shortening condition. In more detail:
\begin{itemize}
    \item \underline{$A_{i=1,2}$ shortening}  (where $i=2$ if and only if $j_L=0$): An $A_i\overline X [j_L,j_R]^{(r)}_\Delta$ multiplet contains the following semi-chiral operator,
        \begin{equation}
            \bO^{(0)}_{\dot\alpha_1\cdots\dot\alpha_{\bar j}} = D_+ \cO_{+\cdots+\dot\alpha_1\cdots\dot\alpha_{\bar j}}\Big|_{\theta_i=\bar\theta_i = 0}\,,
        \end{equation}
        where $\cO$ is the corresponding superfield and $\overline X$ can either be long or denote an anti-chiral shortening condition.
    \item  \underline{$B_1$ shortening:} A $B_1 \overline X [0,j_R]^{(r)}_\Delta$ multiplet contains the following semi-chiral superfield,
        \begin{equation}
            \bO^{(0)}_{\dot\alpha_1\cdots\dot\alpha_{\bar j}} = \cO_{\dot\alpha_1\cdots\dot\alpha_{\bar j}}\Big|_{\theta_i=\bar\theta_i = 0}\,.
        \end{equation}
\end{itemize}
Note that only for the $B_1$-type shortening condition the semi-chiral operator is given by a superconformal primary, the subsector of such operators forms the chiral ring \cite{Lerche:1989uy}. In the twisted theory, the barred fermionic coordinates $\bar\theta^{\dot\alpha}$ transform as anti-holomorphic one-forms and it will be useful to replace them with the differentials $\dd \bz^{\dot\alpha}$. Doing so we introduce the reduced superfield, 
\begin{equation}
    \bO = \e^{\dd\bz^{\dot\alpha}\wti\cQ_{\dot\alpha}} \bO^{(0)} = \bO^{(0)}+\bO^{(1)}+\bO^{(2)}\,,
\end{equation}
Following \cite{Budzik:2023xbr} we call such a superfield a semi-chiral superfield\footnote{See Appendix \ref{app:4dN1SCA} for our superspace conventions and the definition of $D_-$.} if it satisfies 
\begin{equation}
    D_- \bO = \left( \Qbt + \bar\partial \right)\bO = 0\,, \label{eq:semichiraldefinition}
\end{equation}
where $\bar\partial = \frac{\partial}{\partial \bz^{\ald}} \dd \bz^{\ald}$. If $\bO^{(0)}$ is a scalar operator, its descendants $\bO^{(i)}$ are $(0,i)$ forms on $\bbC^2$.
For any semi-chiral operator $\bO^{(0)}$, the higher-form terms satisfy the holomorphic descent relation,
\begin{equation}
    \Qbt \bO^{(k)} + \bar\partial \bO^{(k-1)} = 0 \,.
\end{equation} 
In particular, the first relation $\bar\partial \bO^{(0)} = \Qbt \bO^{(1)}$ states that $\bO^{(0)}$ is a holomorphic operator up to $\Qbt$ exact terms. Note that similar to chiral ring operators, the OPE between two semi-chiral operators is completely regular. This is a consequence of Hartog's theorem, which essentially states that there cannot be isolated singularities on $\bbC^2$. Note however that the OPE between a semi-chiral operator and its descendants, or between two descendants, can be singular. In order to observe more refined algebraic structures in the holomorphically twisted theory it is therefore imperative to carefully keep track of the descendants. Doing so will allow us to extract a plethora of exact detailed OPE data. 

%----------
\subsection{Infinite-dimensional symmetry and the \texorpdfstring{$\lambda$}{lambda}-bracket} \label{subsec:inf-dim-sym}
%----------

The main upshot of the previous discussion is that the holomorphic twist, i.e. the semi-chiral sector (i.e. the $\tfrac{1}{4}$-BPS sector) of any $\cN=1$ supersymmetric theory, can be described by a holomorphic field theory on $\bbC^2$. In particular, the descent relations satisfied by the components of a semi-chiral superfield can be interpreted as conservation laws in the twisted theory. 

Each semi-chiral operator $\bO_1$ gives rise to infinitely many conservation laws. Given any $\bar\partial$-closed $(0,\bullet)$ form $\rho$, the operator $\rho\,\bO_1$ is also conserved. More generally, every operator $\bO_1(z)$ in the twisted theory gives rise to infinitely many symmetry generators $\{ \bO_1\,, \bullet \}_{m,n}$, for $m,n\in \bbZ$, acting on operators at the origin as
\begin{equation}
    \left\{\bO_1 \,, \bO_2 \right\}_{m,n}(0) = \oint_{S^3} \dd^2z \,\rho_{m,n} \ \bO_1(z)\bO(0)\,.
\end{equation}
As already explained in the introduction, the forms $\rho$ are elements of the Dolbeault cohomology $H^{0,\bullet}\left(\bbC^2\backslash \{0\}\right)$, which is known to concentrate in degree $(0,0)$ and $(0,1)$ \cite{Saberi:2019fkq,Gwilliam:2018lpo},
\begin{equation}
    \rho_{m,n} = 
    \left\{\begin{array}{lll}
    (z^\dotp)^m(z^\dotm)^n\,& \in \Omega^{0,0}\left(\bbC^2\backslash \{0\}\right) & m,n \geq 0 \,, \\
    \partial_{\dotp}^{-m-1}\partial_{\dotm}^{-n-1} \omega_{BM}(z)\qquad & \in \Omega^{0,1}\left(\bbC^2\backslash \{0\}\right)\qquad & m,n\leq -1\,,\\
    \phantom{asdf} 0 &  & \text{otherwise}\,.
    \end{array}\right.
\end{equation}
where $\omega_{\rm BM}(z)$ is the Bochner-Martinelli kernel,
\begin{equation}
    \omega_{\rm BM}(z) = 
    \frac{1}{(2\pi \ii)^2}\frac{(\bz,\dd\bz)}{|z|^4} \in \Omega^{0,1}\left(\bbC^2\backslash \{0\}\right)\,,
\end{equation}
and we used the shorthand notation $\partial_\ald=\f{\partial}{\partial z^{\ald}}$ and similarly $\bar\partial_\ald=\f{\partial}{\partial \bz^{\ald}}$. Analogous to two dimensions, the positive modes encode the singular part of the OPE of $\bO_1$ with another operator $\bO_2$, while the negative modes capture the regularised products $\partial_{\dotp}^m\partial_{\dotm}^n\bO_1 \,\bO_2$. It is convenient to collect all the positive modes in a $\lambda$-bracket,
\begin{equation}
    \left\{\bO_1\, {}_{\lambda}\, \bO_2 \right\} = \oint_{S^{3}} \frac{\dd^2 z}{(2\pi i)^2}\   e^{\lambda\cdot z} \, \bO_1(z) \bO_2(0), \label{eq:binarylambdabracket}
\end{equation}
where $\lambda \cdot z = \lambda_\ald z^\ald$. Unlike in one complex dimension where the $\lambda$-bracket satisfies a Jacobi identity, in higher dimensions the Jacobi identity is only satisfied homotopically.\footnote{The simplest examples of such structures arise in $L_\infty$ algebras which are a generalization of Lie algebras. For a brief summary of the relevant properties of $L_\infty$ algebras see Appendix \ref{app:Linfty}. In two dimensions, the triplet $\left(V,T,\left\{\bullet\,{}_\lambda\,\bullet\right\}\right)$ satisfies the axioms of a Lie conformal algebra. In higher dimensions, with the addition of higher brackets one could call the triplet $\left(V,T,\left\{\bullet\,{}_{\lambda_1}\,\cdots\bullet{}_{\lambda_{i-1}}\,\bullet\right\}_{i=2,3,\dots}\right)$ an L$_\infty$ conformal algebra. See related \cite{wang2022nlie}.} Higher homotopies are captured by the higher brackets which roughly capture the extent to which the Jacobi identity is violated. The resulting algebras are holomorphic generalizations of $E_d$, or $d$-disc, algebras and the factorisation algebras discussed in \cite{Costello:2011np,Costello:2016vjw,Costello:2021jvx,Williams:2018ows,Beem:2018fng} in which case the higher $n$-ary brackets are defined through integrals (over specific integration cycles) in the configuration space of $n$ points. While the same is expected to hold for the holomorphic theories under discussion, it is prohibitively impractical to deal with configuration space of more than three points. For this reason, in perturbation theory, it is more convenient to use an alternative definition of $(n+1)$-ary $\lambda$-brackets as proposed in \cite{Budzik:2022mpd, Budzik:2023xbr}.\footnote{The equivalence of this definitions with the definition through integrals over the configuration space of $n$-points is non-trivial and will be explained in more detail in \cite{ourpaper-higher}.}
\begin{equation}\label{eq:nlambdabracket}
    \left\{\bO_1 \, {}_{\lambda_1}\,  \cdots \,\bO_n \, {}_{\lambda_n}\, \bO_0 \right\}\equiv  \Qbt  \left( \prod_{k=1}^n \int_{\bbC^{2}}  \frac{d^2 z_k}{(2 \pi i)^2}e^{\lambda_k \cdot z_k} : \bO_1(z_1, \bar z_1)\cdots  \bO_n(z_n, \bar z_n) \bO_0(0,0):\right) \,,
\end{equation}
where the operators $\bO_i$ are semi-chiral superfields and $::$ refers to the operation of performing all possible Wick contractions. All fields that are not Wick contracted will be collected and shifted to the origin so that the right-hand side of \eqref{eq:nlambdabracket} is a Feynman integral multiplying an operator inserted at origin and never involves integrating operators. See \cite{Budzik:2022mpd} for more a careful explanation of this definition.

For $n=1$, it is clear that this definition coincides with the definition of the binary $\lambda$-bracket as in \eqref{eq:binarylambdabracket} by employing the defining property of semi-chiral superfields \eqref{eq:semichiraldefinition},
\begin{equation}
    \begin{aligned}
        \left\{\bO_1 \, {}_{\lambda_1}\, \bO_0 \right\} &\equiv \Qbt      \int_{\bbC^{2}} \frac{\dd^2 z_1 }{(2 \pi \ii)^2}e^{\lambda_1 \cdot z_1} : \bO_1(z_1)  \bO_0(0) : =  \int_{\bbC^{2}}  \frac{\dd^2 z_1}{(2 \pi \ii)^2} e^{\lambda_1 \cdot z_1} \bar\partial : \bO_1(z_1)  \bO_0(0) : \\
        & = \oint_{S^{3}} \frac{\dd^2 z_1}{(2\pi \ii)^2}\   e^{\lambda_1\cdot z_1} \ :\bO_1(z_1) \bO_0(0):\,, 
    \end{aligned}
\end{equation}
where the third equality follows from Stokes's theorem. Continuing to $n=2$, the ternary $\lambda$-bracket is given by
\begin{equation}\label{eq:3lambdabracket}
    \left\{\bO_1 \, {}_{\lambda_1}\, \bO_2 \, {}_{\lambda_2}\, \bO_0 \right\}\equiv  \Qbt   \int_{(\bbC^{2})^2} \frac{\dd^2 z_1 \dd^2 z_2}{(2\pi\ii)^4}\e^{\lambda_1 \cdot z_1+\lambda_2 \cdot z_2} :\bO_1(z_1)\bO_2(z_2)  \bO_0(0) : \,.
\end{equation}
Similar to the $n$-ary $\lambda$-brackets defined above, one can define higher brackets capturing the negative modes \cite{ourpaper-higher}. We leave a detailed discussion of such brackets for future work. 

Two important instances of ternary $\lambda$-brackets that play a central role in this work are those containing three conserved current semi-chiral operators, $\bJ^a$, or three stress tensors semi-chiral fields, $\bS_{\dot\alpha}$. In free theories, or free field realisations, both these operators are quadratic in free fields. Hence, all fields are Wick contracted in \eqref{eq:3lambdabracket} and the resulting expression consists of a Feynman integral multiplying the identity operator. In this case the expression $\Qbt :\bO_1(z_1)\bO_2(z_2)  \bO_0(0):$ simply reduces to the anti-holomorphic differential $\bar\partial$ acting on a three-point correlation function. For this reason, we can rewrite the ternary $\lambda$-bracket as
\begin{equation}\label{eq:ltreebracket}
      \left\{\bO_1 \, {}_{\lambda_1}\, \bO_2 \, {}_{\lambda_2}\, \bO_0 \right\}\equiv  \int_{(\bbC^{2})^2} \frac{\dd^2 z_1 \dd^2 z_2}{(2\pi\ii)^4}\e^{\lambda_1 \cdot z_1+\lambda_2 \cdot z_2}  \bar\partial \langle \bO_1(z_1)\bO_2(z_2)  \bO_0(0) \rangle \mathbb{1}\,,
\end{equation}
where we explicitly included the identity operator $\mathbb{1}$ to emphasize the fact ternary $\lambda$-brackets $\{\bJ^a \, {}_{\lambda_1}\, \bJ^b \, {}_{\lambda_2}\, \bJ^c \}$ and $\{\bS_{\dot\alpha} \, {}_{\lambda_1}\, \bS_{\dot\beta} \, {}_{\lambda_2}\, \bS_{\dot\gamma} \}$ are both proportional to the identity operator with coefficients being functions of $\lambda$ parameters. In particular, they are central and encode the central extensions of the higher VOA. More generally, this is not always the case and ternary $\lambda$-brackets involving other operators generically contain terms other than the identity.

The goal of this work is to compute these $\lambda$-brackets in interacting $\cN=1$ superconformal field theories, possibly with additional flavour symmetry $\fg$. It is a non-trivial statement that in an interacting theory \eqref{eq:ltreebracket} still holds when inserting three supercurrent semi-chiral operators. We will use this as the definition for the $\lambda$-bracket of three conserved (super)currents, in particular, this definition satisfies the axioms of higher $\lambda$-brackets, see Appendix \ref{app:LieConformal}. Before explicitly computing the ternary $\lambda$-brackets we discuss some general aspects of the expected higher Kac-Moody and higher Virasoro symmetry algebras arising in the holomorphically twisted theory. More details can be found in Appendix \ref{app:LieConformal}.

%----------
\subsection{Higher Kac-Moody algebra}
\label{subsec:2KacMoody}
%----------

In a two-dimensional VOA with continuous global symmetry $\fg$, the symmetry is enhanced to an infinite-dimensional symmetry algebra known as the Kac-Moody algebra (or the Heisenberg algebra, when $\fg = \uu(1)$). For any such symmetry, the theory contains conserved holomorphic currents $\bJ^a(z)$, where $a$ is an adjoint $\fg$ index. The Kac-Moody algebra is then generated by the following action on local operators\footnote{
For more details on the two-dimensional Kac-Moody algebra see Appendix \ref{app:LieConformal}.}
\begin{equation}
    \bJ^a_n \cdot O(0) = \oint\frac{\dd z}{2\pi \ii} z^n \bJ^{a}(z) \bO(0), \quad n \in \mathbb{Z}\,.
\end{equation}
Similarly, in four-dimensional $\cN=1$ theories, whenever the theory enjoys a continuous global symmetry $\fg$, we find an infinite-dimensional symmetry enhancement in the twisted theory. Every such continuous global symmetry necessarily comes with an associated short current multiplet $A_2\overline A_2 [0,0]_2^{(0)}$ in the spectrum. As discussed above, this multiplet contains a semi-chiral primary $\bJ = D_+ \cJ$, where $\cJ$ denotes the current superfield. Analogous to the two-dimensional setting, the modes of $\bJ$ give rise to a higher Kac-Moody symmetry algebra in the holomorphically twisted theory. In this case, the modes can be defined as
\begin{equation}
    \bJ^a_{m,n}\cdot \bO(0) \equiv  \left\{ \bJ^a\,\,,\bO \right\}_{m,n} =  \Qbt \int_{\bbC^2} \dd^2 z \,\rho_{m,n}\, \bJ^{a}(z) \bO(0)\,, \qquad m,n \in \mathbb{Z}\,.
\end{equation}
We can obtain the commutation relations of the higher Kac-Moody algebra through the associativity relations of the $\lambda$-bracket,
\begin{equation}\label{eq:l3associativity}
    \left\{\bJ^a  \, {}_{\lambda_1} \left\{ \bJ^b\, {}_{\lambda_2} \,\bO \right\}\right\} - \left\{\bJ^b  \, {}_{\lambda_2} \left\{ \bJ^a \, {}_{\lambda_1} \,\bO \right\}\right\} - \left\{\left\{ \bJ^a\, {}_{\lambda_1} \,\bJ^b \right\} \, {}_{\lambda_1+\lambda_2} \bO   \right\}=0\,.
\end{equation}
The binary $\lambda$-bracket, $\left\{\bJ^a\,{}_{\lambda}\,\bJ^b\right\}$ can be easily computed, as will be done explicitly in the next section. However, note that $\{\bJ^a \, {}_{\lambda}  \bJ^b\}$ has to be $\SU(2)$ invariant as well as anti-symmetric in the adjoint $\fg$ indices $a,b$. Therefore the only possible expression for the $\lambda$-bracket one can write down is
\begin{equation}\label{eq:JJ2bracket}
     \left\{ \bJ^a\,{}_{\lambda}\,\bJ^b \right\} =  f^{ab}{}_c \bJ^c \,,
\end{equation}
where $f^{ab}{}_c$ are the structure constants of the flavour symmetry group $\fg$ and the normalisation coefficient on the right-hand side has been absorbed in the normalisation of the current operator. Expanding both sides of the associativity relation \eqref{eq:l3associativity} in $\lambda_{1, \dot\pm}$ and $\lambda_{2,\dot\pm}$ we find the commutation relations,
\begin{equation}
    \left[\bJ_{m,n}^a\,,\bJ_{kl}^b\right] = f^{ab}{}_c\, \bJ^c_{m+k,n+l}\,.
\end{equation}
The ternary $\lambda$-bracket on can similarly be constrained through its symmetries. First of all, the only $\SU(2)$ invariant polynomials of $\lambda_{1,\dot\pm}$ and $\lambda_{2,\dot\pm}$ are given by powers of the pairing $(\lambda_1,\lambda_2)$. Moreover, the graded commutativity of the $\lambda$-bracket,
\begin{equation}
    \left\{ \bJ^a \,{}_{\lambda_2}\, \bJ^b \,{}_{\lambda_1}\, \bJ^c \right\} = - \left\{ \bJ^b \,{}_{\lambda_1}\, \bJ^a \,{}_{\lambda_2}\,  \bJ^c \right\} =  -\left\{ \bJ^a \,{}_{\lambda_2}\, \bJ^c \,{}_{-\lambda_1-\lambda_2-\partial}\,\bJ^b \right\}\,
\end{equation}
constrains this to a single possible expression,
\begin{equation}\label{eq:JJJ3brack}
     \left\{\bJ^a \,{}_{\lambda_1}\, \bJ^b \,{}_{\lambda_2}\, \bJ^c \right\} = k\, d^{abc} (\lambda_1,\lambda_2)\,,
\end{equation}
where $d^{abc}$ is the totally symmetric invariant tensor on $\fg$ and the coefficients $k$ is a theory dependent constant. In the next section, we will show that $k$ is given by the flavour central charge $k=k_G$. 

In contrast to the two-dimensional Kac-Moody symmetry algebra, the higher Kac-Moody algebra has a non-trivial ternary $L_\infty$-bracket due to the non-trivial ternary $\lambda$-brackets, encoding its central extensions. The ternary $L_\infty$-bracket is given by
\begin{equation}\label{eq:3commutatorJJJ}
    \left[\bJ_{n,m}^a, \bJ_{k,l}^b,  \bJ_{r,s}^c\right]= K \,d^{abc}(k s-\ell r) \delta_{ k+r+n,0} \delta_{\ell+s+m,0} \,,
\end{equation}
where $K$ is proportional to $k_G$. This ternary bracket should again be related to the ternary $\lambda$-bracket but the precise relationship is not immediate. A proper derivation including the relation between $k$ and $K$ is left to the future. 

%----------
\subsection{Higher Virasoro algebra}
\label{subsec:2Virasoro}
%----------

In a two-dimensional (conformal) VOA, the conformal $\sl(2)$ symmetry is enhanced to the infinite-dimensional Virasoro symmetry algebra. The theory contains a holomorphic stress tensor $\bS(z)$ whose modes generate holomorphic diffeomorphisms and act on local operators as\footnote{In Appendix \ref{app:LieConformal} we review the two-dimensional Virasoro algebra in more detail as well as list the standard commutation relations.}
\begin{equation}
    \bL_n \cdot \bO(0) = \oint\frac{\dd z}{2\pi \ii} z^n \bS(z) \bO(0), \quad n \in \mathbb{Z}\,.
\end{equation}
Similarly, in an $\cN=1$ superconformal field theory the supercurrent multiplet is a short $A_1\overline{A_1}[1,1]_3^{(0)}$ multiplet containing a semi-chiral operator $\bS_\ald \equiv D_{+}\cS_{+\ald}$, where $\cS_{\alpha\ald}$ denotes the supercurrent superfield. Following the discussion above we can define the modes of $\bS_\ald$ through the brackets
\begin{equation}
    \bL^\ald_{m,n}\cdot \bO(0) = \left\{ \bS_\ald , \cO \right\}_{m,n} = \Qbt \int_{\bbC^2} \dd^2 z \,\rho_{m,n}\, \bS_\ald(z)\bO(0)\,.
\end{equation}
More precisely, the negative modes, $m,n \leq -1$, act as
\begin{equation}
    \begin{aligned}
        \bL^{\ald}_{m,n} \cdot \bO 
        & = \frac{(-1)^{-m-1}}{(-m-1)!}\frac{(-1)^{-n-1}}{(-n-1)!} \oint_{S^3} \dd^2z\ \partial_{\dot +}^{-m-1}\partial_{\dot -}^{-n-1}\omega_{\rm BM}\  \bS_\ald \bO\\
        &\equiv \frac{1}{(-m-1)!}\frac{1}{(-n-1)!} \partial_{\dot +}^{-m-1}\partial_{\dot -}^{-n-1} \bS_\ald \bO \,,
    \end{aligned}
\end{equation}
which is nothing but the regularised product of derivatives of $\bS_\ald$ with the operator $\bO$. The positive modes generate holomorphic diffeomorphisms, 
\begin{equation}
	\bL^\ald_{m,n} \cdot \bO = -l_{m,n}^\ald \bO\,,
\end{equation}
where $l_{m,n}^\alpha$ are defined as the following vector fields
\begin{equation}
    l^\ald_{m,n} = -(z^{\dotp})^{m}(z^{\dotm})^{n} \partial_{\ald}, \qquad  m,n \geq 0\,,    
\end{equation}
Following the same logic as above, we can constrain the form of the binary $\lambda$-bracket using the associativity relations for the $\lambda$-bracket and demanding $\SU(2)$ covariance. Absorbing all normalisation factors in $\bS_\ald$ and $\lambda$ the following expression arises as the unique possibility,
\begin{equation} \label{eq:2bracketSSsec2}
     \left\{\bS_\ald\, {}_{\lambda}\, \bS_{\bed}\right\} = \lambda_{\ald} \bS_{\bed}+\lambda_{\bed} \bS_{\ald} + \partial_\ald \bS_{\bed}\,,
\end{equation}
Using this expression we can again expand \eqref{eq:l3associativity} to obtain the commutation relations of the higher Virasoro algebra \cite{Saberi:2019fkq,faonteHigherKacMoodyAlgebras2019,WilliamsThesis},
\begin{equation}\label{eq:Lmncommutator}
    \begin{aligned}
    \comm{\bL^\dotp_{m,n}}{\bL^\dotp_{k,l}} &= (m-k)\, \bL^\dotp_{m+k-1,n+l}\,,\quad & \comm{\bL^\dotp_{m,n}}{\bL^\dotm_{k,l}} &= n\, \bL^\dotp_{m+k,n+l-1} - k\, \bL^\dotm_{m+k-1,n+l}\,,\\
    \comm{\bL^\dotm_{m,n}}{\bL^\dotm_{k,l}} &= (n-l)\, \bL^\dotm_{m+k,n+l-1}
    \,,\quad&
    \comm{\bL^\dotm_{m,n}}{\bL^\dotp_{k,l}} &= -l \, \bL^\dotp_{m+k,n+l-1} + m \, \bL^\dotm_{m+k-1,n+l}\,.
\end{aligned}
\end{equation}
for $m,n,k,l \geq 0$, or $-k > m \geq 0$, $-l > n \geq 0$. In other cases the binary brackets vanish. More details on the precise derivation of these commutation relations are deferred to Appendix \ref{app:Canalysis}. 

The ternary $\lambda$-bracket can again be constrained through its symmetries. It has to be central, i.e. it is a multiple of the identity operator with coefficients proportional to the parameters $\lambda_{1,2}$. It has to transform covariantly under $\SU(2)$ and has to be totally anti-symmetric upon interchanging any pair of indices. Finally, since both sides of the equation have to scale similarly the right-hand side has to be a quintic polynomial in $\lambda_{12}$. These constraints completely determine the ternary $\lambda$-bracket, up to two theory-dependent constants,
\begin{equation}
    \left\{ \cS_{\dot \alpha}\,{}_{\lambda_1}\,\cS_{\dot \beta}\,{}_{\lambda_2}\,\cS_{\dot \gamma} \right\} = A \, \Lambda_{\mathrm{vector}}+ C \,\Lambda_{\mathcal{N}=4}\,,
\end{equation}
where $A$ and $C$ are independent of $\lambda_{1,2}$ and the polynomials $\Lambda_{\rm vector}$ and $\Lambda_{\cN=4}$ are defined as
\begin{align}
     \Lambda_{\mathrm{vector}} &= -\frac{1}{24\pi^2} (\lambda_1,\lambda_2)^2 \Big[ \lambda_{1\dot\alpha} \epsilon_{\dot\beta\dot\gamma}+\lambda_{2\dot\beta} \epsilon_{\dot\gamma\dot\alpha}-(\lambda_{1\dot\gamma}+\lambda_{2\dot\gamma}) \epsilon_{\dot\alpha\dot\beta}\Big]\\
      \Lambda_{\mathcal{N}=4} &= \frac{1}{3\pi^2} \lambda_{1\dot\alpha}\lambda_{2\dot\beta}(\lambda_1+\lambda_2)_{\dot\gamma}(\lambda_1,\lambda_2)\,.
\end{align}
where $(\lambda_1,\lambda_2) \equiv \lambda_{1\dot+} \lambda_{2\dot -}-\lambda_{1\dot-} \lambda_{2\dot +}$. The naming for the polynomials might seem arbitrary at this point but will become clear when we consider some examples. In the next section, we will explicitly compute the $\lambda$-bracket involving three stress tensor semi-chiral superfields and show that $A$ and $C$ are simple linear combinations of the familiar conformal anomalies $a$ and $c$. We refer the impatient reader to equation \eqref{eq:3brackettoac} for the final expression.

In terms of the modes $\mathbf{L}_{m,n}^{\dot\alpha}$, the non-trivial ternary $\lambda$-brackets due to central extensions again forces us to have a non-trivial ternary $L_{\infty}$-bracket, which takes the schematic form
\begin{equation}\label{eq:[LLLmaintext]}
    \left[\bL_{m,n}^{\dot \alpha}, \bL_{k,l}^{\dot \beta},  \bL_{r,s}^{\dot \gamma}\right] = \tilde{A} \psi_1 + \tilde{C} \psi_2\,,
\end{equation}
where $\psi_{1,2}$ are two theory-independent kinematic factors. The precise expression is given in \eqref{eq:[LLL]}. A proper derivation of this bracket from the $\lambda$-brackets, relating $\{\tilde A\,,\, \tilde C\}$ and $\{A\,,\, C\}$ is left to future work.

%------------
\subsection{Non-conformal theories}
%------------

In the above, we considered the situation when the physical four-dimensional $\cN=1$ theory is conformal. Conformal symmetry however is not necessary in order to define the holomorphic twist. Therefore a natural question is to what extent the above discussion generalises in the non-conformal setting. On the level of conserved current multiplets, nothing changes, and we expect to observe a similar infinite-dimensional Kac-Moody algebra in non-conformal theories. 

On the other hand, as could be expected, the Virasoro symmetry is specific to the conformal case. In non-conformal theories, the superconformal supercurrent multiplet is not well-defined and instead, we need to use a more general multiplet, the $\cS$ multiplet \cite{Komargodski:2010rb,Dumitrescu:2011iu}. This multiplet exists for any supersymmetric theory and satisfies the constraint
\begin{equation}
    \overline D^{\dot\alpha}\cS_{\alpha\dot\alpha} = \chi_\alpha + \cY_\alpha\,,
\end{equation}
where $\chi_\alpha$ is a chiral multiplet and $D_{(\alpha} \cY_{\beta)}= \overline D^2 \cY_\alpha = 0$. Whenever $\cY_\alpha$ vanishes, the constraint implies that the bottom component of the $\cS$ multiplet is a conserved current and for the remainder of this work we restrict ourselves to this situation.\footnote{In this case the $\cS$ multiplet is not irreducible but decomposes in a smaller multiplet known as the $\cR$ multiplet \cite{Gates:1981yc,Gates:1983nr,Dienes:2009td,Kuzenko:2009ym}.} The $\cS$ multiplet contains a semi-chiral operator but without conformal symmetry it is given by $\wti\bS = \partial^\ald D_+ \cS_{+\ald}$. This semi-chiral operator still generates an infinite-dimensional symmetry algebra in the twisted theory, but in contrast with the conformal case, the positive modes only generate the sub-algebra of complex symplectomorphisms. We leave a detailed study of this infinite-dimensional symmetry algebra for future work.

Whenever the theory has additional $\UU(1)$ flavour symmetries along the RG flow they might mix with the UV R-symmetry to form the IR R-symmetry. From the point of view of the $\cN=1$ (non-conformal) supersymmetry algebra, all such R-symmetries are allowed but only one of those is consistent with superconformal symmetry, i.e. there is a unique IR (superconformal) R-symmetry which can be selected using $a$-maximisation \cite{Intriligator:2003jj}. This ambiguity is reflected in the fact that we can always improve the $\cS$-multiplet by adding improvement terms
\begin{equation}
\begin{aligned}
    \cS^\prime_{\alpha\dot\alpha} =& \cS_{\alpha\dot\alpha} + \sum_i \zeta_i \comm{D_\alpha}{\bar D_{\dot\alpha}}\cJ_i\,,\\
    \chi^\prime_\alpha =& \chi_\alpha + \sum_i \zeta_i\, \f32 \,\bar D^2 D_\alpha \cJ_i\,,
\end{aligned}
\end{equation}
where $\cJ_i$ are real linear multiplets containing the $\UU(1)_i$ flavour currents. In some of the examples in Section \ref{sec:examples} we find ourselves exactly in this situation and will explicitly see how this phenomenon manifests itself in the framework of the holomorphic twist. 

%%%%%%%%%%%%%%%%%%%%%%%%%%%%%%%%%%%%%%%%%%%%%%%%%%%%%%%%%%%%%%
\section{Conformal anomalies from integrated correlators}    %
\label{sec:lbracket}                                         %
%%%%%%%%%%%%%%%%%%%%%%%%%%%%%%%%%%%%%%%%%%%%%%%%%%%%%%%%%%%%%%

As introduced above, the symmetry algebra of any four-dimensional $\cN \geq 1$ superconformal theory gets enhanced to an infinite-dimensional symmetry algebra after passing to the subsector obtained by performing the holomorphic twist. Continuing the analogy with two-dimensional VOAs we expect the central charges of the four-dimensional untwisted theory to be encoded in the central extensions of the associated higher-VOA. In this section, we make this statement precise by explicitly computing the binary and ternary $\lambda$-brackets involving conserved (super)current semi-chiral superfields. Doing so we recover the expressions obtained in the previous section allowing us to fix the constant coefficients in terms of the conformal anomalies $a$ and $c$ and flavour central charges.

To compute the two- and three-point functions in this section we made use of the Mathematica packages \texttt{CFTs4d} \cite{Cuomo:2017wme} and \texttt{SpinorAlgebra} \cite{Manenti:2019jds}. Although conceptually straightforward, extracting the relevant descendant three-point function from superspace expressions, see \cite{Osborn:1998qu} and \cite{Fortin:2011nq,Manenti:2018xns,Manenti:2019jds}, is a most tedious affair. In order not to obscure the main message, we defer the details of the computations to Appendix \ref{app:Canalysis}.

%----------
\subsection{Flavour current multiplets}
\label{subsec:flavourcurrent}
%----------

Whenever an $\cN=1$ superconformal theory contains a continuous global symmetry $\fg$, this symmetry is enhanced in the holomorphically twisted theory to an infinite-dimensional four-dimensional higher Kac-Moody symmetry $\hat{\fg}_{2,k}$. To see how the central charge $k$ of this algebra can be related to the flavour central charge $k_G$ of the original theory we proceed to explicitly compute the binary and ternary $\lambda$-brackets containing resp. two and three conserved current semi-chiral operators. 

The conserved current $j^a_\mu$ is embedded in a real linear multiplet $\cJ^a$ satisfying 
\begin{equation}\label{eq:constraint}
    D^2\cJ^a = \overline D^2\cJ^a = 0\,,
\end{equation}
where $a$ is an adjoint index of the flavour symmetry $\fg$. It is straightforward to see that the associated semi-chiral superfield is given by $\bJ^a = D_+ \cJ^a$. More explicitly we have
\begin{equation}
    \bJ^a = \ii j^a_+ + \left( j^a_{+\ald} + \f \ii 2 \partial_\ald J^a\right) \dd \bz^{\ald} - \f12 \partial_\ald \bar j^{a\dot\alpha} \dd^2 \bz\,,
\end{equation}
where $j^\alpha$ and $\bar j^\ald$ are chiral and anti-chiral spinors respectively and $J$ is a scalar and the constraints \eqref{eq:constraint} guarantee that the current is conserved, $\partial^\mu j_\mu = 0$. With the explicit semi-chiral superfield at hand, we can compute the binary and ternary $\lambda$-bracket following the prescription laid out in equation \eqref{eq:nlambdabracket} and \eqref{eq:ltreebracket} above. The binary $\lambda$-bracket is computed as follows,
\begin{equation}
    \begin{aligned}
        \left\{ \bJ^a\,{}_{\lambda} \,\bJ^b\right\} =&\,  \int_{\bbC^{2}} \f{\dd^2 z}{(2\pi \ii)^{2}} \,\e^{\lambda \cdot z }\, \bar\partial \left[ \bJ^{(1)a}(z) \bJ^{(0)b} \right]\\
        =&\, \int_{\bbC^{2}} \f{\dd^2 z}{(2\pi \ii)^{2}} \,\e^{\lambda \cdot z }\, \bar\partial \left[ \f{\left(\bar z, \dd \bar z\right)}{|z|^4} \, f^{ab}{}_c \bJ^{(0)c} + \cdots \right]\\
        =&\, f^{ab}{}_c \, \bJ^{(0)c}(0)\,.
\end{aligned}
\end{equation}
To advance from the first to the second line we used the OPE between $\bJ^{(1)}$ and $\bJ^{(0)}$. Unlike for a general multiplet, where the OPE of the descendants is not guaranteed to be fixed in terms of the OPE of the superconformal primary, the semi-chiral shortening conditions precisely allow us to do so. For this reason we can obtain the $\bJ^{(1)}\bJ^{(0)}$ from the $\cJ\cJ$ OPE, see for example \cite{Osborn:1998qu,Fortin:2011nq}. To compute the integral over $\bbC^2$ we recognise the Bochner-Martinelli kernel allowing us to use a higher-dimensional residue formula.\footnote{See Appendix \ref{app:Canalysis} for more details on the Bochner-Martinelli kernel and higher dimensional residues.} The $\bJ^{(1)}\bJ^{(0)}$ OPE contains many more terms indicated by the $\cdots$. However, these are all less singular and therefore will not contribute to the integral. Note that the binary $\lambda$-bracket is proportional to the anti-symmetric structure constant $f^{ab}{}_c$ of the flavour symmetry algebra $\fg$. From this, we immediately see that for abelian symmetries the binary $\lambda$-bracket is trivial, in line with the previous discussion.

Next, we compute the ternary $\lambda$-bracket of three flavour current semi-chiral operators. Following \eqref{eq:ltreebracket} the ternary $\lambda$-bracket is defined as
\begin{equation}\label{eq:flavourlbrak}
    \left\{ \bJ^a \,{}_{\lambda_1}\, \bJ^b \,{}_{\lambda_2}\, \bJ^c \right\} = \int_{(\bbC^2)^2} \f{\dd^2 z_1 \dd^2 z_2}{(2\pi \ii)^4} \e^{\lambda_1\cdot z_1+\lambda_2\cdot z_2} \bar\partial \vev{\bJ^{a(2)}(z_1)\bJ^{b(1)}(z_2)\bJ^{c(0)}(0)}\,.
\end{equation}
Purely on kinematic grounds, the three-point function in this expression takes the following form \cite{Cuomo:2017wme},
\begin{equation}
    \vev{\bJ^{a(2)}(z_1)\bJ^{b(1)}(z_2)\bJ^{c(0)}(0)} = \f{c_1 \, T^1_{23}U^{13}T^1_{13} + c_2 \,T^1_{23}U^{12}U^{23} + c_3\, U^{12}U^{13}U^{21}}{|z_{12}|^6|z_{1}|^4|z_{2}|^2}\,,
\end{equation}
where the tensor structures $T^i_{jk}$ and $U^{ij}$ are defined as
\begin{equation} \label{eq:Tconstituents}
    U^{ij} = s_{j\alpha}x_{ij}^{\alpha\ald} \bar{s}_{i\ald}\,, \qquad T^i_{jk} = \f{|x_{ij}|^2|x_{ik}|^2}{|x_{jk}|^2}\left( \f{s_{j\alpha}x_{ij}^{\alpha\ald} \bar{s}_{i\ald}}{|x_{ij}|^2} - \f{s_{k\alpha}x_{ik}^{\alpha\ald} \bar{s}_{i\ald}}{|x_{ik}|^2} \right)\,,
\end{equation}
where we defined $x_{ij} = x_i - x_j$ and contracted all indices with auxiliary spinors $s_i$ and $\bar s_j$. Written as such this expression is not entirely correct and one needs to contract the indices in line with the index structure of the components of the semi-chiral superfield. The constants $c_{1,2,3}$ can be fixed by imposing conservation of the current and relating it to the supersymmetric three-point function of conserved current multiplets. Doing so uniquely fixed the three coefficients to be $c_{1,2,3} = k_G d^{abc}$. Acting on the resulting three-point function with $\bar\partial$ results in
\begin{equation}\label{eq:flavourintegrand}
    \bar\partial \vev{\bJ^{(2)a}(z_1,\bz_1)\bJ^{(1)b}(z_2,\bz_2)\bJ^{(0)c}(0)} = - k_G\, d^{abc}\, \f{(\bz_{1},\bz_{2})}{|z_{12}|^4|z_{1}|^4|z_{2}|^4}\left( 1+2 \f{(\bz_{12},z_1)}{|z_{12}|^2} \right)\,,
\end{equation}
where $d^{abc} = \Tr(T^a\acomm{T^b}{T^c})$ are the symmetric structure constants of the flavour symmetry group, and we introduced the notation $z_{ij} = z_i - z_j$ and analogously $\bz_{ij} = \bz_i-\bz_j$. Now all that is left is to integrate this function over $(\bbC^2)^2$. This integral is rather non-trivial to perform but can be greatly simplified by employing heat kernel regularisation. In particular, we find that the integral over the heat kernel time is convergent making the result manifestly scheme independent. The details of this procedure are explained and applied to the integral in question in Appendix \ref{subsec:heatKernel} and \ref{subsec:intcorrelators}. Following that appendix, we obtain the following result for the ternary $\lambda$-bracket, 
\begin{equation}\label{eq:flavourres}
    \left\{ \bJ^a\,{}_{\lambda_1}\,\bJ^b\,{}_{\lambda_2}\,\bJ^c \right\} = k_G \, d^{abc} \, (\lambda_1,\lambda_2)\,. 
\end{equation}
This matches with the expression in the previous section starting from the mode algebra and rigorously demonstrates that indeed the central charge of the higher Kac-Moody symmetry algebra encodes the flavour central charge of the physical $\cN=1$ theory. In the next section, we will confirm this more explicitly in various examples.

%----------
\subsection{Supercurrent multiplet}
\label{subsec:supercurrent}
%----------

Next, we proceed analogously to compute the binary and ternary $\lambda$-bracket for supercurrent semi-chiral superfields. Although the computation is identical in spirit, the details are quite a bit more involved and most of the explicit details are referred to Appendix \ref{subsec:intcorrelators}. Whenever a theory has $\cN=1$ superconformal symmetry, we observed that after the holomorphic twist, this symmetry is enhanced to an infinite-dimensional higher Virasoro symmetry algebra. The cohomology classifying the central extensions is two-dimensional \cite{WilliamsThesis} which fits with the expectation that the central charges should be related to the $a$ and $c$ conformal anomaly of the physical theory. 

The higher Virasoro symmetry algebra is generated by the modes of the semi-chiral operator contained in the supercurrent multiplet $\cS_{\alpha\ald}$. There are various supercurrent multiplets, appropriate in different situations. The most general supercurrent multiplet is the $\cS$ multiplet, which can be defined for any $\cN=1$ supersymmetric theory \cite{Komargodski:2010rb,Dumitrescu:2011iu}. In the context of superconformal $\cN=1$ theories there necessarily exists a non-anomalous $\UU(1)_R$ symmetry and the K\"ahler form on the target space is always globally well-defined so we can use a simpler conformal supercurrent multiplet.\footnote{Whenever the theory has a non-anomalous $\UU(1)_R$ symmetry or the K\"ahler form on the target space is globally well-defined there are exist also consistent intermediate multiplets called respectively the $\cR$-multiplet \cite{Gates:1981yc,Gates:1983nr,Dienes:2009td,Kuzenko:2009ym} or the Ferrara-Zumino multiplet \cite{Ferrara:1974pz}.} This is a real multiplet satisfying 
\begin{equation}\label{eq:Sconstraints}
    D^{\alpha} \cS_{\alpha\ald} = \overline D^{\ald} \cS_{\alpha\ald} = 0\,.
\end{equation}
In components, this multiplet is given by
\begin{equation}
    \begin{aligned}
        \cS_\mu =&\, j_\mu + \ii\, \theta^\alpha S_{\mu\alpha} - \ii\,  \bar\theta^{\dot\alpha}\bar S_{\mu\dot\alpha} + (\theta\sigma^\nu \bar\theta)\left( 2T_{\mu\nu} + \f \ii 2 \epsilon_{\mu\nu\rho\sigma}\partial^\rho j^\sigma\right)  \\
        &\,- \f12 \theta^2 \bar\theta^{\dot\alpha}(\bar\sigma^\nu)_{\dot\alpha\alpha} \partial_\nu S_{\mu}^{\alpha} + \f12 \bar\theta^2 \theta^\alpha (\sigma^\nu)_{\alpha\dot\alpha}\partial_\nu \bar S_{\mu}^{\dot\alpha} -\f14 \theta^2\bar\theta^2 \partial^2 j_\mu\,,
    \end{aligned}
\end{equation}
where the constraints \eqref{eq:Sconstraints} impose that the $R$-symmetry current, stress tensor and the supersymmetry currents are conserved, $\partial^\mu j_\mu = \partial^\mu T_{\mu\nu} = \partial^\mu S_{\mu\alpha} = \partial^\mu \bar S_{\mu\dot\alpha} = 0$, as well as the trace relations $T^\mu{}_\mu =(\sigma^\mu)_{\beta\dot\beta} \bar S_{\mu}^{\dot\beta} = (\bar\sigma^\mu)_{\dot\beta\beta} S_{\mu}^{\beta} = 0$. As already noted in Section \ref{sec:holotwist} above, the supercurrent multiplet is a $\overline{A}_2A_2[1,1]_3^{(0)}$ superconformal multiplet which contains a semi-chiral operator $\bS_{\ald} = D_+ \cS_{+\ald}$ given as
\begin{equation}
    \bS_{\dot\alpha} =\, \ii\, S_{+\dot\alpha +} + \dd\bz^\bed \,% (\sigma^\nu)_{+\dot\beta}(\sigma^\mu)_{+\dot\alpha}
    \left(%2 T_{\mu\nu} + \f \ii 2 \epsilon_{\mu\nu\rho\sigma} \partial^\rho j^\sigma
    T_{\ald\bed} + \f \ii 2 \left(\star \dd j \right)_{\ald\bed}\right) + \f 12 \dd\bz^2 \,\partial_{\bed} \bar S_{+\dot\alpha}^{\dot\beta} \,.
\end{equation}
where $T_{\ald\bed}$ and $\left(\star \dd j \right)_{\ald\bed}$ are the holomorphic part of the stress tensor and dual R-symmetry field strength. Having introduced the semi-chiral superfield $\bS_\ald$ we proceed to compute the binary and ternary $\lambda$-brackets. To compute the binary bracket we proceed identically as above. We first use the OPE \cite{Osborn:1998qu} after which we compute the binary bracket by using the higher-dimensional residue theorem
\begin{equation}
\begin{aligned}
    \left\{ \bS_{\ald}\,{}_{\lambda}\, \bS_{\bed}\right\} =&\,  \int_{\bbC^{2}} \f{\dd^2 z}{(2\pi \ii)^{2}} \,\e^{\lambda \cdot z}\, \bar\partial \left[ \bS_{\dot\alpha}^{(1)}(z) \bS_{\dot\beta}^{(0)}(0) \right]\\
    =&\, \int_{\bbC^{2}} \f{\dd^2 z}{(2\pi \ii)^{2}} \,\e^{\lambda \cdot z}\, \bar\partial \left[ \f{\left(\bz,\dd\bz\right)}{|z|^6} z_{(\ald} \bS_{\bed)}^{(0)}(0) + \f{(\bz,\dd\bz)}{|z|^4}\partial_{\ald}\bS_{\bed}^{(0)}(0) + \cdots \right]\\
    =&\, \lambda_{\ald} \bS_{\bed}^{(0)}(0) + \lambda_{\bed} \bS_{\ald}^{(0)}(0) +  \partial_{\ald} \bS_{\bed}^{(0)}(0)\,,
\end{aligned}
\end{equation}
in line with the discussion in the previous section. The dots above denote additional terms in the OPE that do not contribute when computing the residue in the third line. Next, we proceed to the ternary bracket,
\begin{equation}\label{eq:SSSlbrak}
    \left\{ \bS_{\ald}\,{}_{\lambda_1}\,\bS_{\bed}\,{}_{\lambda_2}\,\bS_{\dot\gamma} \right\} = \int_{(\bbC^2)^2} \f{\dd^2 z_1}{(2\pi \ii)^{2}}\f{\dd^2 z_2}{(2\pi \ii)^{2}} \e^{\lambda_1\cdot z_1+\lambda_2\cdot z_2} \bar\partial \vev{\bS_{\ald}^{(2)}(z_1,\bz_1)\bS_\bed^{(1)}(z_2,\bz_2)\bS_{\dot\gamma}^{(0)}(0)}\,.
\end{equation}
As shown in \cite{Osborn:1998qu} the three-point function is completely determined as a sum over tensor structures, where the only theory-dependent data is contained in the prefactors which are linear functions of the conformal anomalies $a$ and $c$. In particular, we can use these results to compute the three-point function with the specific components as indicated above. In comparison with the expression above for conserved currents, in this case, there are eleven possible tensor structures appearing as explicitly listed in Appendix \ref{app:Canalysis}. In Appendix \ref{app:Canalysis} we explicitly list the tensor structures and determine the correct coefficients. Since the explicit resulting expression for the integrand is rather lengthy we refrain from listing it here. The evaluation of the remaining integral in \eqref{eq:SSSlbrak} is very tedious and the details are relegated to Appendix \ref{subsec:intcorrelators}. Note however that just like for the conserved currents, the integration over the heat kernel time is completely convergent. Note however that the condition for convergence crucially depends on some non-trivial cancellations between different tensor structures. Carefully performing all the steps outlined above we arrive at the final result for the stress tensor ternary $\lambda$-bracket,
\begin{equation}\label{eq:3brackettoac}
     \left\{ \bS_\ald \, {}_{\lambda_1}\, \bS_\bed \, {}_{\lambda_2}\, \bS_{\dot\gamma} \right\}  = \frac{4}{3}\,\left( 3 c - 2 a \right)\,\Lambda_{\mathcal{N}=4} + 16\,\left( a - c \right)\,\Lambda_{\mathrm{vector}}\,.
\end{equation}
This expression contains the (theory independent) kinematic tensors $\Lambda_{\cN=4}$ and $\Lambda_{\rm vector}$ given by
\begin{align}\label{eq:Lambdaexpression}
     \Lambda_{\mathrm{vector}} &= -\frac{1}{24\pi^2} (\lambda_1,\lambda_2)^2 \Big[ \lambda_{1\dot\alpha} \epsilon_{\dot\beta\dot\gamma}+\lambda_{2\dot\beta} \epsilon_{\dot\rho\dot\alpha}-(\lambda_{1\dot\gamma}+\lambda_{2\dot\gamma}) \epsilon_{\dot\alpha\dot\beta}\Big]\,,\\
      \Lambda_{\mathcal{N}=4} &= \frac{1}{3\pi^2} \lambda_{1\dot\alpha}\lambda_{2\dot\beta}(\lambda_1+\lambda_2)_{\dot\gamma}(\lambda_1,\lambda_2)\,.
\end{align}
All theory-dependent data is contained in the two prefactors as linear combinations of the conformal anomalies $a$ and $c$. This computation therefore presents us with a precise understanding of how the conformal anomalies $a$ and $c$ are encoded in central extensions of the infinite-dimensional higher Virasoro symmetry algebra.

%%%%%%%%%%%%%%%%%%%%%%%%%%%
\section{Examples}        %
\label{sec:examples}      %
%%%%%%%%%%%%%%%%%%%%%%%%%%%

Having established our main results, \eqref{eq:flavourres} and \eqref{eq:3brackettoac}, we continue with a variety of explicit examples to demonstrate their validity. Our examples consist of Lagrangian $\cN=1$ SCFTs for which an explicit free field realisation in terms of coupled $\mathbf{b}\mathbf{c}$-$\bfbeta\bfgamma$ system has been formulated \cite{Costello:2011np,Costello:2013zra,Saberi:2019ghy,Elliott:2020ecf,Budzik:2023xbr}. More concretely, consider an $\mathcal{N}=1$ supersymmetric gauge theory with a vector multiplet for the gauge Lie algebra $\fg$ and a collection of chiral multiplets transforming in the representations $R_i$ of $\fg$ and possibly a non-trivial superpotential $W$, which is a gauge invariant holomorphic function of the chiral superfields. It was shown in \cite{Costello:2013zra} that in the BV-BRST formalism, by adding the supercharge $\Qbt$ to the BRST charge $Q_{\mathrm{BRST}}$, one obtains a holomorphically twisted theory equivalent to the following holomorphic field theory with BV action
\begin{equation}
    S_{\rm BV} = \int_{\bbC^2} \dd^2z\ \Tr \mathbf{b}\left(\bar{\partial} \mathbf{c} -\frac12  [\mathbf{c},\mathbf{c}]\right) +  \sum_i \bfbeta_i \left(\bar \partial \bfgamma^i + [\mathbf{c},\bfgamma^i] \right) + W(\bfgamma), \label{eq:generalBVaction}
\end{equation}
where $\bar{\partial} = \dd \bar{z}^{\dot +} \bar{\partial}_{\dot +}+\dd \bar{z}^{\dot -} \bar{\partial}_{\dot -}$ is the Dolbeault differential on $\bbC^2$ and
\begin{equation}
    \begin{aligned}
    \mathbf{b}& =\mathbf{b}^{(0)}+\mathbf{b}^{(1)}+\mathbf{b}^{(2)} \in \Omega^{0,\bullet}(\bbC^2,\fg^\vee), & \mathbf{c}&=\mathbf{c}^{(0)}+\mathbf{c}^{(1)}+\mathbf{c}^{(2)} \in \Omega^{0,\bullet}(\bbC^2,\fg)[1] \, ,\\
    \bfbeta_i&=\bfbeta_i^{(0)}+\bfbeta^{(1)}_i+\bfbeta_i^{(2)} \in \Omega^{0,\bullet}(\bbC^2,R_i^\vee)[1], &  \bfgamma^i&=\bfgamma^{i(0)}+\bfgamma^{i(1)}+\bfgamma^{i(2)} \in \Omega^{0,\bullet}(\bbC^2,R_i) \, ,\label{eq:mattercontent}
\end{aligned}
\end{equation}
The action \eqref{eq:generalBVaction} in terms of the superfields should be thought of as substituting \eqref{eq:mattercontent} and selecting the correct form degree, i.e. the $(2,2)$-form part, necessary to integrate over $\bbC^2$. In particular, the kinetic (BF) terms define Wick contractions between $\mathbf{b}^{(1)}$ and $\mathbf{c}^{(0)}$, $\mathbf{b}^{(0)}$ and $\mathbf{c}^{(1)}$ and similarly between $\bfbeta^{(1)}$ and $\bfgamma^{(0)}$ and $\bfbeta^{(1)}$ and $\bfgamma^{(0)}$ where each contraction uses the Green's function pairing resulting in a factor $\omega_{\rm BM}$, where $\omega_{\rm BM}$ is the kernel for $\bar{\partial}$ as reviewed in Appendix \ref{app:Canalysis}. We will not review the precise derivation of this action but content ourselves with presenting an intuitive picture explaining the origin of the various parts of the resulting BV action. For more details, we refer the reader to the references above.

Each free chiral multiplet in the untwisted theory gives rise to two semi-chiral superfields. A chiral multiplet corresponds to a superfield $\chi$ satisfying $D_{\alpha} \chi = 0$ and the equation of motion $\overline D^2 \chi = 0$. We immediately recognise the bosonic semi-chiral superfield
\begin{equation}
    \bfgamma = \chi\,.
\end{equation}
The complex conjugate superfield $\bar\chi$ contains a fermionic semi-chiral superfield given by
\begin{equation}
    \bfbeta= D_+ \bar\chi\,.
\end{equation}
These two superfields form a $\bfbeta\bfgamma$-system very similar to the more familiar two-dimensional $\beta\gamma$-system. In addition, one can add a superpotential $W(\chi)$ changing the equations of motion to $\overline D^2 \chi = \partial_\chi W(\chi)$ which appears unchanged in the BV action \eqref{eq:generalBVaction}.

When the $\cN=1$ theory contains gauge fields, the corresponding holomorphic gauge theory has been identified with a holomorphic version of the $\mathbf{b}\mathbf{c}$ BF theory. The elementary fields in the twisted theory are a bosonic superfield $\mathbf{b}$ and a fermionic superfield $\mathbf{c}$ with ghost degree $1$. The identification of the twisted fields with the $\Qbt$-cohomology of the original theory is non-trivial as it involves various simplifications which modify the field content without changing the cohomology. The vector multiplet BPS letters with respect to the supercharge $\Qbt$ are given by the self-dual $(2,0)$ part of the field strength, $F_{++}$, and the gaugino $\bar\lambda_{\ald}$. Roughly speaking these fields can be identified with the bottom components of the superfields $\dd^2 z \, \mathbf{b}^{(0)}$ and $\partial_\ald \mathbf{c}^{(0)}$ respectively. For a more extended discussion of the correspondence to the physical fields see for example \cite{Budzik:2023xbr}. Finally, entirely analogous to the 2d $\beta\gamma$-system, the gauging of a flavour symmetry proceeds by coupling the $\bfbeta\bfgamma$ system to an adjoint valued $\mathbf{b}\mathbf{c}$ system describing the gauge sector.   

Throughout this section we choose the following normalisation for the $\mathbf{b}\mathbf{c}$ and $\bfbeta\bfgamma$ systems,
\begin{equation}
    \left\{\mathbf{b}_A\, {}_{\lambda}\, \mathbf{c}^B \right\} = \delta_A^B \,,\qquad \qquad \{\bfbeta_{ia} \, {}_{\lambda}\, \bfgamma^{jb}\} = \delta_a^b\delta_i^j\,,
\end{equation}
where $A,B$ denote adjoint indices, $i,j$ label the different chiral multiplets and $a,b$ are indices for the representation $R_i$. When the theory contains a flavour symmetry, the conserved currents in the twisted theory take the simple form $\bJ \propto \bfbeta_i\bfgamma^j$ where all gauge indices are contracted and the normalisation is fixed by demanding that the binary $\lambda$-bracket takes the following form\,
\begin{equation}\label{eq:universalJJ}
    \left\{ \bJ^a\,{}_\lambda \,\bJ^b \right\} = f^{ab}{}_{c}\, \bJ^c\,.
\end{equation}
A first guess for the stress tensor would be given by the sum $\bS_{\dot\alpha,0} = \bfbeta_i\partial_{\dot\alpha} \bfgamma^i-\mathbf{b}\partial_{\dot\alpha} \mathbf{c}$. However, this naive expression is not $\Qbt$-closed. Since the the stress tensor is quadratic in free fields we only have to require $\bS_\ald$ to be annihilated by the tree level and one loop supercharge,
\begin{equation}
    \Qbt\ \mathbf{S_\ald}^{(0)} = (Q_0 + Q_1)\ \mathbf{S_\ald}^{(0)} =  0, \label{eq:QS=0}\,.
\end{equation}
where the supercharges $Q_0$ and $Q_1$ are defined in terms of the binary and ternary $\lambda$-brackets as \cite{Budzik:2023xbr},
\begin{equation}
    Q_0 \bS_\ald^{(0)} \equiv \{\mathbf{I} \, {}_{0}\, \bS \}\,, \qquad\qquad Q_1 \bS \equiv \{\mathbf{I} \, {}_{0}\, \mathbf{I} \, {}_{0}\, \bS^{(0)} \}\,,
\end{equation}
where $\mathbf{I}$ is given by the interaction terms in \eqref{eq:generalBVaction} and all $\lambda$ parameters in brackets are put to zero. Demanding that the stress tensor defines a semi-chiral superfield fixes its form to 
\begin{equation}
    \bS_{\dot\alpha} \propto \bS_{\ald,0} - \zeta \, \partial_{\dot\alpha} \bJ\,,
\end{equation}
where $\bJ$ is the superconformal R-symmetry current superfield and the coefficient $\zeta$ is fixed by the $r$-charges of the matter fields. Whenever the theory is defined as the IR fixed of an RG flow, along the flow the R-symmetry might mix with abelian flavour symmetries. At the IR fixed point, however, there is a unique superconformal R-symmetry which can be selected through $a$-maximisation. This step is crucial to reproduce the correct symmetry algebra. Similar to the current superfield above, we fix the normalisation of the supercurrent semi-chiral superfield by demanding that the binary $\lambda$-bracket takes the following universal form, \eqref{eq:2bracketSSsec2},
\begin{equation}
    \left\{\bS_{\ald}\, {}_{\lambda}\, \bS_{\bed}\right\} = \lambda_{\ald} \bS_{\bed}+\lambda_{\bed} \bS_{\ald} + \partial_{\ald} \bS_{\bed}\,.
\end{equation}
Having introduced the general properties of the twisted theories, let us proceed to some explicit examples.

%----------
\subsection{Free vector multiplet} 
\label{subsec:freeVec}
%----------

As a first example we consider the free vector multiplet. Its holomorphic twist is given by a $\mathbf{b}\mathbf{c}$ system with BV action
\begin{equation}
     S_{\rm BV} = \int_{\bbC^2} \dd^2z\  \mathbf{b}\bar{\partial} \mathbf{c}\,.
\end{equation}
The stress tensor satisfying the closure condition, \eqref{eq:QS=0}, is given by
\begin{equation}
    \bS_{\dot\alpha} = - \mathbf{b} \partial_{\dot\alpha} \mathbf{c}\,,
\end{equation}
and we used the binary $\lambda$-bracket to fix the normalisation. With this stress tensor at hand, we are ready to compute the ternary $\lambda$-bracket. Following the definition \eqref{eq:3lambdabracket} above we compute the $\lambda$-bracket by performing all the Wick rotations and integrating the remaining product of propagators, where as mentioned above each propagator is given by a Bochner-Martinelli kernel. The computation is somewhat involved but entirely algorithmic and was explained in all generality in \cite{Budzik:2022mpd}. For the details of this particular computation, we refer the reader to section 4.3 of that paper. In Appendix \ref{app:Canalysis} we provide an explicit map from the expression in that paper to the ternary $\lambda$-bracket. The resulting expression is given by
\begin{align}
     \left\{\bS_{\dot\alpha} \, {}_{\lambda_1}\, \bS_{\dot\beta}\, {}_{\lambda_2}\, \bS_{\dot\gamma}\right\}  =&\, -\frac{1}{24\pi^2} (\lambda_1,\lambda_2)^2 \Big[ \lambda_{1\dot\alpha} \epsilon_{\dot\beta\dot\gamma}+\lambda_{2\dot\beta} \epsilon_{\dot\rho\dot\alpha}-(\lambda_{1\dot\gamma}+\lambda_{2\dot\gamma}) \epsilon_{\dot\alpha\dot\beta}\Big]\\
     =&\, \Lambda_{\mathrm{vector}}\,,
\end{align}
explaining the origin of the naming in \eqref{eq:3brackettoac}. Comparing with this equation we find the correct values for the central charges of a single free vector,
\begin{equation}
    a = \frac{3}{16}\,, \qquad\qquad c=  \frac{1}{8}\,.
\end{equation}
%

%----------
\subsection{\texorpdfstring{$N_f$}{Nf} free chirals}
\label{subsec:freeChiral}
%----------

As a second example, consider a theory of $N_f$ free chiral multiplets. The BV action of the holomorphically twisted theory is given by
\begin{equation}
    S_{\rm BV} = \int_{\bbC^2} \dd^2 z \sum_{i=1}^{N_f} \bfbeta_i \bar{\partial} \bfgamma^i\,.
\end{equation}
The theory has a global $\UU(N_f)$ symmetry rotating the chirals, with an associated conserved current
\begin{equation}
    \bJ_{i}^j = \bfbeta_i\bfgamma^j, \quad i,j = 1,\dots, N_f\,,
\end{equation}
where the trace $\Tr\,\bJ$ generates the $\UU(1)_R$ symmetry and the remaining generators generate an $\SU(N_f)$ flavour symmetry. In this section, it is more convenient to work with fundamental indices but one can straightforwardly convert the results to adjoint indices using the relevant intertwining matrices. To compute the binary $\lambda$-brackets of two currents we first compute the bracket of a current with (derivatives of) free fields,
\begin{equation}
\begin{aligned}
    \left\{\bJ_i^j \,{}_{\lambda}\, \partial_{\dot +}^{n_1} \partial_{\dot -}^{n_2} \bfgamma^k \right\} &= 
    +\delta_i^k\sum_{m_1=0 }^{n_1}\sum_{m_2 =0}^{n_2} \left(\begin{array}{c}
         n_1 \\
          m_1
    \end{array}\right)\left(\begin{array}{c}
         n_2 \\
          m_2
    \end{array}\right)\lambda_{\dot +}^{m_1}\lambda_{\dot -}^{m_2} \partial_{\dot +}^{n_1-m_1}\partial_{\dot -}^{n_2-m_2}\bfgamma^j\\
    &= +\delta_i^k (\partial_{\dot +}+\lambda_{\dot +})^{n_1} (\partial_{\dot -}+\lambda_{\dot -})^{n_2} \bfgamma^j\\
    \{\bJ_i^j \,{}_{\lambda}\, \partial_{\dot +}^{n_1} \partial_{\dot -}^{n_2} \bfbeta_k\} &= 
    -\delta_k^j \sum_{m_1=0 }^{n_1}\sum_{m_2 =0}^{n_2} \left(\begin{array}{c}
        n_1 \\
        m_1
    \end{array}\right)\left(\begin{array}{c}
         n_2 \\
         m_2
    \end{array}\right)\lambda_{\dot +}^{m_1}\lambda_{\dot -}^{m_2} \partial_{\dot +}^{n_1-m_1}\partial_{\dot -}^{n_2-m_2}\bfbeta_i \\
    =& -\delta_k^j (\partial_{\dot +}+\lambda_{\dot +})^{n_1} (\partial_{\dot -}+\lambda_{\dot -})^{n_2} \bfbeta_i
\end{aligned}
\end{equation}
Combining these equations we can straightforwardly compute the binary $\lambda$-bracket,
\begin{equation}
    \left\{\bJ_{i}^j \,{}_\lambda\, \bJ_{k}^l \right\} = \delta_{i}^l \bJ_{k}^j - \delta_{k}^j \bJ_{i}^l\,,
\end{equation}
to confirm that this current indeed has the correct normalisation to reproduce the universal form \eqref{eq:universalJJ}. Following similar steps as above, we compute the ternary $\lambda$-bracket of three currents resulting in
\begin{equation}
    \left\{\bJ_{i}^j \,{}_{\lambda_1}\, \bJ_{k}^l \,{}_{\lambda_2}\, \bJ_{m}^n \right\} =  d_{ij,kl,mn} \left(\lambda_1 , \lambda_2\right)
\end{equation}
where the coefficients are precisely the totally symmetric tensor for $U(N_f)$, which matches precisely with equation \eqref{eq:flavourres}. 

As explained above, in the presence of additional abelian flavour symmetries the stress tensor can mixed with the derivatives of the current. In this case, the only available symmetry is the overall $\UU(1) \subset \UU(N_f)$ which in fact can be identified with the superconformal R-symmetry. The stress tensor semi-chiral superfield then takes the form of a one-parameter family,
\begin{equation}
    \bS_\ald = \sum_{i=1}^{N_f} \bfbeta_i \partial_{\ald} \bfgamma^i - \zeta \partial_{\ald} \bJ_i^i 
\end{equation}
with an a priori arbitrary coefficient $\zeta$. Computing the ternary $\lambda$-bracket results in,
\begin{align}
     \left\{\bS_\ald \, {}_{\lambda_1}\, \bS_\bed\, {}_{\lambda_2}\, \bS_{\dot\gamma} \right\}  = \frac32 N_f \zeta(1-\zeta)(1- 2\zeta) \Lambda_{\cN=4}-N_f (1-2\zeta) \Lambda_{\mathrm{vector}}
\end{align}
Comparing this equation to \eqref{eq:3brackettoac}, we can identify the central charges 
\begin{equation}
    \begin{aligned}
        a &= -\frac{3}{16} N_f\left(1-8 \zeta+ 18 \zeta^2-12 \zeta^3\right)\,,\\
        c &= -\frac{1}{8} N_f\left(1-11 \zeta+27 \zeta^2-18 \zeta^3\right)\,,
    \end{aligned}
\end{equation}
Note that $a$ is maximised at $\zeta = \f13$, which is precisely $\f12 r$, where $r=\f23$ is the $r$-charge of a free chiral. Plugging this back into the equation above we reproduce the correct values of the central charges for a collection of $N_f$ free chirals, 
\begin{equation}
    a= \frac{N_f}{48}, \qquad\qquad c = \frac{N_f}{24}\,.
\end{equation}
%

%----------
\subsection{\texorpdfstring{$\cN=4$}{N=4} SYM}
\label{subsec:N4SYM}
%----------

Viewed as a $\cN=1$ SCFT, $\cN=4$ super Yang-Mills (SYM) is a gauge theory with $N_f = 3$ chiral multiplets in the adjoint representation of the gauge Lie algebra $\fg$. The BV action for the twisted theory is given by
\begin{equation}
    S_{\rm BV} = \int_{\bbC^2} d^2z\ \Tr \mathbf{b}\left(\bar{\partial} \mathbf{c} - \f12 [\mathbf{c},\mathbf{c}]\right)  + \bfbeta_i \bar{\partial} \bfgamma^i + \mathbf{c}[\bfbeta_i,\bfgamma^i] + \bfgamma^1 [\bfgamma^2,\bfgamma^3]\,.
\end{equation}
The choice of $\cN=1$ subalgebra of $\cN=4$ breaks the $\SU(4)_R$ symmetry into $\UU(1)_R\times \SU(3)_f$ where the latter is a global symmetry from an $\cN=1$ point of view. The current operator of the $\SU(3)_f$ flavour symmetry is given by
\begin{equation}
    \bJ^a = \sum_{i,j} \beta_i (t^a)^i_j \gamma^j, \quad a = 1, \cdots, 9
\end{equation}
where $t^a$ are orthonormal basis of $SU(3)$, given explicitly in the appendix \ref{app:sec:liealgebra}.
%
% \begin{equation}
%     \bJ_i^j = \bfbeta_i \bfgamma^j - \frac13 \delta_i^j \bfbeta_k\bfgamma^k, \quad i,j = 1, 2, 3
% \end{equation}
%
The computation proceeds very similarly as above and we find the binary $\lambda$-brackets of the current as
\begin{equation}
    \left\{\bJ^a, \bJ^b \right\} = f^{ab}{}_{c} J^c\,,
\end{equation}
% \begin{equation}
%     \left\{\bJ_{i}^j, \bJ_{k}^l \right\} = \delta_{i}^l \bJ_{k}^j - \delta_{k}^j \bJ_{i}^l\,,
% \end{equation}
%
matching with the general form \eqref{eq:universalJJ}. Next, we compute the ternary $\lambda$-bracket of currents, resulting in (for $\SU(2)$ gauge group)
\begin{equation}
    \left\{\bJ^a \,{}_{\lambda_1}\, \bJ^b \,{}_{\lambda_2}\, \bJ^c \right\} = \dim_\fg\, d^{abc} \,\left(\lambda_1, \lambda_2\right)\,.
\end{equation}
%
% \begin{equation}
%     \left\{\bJ_{i}^j \,{}_{\lambda_1}\, \bJ_{k}^l \,{}_{\lambda_2}\, \bJ_{m}^n \right\} = \text{\TODO{}{Add result}} d_{ij,kl,mn} \lambda_1 \wedge \lambda_2\,.
% \end{equation}
%
The stress tensor, satisfying \eqref{eq:QS=0}, takes a form very similar to the examples above,
\begin{equation}
    \bS_{\dot\alpha} = -\mathbf{b}\partial_{\dot\alpha} \mathbf{c} + \sum_{i=1}^{N_f=3}\bfbeta_i\partial_{\dot\alpha} \bfgamma^i- \frac13\sum_{i=1}^{N_f=3} \partial_{\dot\alpha}(\bfbeta_i\bfgamma^i)
\end{equation}
where gauge indices are contracted. Note that the coefficient in front of the $R$-symmetry current matches with half the $r$-charge $r=\f23$ of the chiral multiplets in $\cN=4$ SYM. After checking that this stress tensor indeed is correctly normalised we compute the ternary $\lambda$-bracket with the following result,
\begin{align}
     \left\{\bS_\ald \, {}_{\lambda_1}\, \bS_\bed\, {}_{\lambda_2}\, \bS_{\dot\gamma} \right\}  =&\, \frac{\dim_\fg}{9\pi^2} \lambda_{1\dot\alpha}\lambda_{2\dot\beta}(\lambda_1+\lambda_2)_{\dot\gamma}(\lambda_1,\lambda_2)\\
     =& \, \f13 \dim_\fg \Lambda_{\mathcal{N}=4}
\end{align}
where $\dim_\fg$ is the dimension of the lie algebra $\fg$. Comparing this to \eqref{eq:3brackettoac}, we find the following values of the central charges,
\begin{equation}
    a = c = \frac{1}{4}\dim_\fg \,, 
\end{equation}
which precisely matches with the expected values for the $\cN=4$ SYM$_\fg$ theory. This example thus clarifies our choice of naming for the second kinematic tensor in \eqref{eq:3brackettoac}.

%----------
\subsection{Fundamental \texorpdfstring{$\SU(2)$}{SU(2)} SQCD}
\label{subsec:fundSQCD}
%----------

As our next example, we consider $\SU(2)$ SQCD with $N_f$ pairs of fundamental chiral multiplet and anti-fundamental chiral multiplets and vanishing superpotential. The BV action for the twisted theory is 
\begin{equation}
     S_{\rm BV} = \int_{\bbC^2} \dd^2 z\ \Tr \mathbf{b}\left(\bar{\partial} \mathbf{c} -\f12 [\mathbf{c},\mathbf{c}]\right) +\sum_{i=1}^{N_f} \bfbeta_i \bar{\partial} \bfgamma^i + \bfbeta_i\mathbf{c}\bfgamma^i+ \tilde{\bfbeta}^i \bar{\partial} \tilde{\bfgamma}_i + \tilde{\bfgamma}_i\mathbf{c}\tilde{\bfbeta}^i \,.
\end{equation}
Since the fundamental representation of $\SU(2)$ is pseudoreal, the global symmetry is enhanced from $\SU(N_f)\times \SU(N_f)\times \UU(1)_b$ to $\SO(4N_f)\times \UU(1)_b$, where $U(1)_b$ is the baryonic symmetry under which $\bfgamma^i$ has charge 1 and $\tilde{\bfgamma}_i$ has charge -1. The current operators for the $\SU(N_f)\times \SU(N_f)$ symmetry are given by
\begin{equation}
    \bJ_i^j = \bfbeta_i\bfgamma^j, \qquad\qquad \tilde{\bJ}^i_j = \tilde{\bfbeta}^i \tilde{\bfgamma}_j\,.
\end{equation}
By combining these currents with additional currents with one tilded field we find the currents for the full $\SO(4N_f)$ flavour symmetry. As in the previous examples we can compute the $\lambda$-brackets containing three currents and read off the flavour central charge $k_G = 2$, independent of $N_f$. The semi-chiral superfield corresponding to the baryonic $U(1)_b$ current is 
\begin{equation}
    \bJ_b = \bfbeta_{i}\bfgamma^{i} - \tilde{\bfbeta}^i \tilde{\bfgamma}_i\,,
\end{equation}
which has vanishing ternary $\lambda$-bracket. 

The correctly normalised stress tensor satisfying \eqref{eq:QS=0} is given by
\begin{equation}
    \bS_\alpha = -\mathbf{b}\partial_{\dot\alpha} \mathbf{c} + \sum_{i=1}^{N_f}\bfbeta_i\partial_{\dot\alpha} \bfgamma^i + \sum_{i=1}^{N_f}\tilde{\bfbeta}_i\partial_{\dot\alpha} \tilde{\bfgamma}^i - \frac{N_f-2}{2N_f}\sum_{i=1}^{N_f} \partial_{\dot\alpha}\left(\bfgamma_i \bfbeta^i + \tilde{\bfgamma}_i \tilde{\bfbeta}^i\right) \,.
\end{equation}
where again the coefficient matches with half the $r$ charge of the chirals. With this expression at hand, we can compute the ternary $\lambda$-bracket,
\begin{align}
     \left\{\bS_{\dot\alpha} \, {}_{\lambda_1}\, \bS_{\dot\beta}\, {}_{\lambda_2}\, \bS_{\dot\gamma}\right\}  = 3\left(1-\frac{4}{N_f^2}\right)\Lambda_{\mathcal{N}=4} - 5\Lambda_{\mathrm{vector}}\label{eq:triplebracketQCD}
\end{align}
Comparing with \eqref{eq:3brackettoac}, we find the value of the central charges to be
%$
\begin{equation}
    a = \frac{21}{16} - \frac{9}{N_f^2}, \quad c = \frac{13}{8}- \frac{9}{N_f^2}\,,
\end{equation}
again reproducing the expected result. In this case, there is no free parameter and no need to perform $a$-maximisation since baryonic symmetries cannot mix with the $R$-symmetry. In the next example, however, we consider adjoint SQCD where $a$-maximisation will be necessary.

\subsection{Adjoint \texorpdfstring{$\SU(2)$}{SU(2)} SQCD} 
As our final example, we continue with $\SU(2)$ SQCD but introduce an additional chiral multiplet $\chi$ in the adjoint representation of the gauge algebra, which couples to the fundamental chirals with vanishing superpotential. In the range $3\leq N_f \leq 6$ this theory is believed to flow to an $\cN=1$ SCFT in the IR. To distinguish the $\bfbeta\bfgamma$ system originating from the adjoint chiral multiplet from the others we call the fields 
\begin{equation}
    \bfbeta_{\rm adjoint} = \bfpsi \,,\qquad \qquad \bfgamma_{\rm adjoint} = \bfphi\,.
\end{equation}
The BV action of the twisted theory is given by
\begin{equation}
      S_{\rm BV} = \int_{\bbC^2} d^2z\ \Tr \mathbf{b}\left(\bar{\partial} \mathbf{c} -\f12 [\mathbf{c},\mathbf{c}]\right) + \bfpsi (\bar{\partial}\bfphi +  [\mathbf{c}, \bfphi]) + \sum_{i=1}^{N_f} \bfbeta_i \bar{\partial} \bfgamma^i + \bfbeta_i\mathbf{c}\bfgamma^i+ \tilde{\bfbeta}^i \bar{\partial} \tilde{\bfgamma}_i + \tilde{\bfgamma}_i \mathbf{c} \tilde{\bfbeta}^i\,.
\end{equation}
By adding the additional adjoint chiral, the global symmetry is broken to $\SU(N_f)\times \UU(1)_b\times \UU(1)_c$ where the last $\UU(1)_c$ acts on the adjoint chiral with charge $1$ and on the (anti-)fundamental chirals with charge $\f1{2N_f}$. The surviving $\SU(N_f)$ symmetry is the diagonal subgroup of the two $\SU(N_f)$s,
\begin{equation}
    \bJ_{\rm diag,i}^j = \bJ^j_i+\tilde\bJ^j_i\,,
\end{equation}
Computing the $\bJ\bJ\bJ$ $\lambda$-bracket we can read off the flavour central charge which in this case is given by $k_G = 4$.

In this case, the R-symmetry can mix with the additional non-baryonic $\UU(1)_c$ symmetry. For this reason, we find a one-parameter family of (correctly normalised) stress tensors that satisfy the $\Qbt$-closure condition \eqref{eq:QS=0},
\begin{equation}
    \begin{aligned}
    \bS_{\ald} =\, -\mathbf{b}\partial_{\dot\alpha} \mathbf{c} &+ \bfpsi\partial_{\dot\alpha} \bfphi+ \sum_{i=1}^{N_f} \left(\bfbeta_i\partial_{\dot\alpha} \bfgamma^i+\tilde{\bfbeta}_i\partial_{\dot\alpha} \tilde{\bfgamma}^i\right)\\
    & -\frac{\zeta}{2} \partial_{\dot\alpha}\left( \bfphi \bfpsi\right) - \left(\frac12 -\frac{\zeta}{N_f}\right)\sum_{i=1}^{N_f}\partial_{\dot\alpha}(\bfgamma_i \bfbeta^i+\tilde{\bfgamma}_i \tilde{\bfbeta}^i)  
\end{aligned}
\end{equation}
With this stress tensor at hand, we once more compute the ternary $\lambda$-bracket,
\begin{align}
     \left\{\bS_{\dot\alpha} \, {}_{\lambda_1}\, \bS_{\dot\beta}\, {}_{\lambda_2}\, \bS_{\dot\gamma} \right\}  = \frac{3}{8}\zeta\left(14-9\zeta+\left(3-\frac{32}{N_f^2}\right)\zeta^2\right)\Lambda_{\mathcal{N}=4} - 5\zeta \Lambda_{\mathrm{vector}}\,.
\end{align}
Comparing this with \eqref{eq:3brackettoac}, we find the values of the central charges to be
\begin{equation}
\begin{aligned}
    a =&\, \frac{3}{32} \zeta\left(32-27 \zeta+\left(9-\frac{96}{N_f^2}\right) \zeta^2\right)\,,\\
    c =&\, \frac{1}{32} \zeta\left(106-81 \zeta+\left(27-\frac{288}{N_f^2}\right) \zeta^2\right)\,.
\end{aligned}
\end{equation}
We find the superconformal $R$-symmetry maximising the central charge $a$ at
\begin{equation}
    \zeta_* = \frac{N_f\left(9 N_f-\sqrt{1024-15 N_f^2}\right)}{-96+9 N_f^2}\,.
\end{equation}
At this value of the parameter we find the central charges to be
\begin{equation}
    \begin{aligned}
        a_* &= \frac{N_f\left(1024 \sqrt{1024-15 N_f^2}+3 N_f \left(-4608+189 N_f^2-5 N_f\sqrt{1024-15 N_f^2}\right)\right)}{48\left(32-3 N_f^2\right)^2}\,,\\
        c_* & = \frac{N_f\left(592 \sqrt{1024-15 N_f^2}+3 N_f\left(-2544+117 N_f^2-5 N_f\sqrt{1024-15 N_f^2}\right)\right)}{24\left(32-3 N_f^2\right)^2}\,,
    \end{aligned}
\end{equation}
In line with results from the literature \cite{Intriligator:2003jj,Tachikawa:2018sae}. With these examples, we have covered a wide variety of different theories and phenomena firmly establishing the validity of our results. Turning on the superpotential $W=\bfgamma\bfphi\tilde\bfgamma$ triggers a relevant deformation towards $\cN=2$ SQCD. For $N_f=4$, we find that with this superpotential turned on, our result reproduces the correct conformal anomalies as well as the flavour central charges. This is the only $\cN=2$ case in which the resulting IR theory is conformal and indeed we find that the stress tensor $\bS_\ald$ is not semi-chiral for any other values of $N_f$.

%%%%%%%%%%%%%%%%%%%%%%%%%%%%%
%	ACKNOWLEDGEMENTS
%%%%%%%%%%%%%%%%%%%%%%%%%%%%%
%\newpage
\bigskip\bigskip\bigskip

\leftline{\bf Acknowledgements}
\smallskip

\noindent It is with pleasure that we thank Christopher Beem, Davide Gaiotto and Brian Williams for useful and inspiring discussions. We also gratefully acknowledge Andreas Stergiou for pointing out useful Mathematica packages which facilitated some of the computations done in this work. PB and JW are grateful to the organizers of the Pollica Summer Workshop, supported by the Regione Campania, Universit\'a degli Studi di Salerno, Universit\'a degli Studi di Napoli “Federico II”, the Physics Department “Ettore Pancini” and “E. R. Caianiello”, and Istituto Nazionale di Fisica Nucleare, where part of this work was performed. The contributions of PB were made possible through the support of grant No. 494786 from the Simons Foundation and the ERC Consolidator Grant No. 864828, titled “Algebraic Foundations of Supersymmetric Quantum Field Theory'' (SCFTAlg). The work of JW is supported by the UKRI Frontier Research Grant, underwriting the ERC Advanced Grant titled “Generalized Symmetries in Quantum Field Theory and Quantum Gravity” and by the Simons Foundation Collaboration on “Special Holonomy in Geometry, Analysis, and Physics”, Award ID: 724073, Schafer-Nameki.

%%%%%%%%%%%%%%%%
\newpage
\appendix
%%%%%%%%%%%%%%%%

%%%%%%%%%%%%%%%%%%%%%%%%%%%%%%%%%%%%%%%%%%%%%%%%%%%%%%%%%%%%%%%%%%%
\section{4d \texorpdfstring{$\cN=1$}{N=1} superconformal algebra} %
\label{app:4dN1SCA}                                               %
%%%%%%%%%%%%%%%%%%%%%%%%%%%%%%%%%%%%%%%%%%%%%%%%%%%%%%%%%%%%%%%%%%%

In this appendix, we introduce the four-dimensional $\cN=1$ superconformal algebra while simultaneously establishing our notation and conventions. We exhibit the commutation relations for the complexified algebra but throughout this work, we always specialise to Euclidean signature.

The spacetime symmetry algebra for $\cN=1$ superconformal field theories is the superalgebra $\sl(4|1)$. The maximal bosonic subalgebra is $\so(6,\mathbf{C}) \times \gl(1,\mathbf{C})$. The four-dimensional conformal algebra $\so(6,\mathbf{C})$ is generated by translations, special conformal transformations, rotations, and dilatations. The generators for these transformations are given by
\begin{equation}
\cP_{\alpha\ald}\,,\qquad \cK^{\ald\alpha}\,,\qquad {\cM_\alpha}^\beta\,,\qquad {\cM^{\ald}}_{\bed}\,,\qquad \cH\,,
\end{equation}
where $\alpha,\beta= \pm$ and $\ald,\bed = \dot{\pm}$ are fundamental $\SU(2)_L\times \SU(2)_R$ indices. Adding 4 Poincar\'e supercharges $\cQ_\alpha$, $\widetilde{\cQ}_{\dot{\alpha}}$ and 4 conformal supercharges $\cS^\alpha$, $\widetilde{\cS}^{\dot{\alpha}}$ to this algebra we obtain the full 4d $\cN=1$ superalgebra $\sl(4|1)$. These supercharges are acted upon by the $\UU(1)$ R-symmetry with generator $\hat r$.\footnote{We put the hat to distinguish from the charge $r$ under this symmetry.} 

The commutation relations for the $\so(6,\mathbf{C})$ conformal algebra are
\begin{equation}
	\begin{aligned}
	\left[ {{\cM}_{\alpha}}^{\beta} , {{\cM}_{\gamma}}^{\delta} \right] &= c_1 \left(\delta_{\gamma}^{\beta} {{\cM}_{\alpha}}^{\delta}-\delta_\alpha^\delta {{\cM}_{\gamma}}^\beta\right) \,,
    &
	\left[{\cM^{\dot{\alpha}}}_{\dot{\beta}} , {\cM^{\dot{\gamma}}}_{\dot{\delta}}\right] &= c_1\left(\delta^{\dot{\alpha}}_{\dot{\delta}} {\cM^{\dot{\gamma}}}_{\dot{\beta}}-\delta^{\dot{\gamma}}_{\dot{\beta}} {\cM^{\dot{\alpha}}}_{\dot{\delta}}\right) \,,\\
	\left[ {{\cM}_{\alpha}}^{\beta} , {{\cP}_{\gamma\dot{\gamma}}} \right] &= c_1 \left(\delta_\gamma^\beta {\cP}_{\alpha\dot{\gamma}} -\f12  \delta_\alpha^\beta {\cP}_{\gamma\dot{\gamma}}\right)\,,
    &
	\left[{\cM^{\dot{\alpha}}}_{\dot{\beta}} , {{\cP}_{\gamma\dot{\gamma}}} \right] &= c_1 \left(\delta^{\dot{\alpha}}_{\dot{\gamma}}\cP_{\gamma\dot{\beta}}- \f12\delta^{\dot{\alpha}}_{\dot{\beta}} {\cP}_{\gamma\dot{\gamma}}\right)\,,\\
	\left[ {{\cM}_{\alpha}}^{\beta} , {{\cK}^{\dot{\gamma}\gamma}} \right] &= -c_1 \left(\delta_\alpha^\gamma {\cK}^{\dot{\gamma}\beta} - \f12\delta_\alpha^\beta {\cK}^{\dot{\gamma}\gamma}\right)\,,
    &
	\left[{\cM^{\dot{\alpha}}}_{\dot{\beta}} , {{\cK}^{\dot{\gamma}\gamma}} \right] &= -c_1 \left(\delta^{\dot{\gamma}}_{\dot{\beta}}\cK^{\dot{\alpha}\gamma}-\f12 \delta^{\dot{\alpha}}_{\dot{\beta}} {\cK}^{\dot{\gamma}\gamma}\right)\,,\\
	\left[ \cH , \cP_{\alpha\dot{\alpha}} \right] &= \cP_{\alpha\dot{\alpha}}\,,
    &
	\left[ \cH , \cK^{\dot{\alpha}\alpha} \right] &= -\cK^{\dot{\alpha}\alpha}\,,\\
	\left[ \cK^{\dot{\alpha}\alpha} , \cP_{\beta\dot{\beta}} \right] &= 4c_2^2\left(\delta_\beta^\alpha \delta^{\dot{\alpha}}_{\dot{\beta}}\cH + \f{1}{c_1}\delta_\beta^\alpha {\cM^{\dot{\alpha}}}_{\dot{\beta}} + \f{1}{c_1} \delta^{\dot{\alpha}}_{\dot{\beta}} {{\cM}_{\beta}}^{\alpha}\right) \,.&&
	\end{aligned}
\end{equation}
The non-vanishing commutators between the supercharges are
\begin{equation}\label{eq:susycommutators}
	\begin{aligned}
	\left\{ \cQ_\alpha,\widetilde{\cQ}_{\dot{\alpha}} \right\} &=  c_3 \cP_{\alpha\dot{\alpha}}\,,\\
	\left\{ \widetilde{\cS}^{\dot{\alpha}} , \cS^\alpha\right\} &= c_3 \cK^{\dot{\alpha}\alpha}\,,\\
	\left\{ \cQ_\alpha,\cS^\beta \right\} &= c_2 c_3 \left( \delta_\alpha^\beta\cH + \f{2}{c_1}{\cM_\alpha}^\beta - \f3{2q} \delta_\alpha^\beta \,\hat r\right)\,,\\
	\left\{\widetilde{\cS}^{\dot{\alpha}},\widetilde{\cQ}_{\dot{\beta}} \right\} &= c_2c_3 \left( \delta^{\dot{\alpha}}_{\dot{\beta}}\cH + \f{2}{c_1}{\cM^{\dot{\alpha}}}_{\dot{\beta}} + \f3{2q} \delta^{\dot{\alpha}}_{\dot{\beta}} \,\hat r\right)\,,
	\end{aligned}
\end{equation}
Finally, the bosonic generators act on the supercharges as
\begin{equation}
\begin{aligned}
	\left[{\cM_{\alpha}}^{\beta} , \cQ_{\gamma} \right] &= c_1\left(\delta_{\gamma}^{\beta} \cQ_{\alpha} - \f12 \delta_{\alpha}^{\beta} \cQ_{\gamma}\right)\,, 
    &
	\left[ {\cM^{\dot{\alpha}}}_{\dot{\beta}} , \widetilde{\cQ}_{\dot{\gamma}} \right] &= c_1\left(\delta_{\dot{\gamma}}^{\dot{\alpha}} \widetilde{\cQ}_{\dot{\beta}} -\f12  \delta^{\dot{\alpha}}_{\dot{\beta}}  \widetilde{\cQ}_{\dot{\gamma}}\right)\,,\\
	\left[{\cM_{\alpha}}^{\beta} , \cS^{\gamma} \right] &= -c_1\left(\delta_{\alpha}^{\gamma} \cS^{\beta} -\f12 \delta_{\alpha}^{\beta} \cS^{\gamma}\right)\,,\quad 
    &
	\left[ {\cM^{\dot{\alpha}}}_{\dot{\beta}} , \widetilde{\cS}^{\dot{\gamma}} \right] &= -c_1\left(\delta^{\dot{\gamma}}_{\dot{\beta}} \widetilde{\cS}^{\cI\dot{\alpha}} -\f12  \delta^{\dot{\alpha}}_{\dot{\beta}}  \widetilde{\cS}^{\dot{\gamma}}\right)\,,\\
	\left[ \cH , \cQ_\alpha \right] &= \f12 \cQ_\alpha \,,
    &
	\left[ \cH , \widetilde{\cQ}_{\dot{\alpha}} \right] &= \f12 \widetilde{\cQ}_{\dot{\alpha}} \,.\\
	\left[ \cH , \cS^\alpha \right] &= -\f12 \cS^\alpha \,,
    &
	\left[ \cH , \widetilde{\cS}^{\dot{\alpha}} \right] &= -\f12 \widetilde{\cS}^{\dot{\alpha}} \,.\\
	\left[{\cK^{\dot{\alpha}\alpha}} , \cQ_{\beta} \right] &= 2 c_2 \delta_{\beta}^{\alpha}\widetilde{\cS}^{\dot{\alpha}}\,,
    &
	\left[ {\cK^{\dot{\alpha}\alpha}} , \widetilde{\cQ}_{\dot{\beta}} \right] &= 2 c_2 \delta_{\dot{\beta}}^{\dot{\alpha}} \cS^\alpha\,,\\
	\left[{\cP_{\alpha\dot{\alpha}}} , \cS^{\beta} \right] &= -2 c_2 \delta_{\alpha}^{\beta} \widetilde{\cQ}_{\dot{\alpha}}\,,
    &
	\left[ {\cP_{\alpha\dot{\alpha}}} , \widetilde{\cS}^{\dot{\beta}} \right] &= -2 c_2 \delta^{\dot{\beta}}_{\dot{\alpha}} \cQ_\alpha\,,\\
	\left[ \hat r , \cQ_\alpha \right] &= q\cQ_\alpha\,,&
	\left[ \hat r , \widetilde{\cQ}_{\dot{\alpha}} \right] &= -q \widetilde{\cQ}_{\dot{\alpha}}\,,\\
    \left[ \hat r , \cS^\alpha \right] &= - q\cS^\alpha\,,&
    \left[ \hat r , \widetilde{\cS}^{\dot{\alpha}} \right] &= q\widetilde{\cS}^{\dot{\alpha}}\,.
	\end{aligned}
\end{equation}	
All other commutators vanish. In order to easily compare with other references we left explicit the $r$-charge of the supercharges as well as the normalisation of the translation, special conformal and rotation generators. In the main text, we pick our preferred choice of constants,
\begin{equation}
    c_1 = -2\,,\qquad\qquad c_2 = c_3 = 1 \qquad\qquad \text{and} \qquad\qquad q = -1\,.
\end{equation}
In radial quantization, the various generators satisfy the following hermiticity conditions
\begin{equation}
	\begin{aligned}
	\cH^\dagger = \cH\,,& \qquad (\cP_{\alpha\dot{\alpha}})^\dagger = \cK^{\dot{\alpha}\alpha}\,,\qquad ({{\cM}_\alpha}^\beta)^\dagger = {{\cM}_\beta}^\alpha\,,\qquad ({{\cM}^{\dot{\alpha}}}_{\dot{\beta}})^\dagger = {{\cM}^{\dot{\beta}}}_{\dot{\alpha}}\,,\\
	&\hat r^\dagger = \hat r\,,\qquad ({\cQ_{\alpha}})^\dagger = {\cS^{\alpha}}\,,\qquad ({{\widetilde{\cQ}}_{\dot{\alpha}}})^\dagger = {{\widetilde{\cS}}^{\dot{\alpha}}}\,.
	\end{aligned}
\end{equation}

In the notation of \cite{Cordova:2016emh}, the supercharges transform in the representations,  $\cQ \in [1,0]_{\f12}^{(q)}$ and $\wti{\cQ} \in [0,1]_{\f12}^{(-q)}$,
where the representation $[L]^{(q)}_\Delta$ has Lorentz representation $L$ specified by the Dynkin labels of the two $\su(2)$s, $\Delta$ is the scaling dimension and $q$ is the $\UU(1)_r$ charge.

In our conventions fundamental $\SU(2)$ indices are raised and lowered as follows,
\begin{equation}
    \phi^\alpha = \epsilon^{\alpha\beta}\,\phi_\beta\,,\qquad \phi_\alpha = \epsilon_{\alpha\beta}\,\phi^\beta\,,
\end{equation}
where the anti-symmetric $\SU(2)$ invariant tensor $\epsilon$ is defined as
\begin{equation}
    \epsilon_{12} = \epsilon^{21} = 1\,.
\end{equation}
To contract $\SU(2)$ tensors we use the following notation. For two covariant or two contravariant $\SU(2)$ vectors we define the contraction as follows,
\begin{equation}
    (x,y) = x^\ald \,y^\bed \,\epsilon_{\ald\bed} = \epsilon^{\bed\ald}\, x_\ald \, y_\bed\,.
\end{equation}
On the other hand, for one covariant and one contravariant vector, we define the contraction as
\begin{equation}
    x \cdot y = x_\ald\, y^\ald \,.
\end{equation}
%

%-----------
\subsection{\texorpdfstring{$\cN=1$}{N=1} superspace}
\label{subsec:superspace}
%-----------

In some computations in this paper, we employ $\cN=1$ superspace. In order to accommodate for our conventions of the superconformal algebra introduced above, our conventions differ slightly from the usual conventions of Wess and Bagger \cite{Wess:1992cp}. In particular, we define the supercovariant derivatives as
\begin{equation}
\begin{aligned}
    D_\alpha =&\, \f{\partial}{\partial \theta^\alpha} + \f \ii 2 \sigma^\mu_{\alpha\ald}\bar\theta^\ald\f{\partial}{\partial x^\mu}\,,\\
    \overline D_\ald =&\, -\f{\partial}{\partial \bar\theta^\ald} - \f \ii 2 \theta^\alpha \sigma^\mu_{\alpha\ald}\f{\partial}{\partial x^\mu}\,.
\end{aligned}
\end{equation}
The differential operators for the supercharges on the other hand are given by
\begin{equation}
    \begin{aligned}
        Q_\alpha =&\, \f{\partial}{\partial \theta^\alpha} - \f \ii 2 \sigma^\mu_{\alpha\ald}\bar\theta^\ald\f{\partial}{\partial x^\mu}\,,\\
        \overline Q_\ald =&\, -\f{\partial}{\partial \bar\theta^\ald} + \f \ii 2 \theta^\alpha \sigma^\mu_{\alpha\ald}\f{\partial}{\partial x^\mu}\,.
    \end{aligned}
\end{equation}
These differential operators satisfy the following anti-commutation relations,
\begin{equation}
\begin{aligned}
    \acomm{D_\alpha}{\overline D_\ald} =&\, \ii \sigma^\mu_{\alpha\ald}\f{\partial}{\partial x^\mu} = P_{\alpha\ald}\,,\\
    \acomm{Q_\alpha}{\overline Q_\ald} =&\, - \ii \sigma^\mu_{\alpha\ald}\f{\partial}{\partial x^\mu} = - P_{\alpha\ald}\,,\\
    \acomm{D_\alpha}{D_\beta} =& \acomm{\overline D_\ald}{\overline D_\bed} = \acomm{Q_\alpha}{Q_\beta} = \acomm{\overline Q_\ald}{\overline Q_\bed} = 0 \,,\\
    \acomm{D_\alpha}{Q_\beta} =& \acomm{D_\alpha}{\overline Q_\bed} = \acomm{\overline D_\ald}{Q_\beta} = \acomm{\overline D_\ald}{\overline Q_\bed} = 0\,.
\end{aligned}
\end{equation}
At first sight, the sign in the first two equations might seem off but this is consistent with the supersymmetry algebra \eqref{eq:susycommutators} due to the definition of the differential operators involving an additional minus sign. See for example Appendix A of \cite{Fortin:2011nq} for a detailed discussion of this issue.

%-------------------
\subsection{Lie algebra conventions}
\label{app:sec:liealgebra}
%-------------------

We work with orthonormal basis $t^a$, $a = 1, \cdots, \dim N^2$ of $\mathfrak{gl}(N)$, so that adjoint indices can be raised and lowered freely.\footnote{The elements are defined as are $\frac{\sqrt{2}}{2}$ times the generalised Gell-Mann matrices. For explicit expressions, see for example \cite{Bertlmann_2008}.} Our normalisation is such that the Killing form is given by,
\begin{equation}
    \Tr t^a\, t^b = \delta^{ab}\,,
\end{equation}
while the commutation relation are given by
\begin{equation}
    \comm{t^a}{t^b} = f^{ab}{}_c \,t^c\,.
\end{equation}
We define the completely anti-symmetric and symmetric structure constants as follows,
\begin{equation}
    \begin{aligned}
        f^{abc} = \Tr \left(\comm{t^a}{t^b}\, t^c \right) \,, \qquad\qquad  d^{abc} = \Tr \left(\acomm{t^a}{t^b}\,t^c\right)\,.
    \end{aligned}
\end{equation}
With this definition, the anti-symmetric structure constants satisfy the following identity,
\begin{equation}
    f^{abc}f_{ab}{}^{d} = 2h^\vee \delta^{cd}\,,
\end{equation}
where $h^\vee$ denotes the dual Coxeter number.

Another basis for $\mathfrak{sl}(N)$ that is useful in some application is given by $\{ E_{i,j}| i\neq j\} \cup \{ E_{i,i}-E_{1,1}| i=2,\dots N \}$ where $E_{i,j}$ denotes the coordinate matrix with all entries zero except one $1$ at position $\{i,j\}$. The commutation relations for the nilpotent elements are,
\begin{equation}
    \left[E_{i,j}, E_{k,l}\right]=\delta_{k,j} E_{i,l}-\delta_{i,l} E_{k,j}    
\end{equation}
These are related to the Chevalley generators $F^{\pm}_i$ for $\mathfrak{n}_\pm$ as $F^{\pm}_i = E_{i\pm 1,i}$.

%%%%%%%%%%%%%%%%%%%%%%%%%%%%%%%%%%%%%%%%%%%%%%%%%%%%%%%%%%%%%
\section{Residues, heat kernels and integrated correlators} %
\label{app:Canalysis}                                       %
%%%%%%%%%%%%%%%%%%%%%%%%%%%%%%%%%%%%%%%%%%%%%%%%%%%%%%%%%%%%%

In this appendix, we collect various technical details that were omitted from the main text. We start with a discussion of complex analysis in higher dimensions and in particular introduce a higher-dimensional residue theorem. This will be a crucial tool for many of the calculations in this work. We proceed by introducing some general aspects of heat kernel regularisation, after which we provide a detailed computation of the ternary $\lambda$-bracket both in the general case as derived in Section \ref{sec:lbracket} as well as for the various examples discussed in Section \ref{sec:examples}.

%----------
\subsection{Higher-dimensional residues}
\label{subsec:residues}
%----------

Cauchy's integral formula for a disc $D_1\subset \bbC$, states that for any smooth function $f:D_1\rightarrow \bbC$ and a point $z\in \bbC$, we have
\begin{equation}\label{eq:cauchy}
    f(w) = \f{1}{2\pi \ii}\oint_{\partial D_1} \f{\dd z}{z-w} f(z) + \f{1}{2\pi \ii}\int_{D_1} \f{\dd z}{z-w} \wedge \bar\partial f\,.
\end{equation}
If $f$ is a holomorphic function the last term drops out and we find the well-known residue formula. The goal of this subsection is to introduce a generalisation of this result to higher dimensions. 

A special role in \eqref{eq:cauchy} is played by the integration kernel $\f{1}{2\pi \ii}\f{\dd z}{z-w}$. We can generalise Cauchy's formula to $d$ complex dimensions by introducing the Bochner-Martinelli kernel,
\begin{equation}
    \omega_{\rm BM}(z,w) = \f{(-1)^d(d-1)!}{(2\pi\ii)^d|z-w|^{2d}}\sum_{i=1}^d (\bz_{i}-\bar{w}_i)\wedge \dd \bz_1\wedge\cdots \wedge \widehat{\dd \bz}_i \wedge \cdots \wedge \dd \bz_d\,.
\end{equation}
where the $i$\textsuperscript{th} anti-holomorphic differential is omitted in the sum and $|z|^2 = \sum_{i=1}^d (z_i,\bz_i)$. This is a differential form of type $(0,d-1)$ on $\bbC^d\backslash \{w\}$ which upon setting $d=1$ precisely reduces to the 'Cauchy kernel' introduced above. The Bochner-Martinelli kernel is a Green's function for the Dolbeault differential $\bar\partial$, i.e. 
\begin{equation}
    \bar\partial \left(\dd^d z\, \omega_{\rm BM}(z,0)\right) = \delta^{(2d)}(z,\bz) \dd^d z \wedge\dd^d \bz\,,    
\end{equation}
and therefore is closely related to the harmonic $(n,n)$-form on $\bbC^d$. Having introduced this kernel, we can generalise the Cauchy integration formula as follows for any smooth function $f: D_d \rightarrow \bbC$ on a $d$-dimensional disc,
\begin{equation}
    f(w) = \oint_{\partial D_d} \dd^d z\wedge \omega_{\rm BM}(z,w) f(z) + \int_{D_d} \dd^d z\wedge\omega_{\rm BM}(z,w)\wedge \bar\partial f(z)\,.
\end{equation}
For holomorphic functions, the second term vanishes, while for cohomologically holomorphic functions it is $\Qbt$-exact. 

In this work, we are mainly interested in the case $d=2$ for which the Bochner-Martinelli kernel takes the form,
\begin{equation}\label{eq:BochnerMartinelli}
    \omega_{\rm BM}(z,w) = \frac{1}{(2\pi \ii)^2} \frac{\left(\bz-\bar w, \dd\bar{z}\right)}{|z-w|^4} \,.
\end{equation}
%

%----------
\subsection{Heat kernel regularisation}
\label{subsec:heatKernel}
%----------

\begin{figure}[!htb]
    \centering
    \includegraphics[width=0.25\textwidth]{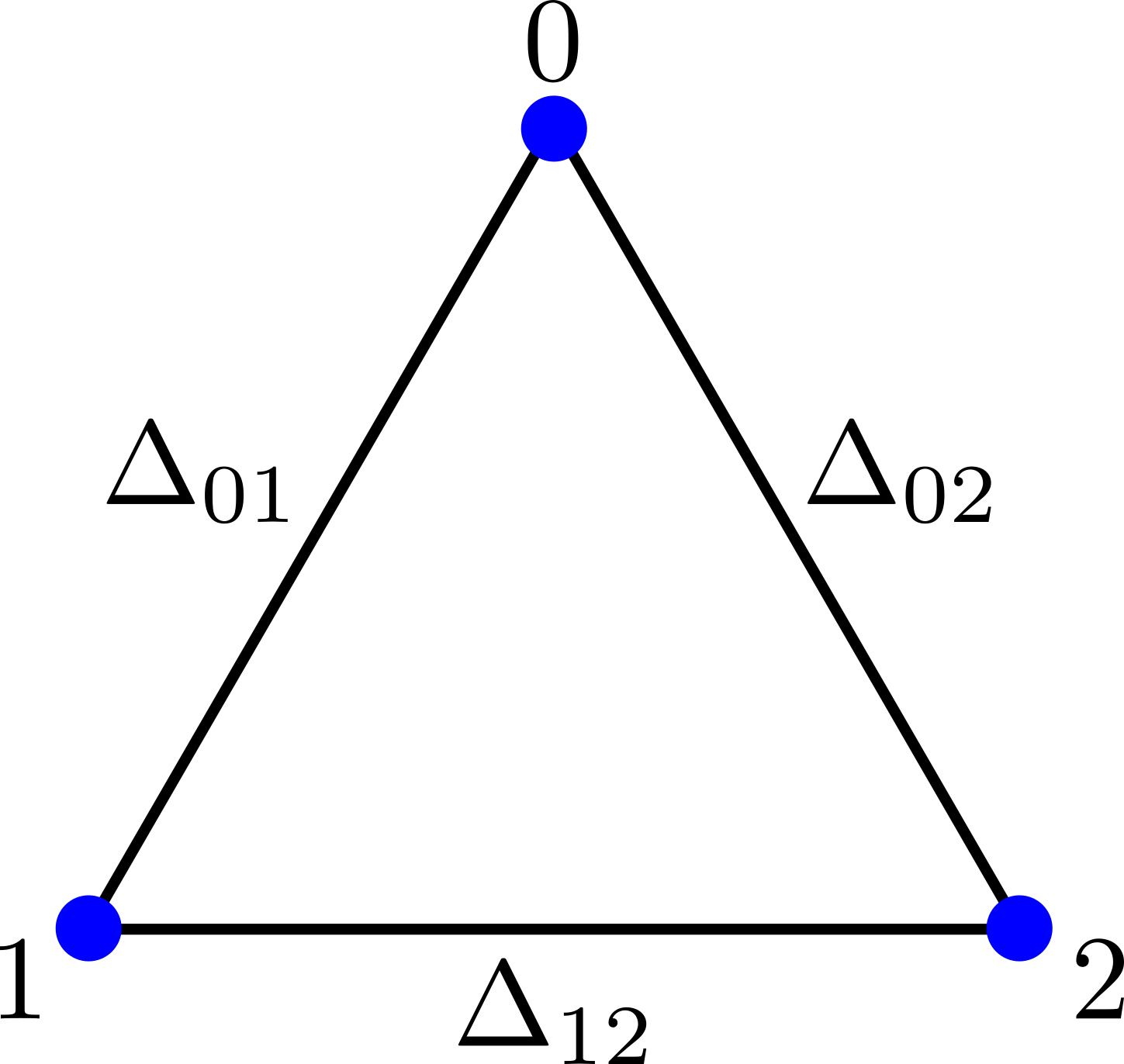}
    \caption{A triangle diagram encoding the integrand \eqref{eq:genericInt}. Every vertex is associated with a variable while every edge is associated with the difference between two vertices. Every edge variable appears in the integrand with an inverse power $2\Delta_{e}$.}
    \label{fig:triangle}
\end{figure}

The integrated correlators encountered in this work are often rather complicated to compute directly. However, we can make progress by introducing heat kernel regularisation. Recall the definition of the heat kernel in $d$ complex dimensions,
\begin{equation}
    K_{t}(x,y) = \frac{1}{(\pi t)^d}e^{-\frac{|x-y|^{2d}}{t}}\,,\qquad \qquad \text{where }\,x,y \in \bbC^d\,, 
\end{equation}
which satisfies the heat equation,
\begin{equation}
    \nabla_x^2 K_t(x,y) = \partial_t K_t(x,y)\,,
\end{equation}
where $\nabla_x^2 = \sum_{i=1}^d \partial_{x_i}\partial_{\bar{x}_i}$ for all $t>0$, with the initial condition 
\begin{equation}
    \lim\limits_{t\rightarrow 0} K_t(x,y) = \delta(x-y)\,.
\end{equation}
From now on we specify to $d=2$ where we can use the heat kernel to regularise the Bochner-Martinelli kernel as follows,
\begin{equation}
    \omega_{{\rm BM},\epsilon}(z,0) = \frac{(\bar{z},\dd\bz)}{(2\ii)^2} \int_\epsilon^{\infty}\frac{\dd t}{t}\  K_t(z)\,,
\end{equation}
where we have introduced a UV regulator $\epsilon$. Sending $\epsilon \rightarrow 0$, we recover the standard Bochner-Martinelli kernel as introduced in \eqref{eq:BochnerMartinelli}. Acting on the regularised Bochner-Martinelli kernel with $\bar\partial$ we find,
\begin{equation}
\bar\partial \omega_{{\rm BM},\epsilon}(z) = \frac{(\dd\bz, \dd\bz)}{(2\ii)^2} K_\epsilon(z),
\end{equation}
where we recognise the regularised $\delta$-function $\delta^{(4)}_\epsilon(z)=K_\epsilon(z)$. More generally, for arbitrary powers of $z_e$, we can rewrite inverse powers as follows
\begin{equation}\label{eq:heatkernelformula}
    \frac{1}{|z_e|^{2\Delta}} = \lim\limits_{\epsilon\rightarrow 0} \frac{\pi^2}{\Gamma(\Delta)} \int_\epsilon^\infty \frac{\dd t}{t^{\Delta-1}}  K_t(z)\,.
\end{equation}
The integrals resulting from the $\lambda$-brackets in Section \ref{sec:lbracket} are always of the form, 
\begin{equation}\label{eq:genericInt}
    I\left(\Gamma_{\Delta_e}\right) = \int_{\bbC^{2n}}\left(\prod_{v=1}^{n}  \f{\dd^2 z_v\dd^2 \bz_v}{(2\pi \ii)^2} \e^{\lambda_v \cdot z_v}\right) \left(\prod_{e} \f{1}{|z_e|^{2\Delta_e}}\right) f(z_v,\bz_v)\,,
\end{equation}
where $f$ is a homogeneous function, $f(a z_v\,,\,a \bz_v )= a^\Lambda f(z_v,\bz_v)$, and $n$ indicates that the integral results from an $n$-ary bracket. These integrals can be associated with a (decorated) graph $\Gamma_{\Delta_e}$ with $n$ vertices $v$ and a collection of edges $e$ connecting them. For example, for the ternary $\lambda$-bracket the graph is given by a triangle as indicated in Figure \ref{fig:triangle}. In this case $v=0,1,2$ and the edges take value $e\in\left\{01\,,\, 02 \,,\, 12\right\}$. The $\Delta_e$ in the diagram indicate the inverse power appearing in the integrand.

Using the formula \eqref{eq:heatkernelformula} we can rewrite this integral as
\begin{equation}
    I\left(\Gamma_{\Delta_e}\right) = \int \left(\prod_{e} \f{\dd t_e}{\Gamma(\Delta_e)t^{\Delta_e + 1}}\right) \int_{\bbC^{2n}} \left( \prod_{v=1}^n \f{\dd^2 z_v\dd^2 \bz_v}{(2\pi \ii)^2}\right) \e^{-S} f(z_v,\bz_v)\,.
\end{equation}
where the function $S$ in the exponent is given by
\begin{equation}
    S = z_v A^{vv^\prime} \bz_{v^\prime} - \lambda^v z_v =  z_v  A^{vv^\prime} \left(\bz_{v^\prime} - (A^{-1}\lambda)_{v^\prime}\right)\,,
\end{equation}
with the matrix $A$ defined as
\begin{equation}
    A^{vv^\prime} = \sum_{e} l^{v,e}l^{v^\prime,e}\,,
\end{equation}
and $l^{v,e}$ is defined as follows. If the edge $e$ runs from $v_1$ to $v_2$, the vector $l^{v,e}$ takes value $1$ at $v=v_1$ and $-1$ if $v=v_2$. Following these rules in the example of the triangle in Figure \ref{fig:triangle} we find the matrix
\begin{equation}
    A^{vv^\prime} = \begin{pmatrix}
        \f{1}{t_{01}} + \f{1}{t_{12}} & -\f{1}{t_{12}} \\
        -\f{1}{t_{12}} & \f{1}{t_{02}} + \f{1}{t_{12}}
    \end{pmatrix}\,.
\end{equation}
After changing coordinates, $\bz_v = \bar w_v +\left(A^{-1}\lambda\right)_v$, the integrals over $z_v$ and $\bar w_v$ reduce to Gaussian integrals and can straightforwardly be performed. In particular, the only contributions come from terms where $z_v$ and $\bar w_v$ are pairwise combined. In that case, we have the 'propagator' 
\begin{equation}
    \vev{z_{v\alpha}\bar w_{v^\prime \beta}} = \left( A^{-1}\right)_{vv^\prime} \epsilon_{\alpha\beta}\,.
\end{equation}
Having done so we are left with a set of scale invariant integrals over the heat kernel parameters $t_e$. These integrals are invariant under rescaling $t_e \rightarrow \Lambda t_e$ and therefore should be divided by the volume $\vol(\bbR_+) = \int_{\bbR_+} \frac{\dd t}{t}$ in order to account for this gauge redundancy and render the integral finite. 

%----------
\subsection{\texorpdfstring{$\lambda$}{l}-bracket from integrated correlators}
\label{subsec:intcorrelators}
%----------

Having introduced the heat kernel regularisation procedure, we can proceed to work out the integrals arising from the integrated correlators in Section \ref{sec:lbracket}. In all such cases, the graph takes the same form as in Figure \ref{fig:triangle}.

\paragraph{Flavour current semi-chiral superfields}~

First, let us discuss the ternary $\lambda$-bracket of three flavour current semi-chiral superfields. As discussed in the main text, there are three possible tensor structures appearing in the $\vev{\bJ^{(2)}\bJ^{(1)}\bJ^{(0)}}$ three-point function. Supersymmetry and current conservation fix the coefficients and after acting with $\bar\partial$ only a single tensor structure remains, see \eqref{eq:flavourintegrand}. The integral we are left to compute is therefore given by,
\begin{equation}
    I_{F} = \int_{(\bbC^2)^2} \f{\dd^2 z_1}{(2\pi \ii)^{2}}\f{\dd^2 z_2}{(2\pi \ii)^{2}} \e^{\lambda_1\cdot z_1+\lambda_2\cdot z_2}\f{(\bz_{1},\bz_{2})}{|z_{12}|^4|z_{1}|^4|z_{2}|^4}\left( 1+2 \f{(\bz_{12},z_1)}{|z_{12}|^2} \right)\,.
\end{equation}
Introducing the heat kernel regularisation we find
\begin{equation}
    I_{F} = \int \left( \f{\dd t_{01}\,\dd t_{02}\,\dd t_{12}}{t_{01}^{3}t_{02}^{3}t_{12}^{3}}\right) \int_{(\bbC^2)^2} \f{\dd^2 z_1}{(2\pi \ii)^{2}}\f{\dd^2 z_2}{(2\pi \ii)^{2}} \e^{-S}\left( (\bz_{1},\bz_{2})+\f{(\bz_{1},\bz_{2})(\bz_{12},z_1)}{t_{12}} \right)\,.
\end{equation}
Following the steps above we immediately see that this integral only gets contributions at $\cO(\lambda^2)$. Indeed, after changing coordinates as specified above we can perform the integration along $z_i$ and $\bar w_i$ to obtain the following integral over $t_e$,
\begingroup
\allowdisplaybreaks
\begin{align}
    I_F^{(1)} =&\, \int \left( \f{\dd t_{01}\,\dd t_{02}\,\dd t_{12}}{t_{01}^{3}t_{02}^{3}t_{12}^{3}}\right)\f{\epsilon^{\ald\bed}}{\left(\det A \right)^{2}}\left( A^{-1} \lambda \right)_{1\ald} \left( A^{-1} \lambda \right)_{2\bed}\,,\\
    =& \, -\int_X \dd t_{01}\,\dd t_{02}\,\dd t_{12} \f{1}{\left( t_{01} + t_{02} + t_{12} \right)^3}\left(\lambda_1,\lambda_2\right)\,,\nn \\
    I_F^{(2)} =&\, \int \left( \f{\dd t_{01}\,\dd t_{02}\,\dd t_{12}}{t_{01}^{3}t_{02}^{3}t_{12}^{4}}\right)\f{\epsilon^{\ald\bed}\epsilon^{\dot\gamma\dot\delta}}{\left(\det A \right)^{2}}\\
    &\qquad\bigg[ \epsilon_{\dot\gamma\dot\delta}\left( (A^{-1})_{12} - (A^{-1})_{22}\right)\left( A^{-1} \lambda \right)_{1\ald}\left( A^{-1} \lambda \right)_{2\bed}\nn \\
    &\qquad\,\,\,+ \epsilon_{\dot\gamma\bed}\left( (A^{-1})_{12} - (A^{-1})_{22}\right)\left( A^{-1} \lambda \right)_{1\ald}\left( A^{-1} \lambda \right)_{2\dot\delta}\nn\\
    &\qquad\,\,\,+ \epsilon_{\dot\gamma\ald}\left( (A^{-1})_{11} - (A^{-1})_{21}\right)\left( A^{-1} \lambda \right)_{2\bed}\left( A^{-1} \lambda \right)_{2\dot\delta} \bigg]\nn\\
    =&\, -\int_X \dd t_{01}\,\dd t_{02}\,\dd t_{12} \f{3 t_{02} }{\left( t_{01} + t_{02} + t_{12} \right)^4}\left(\lambda_1,\lambda_2\right)\,.\nn
\end{align}
\begingroup
Changing coordinates to $T_1= \f{t_{02}}{t_{01}}$ and $T_2 = \f{t_{12}}{t_{01}}$ and dividing by the volume $\vol(\bbR_+)$ we can simply integrate the resulting expressions with the final result,
\begin{equation}
    I_{F} = I_{F}^{(1)} + I_{F}^{(2)} = \left(-\f12 - \f12 \right)\left( \lambda_1 , \lambda_2 \right)  = - \left( \lambda_1 , \lambda_2 \right) \,.
\end{equation}
Substituting this in the expression \eqref{eq:flavourlbrak} we find the result presented in \eqref{eq:flavourres} in the main text. Let us stretch the fact that this integral is completely UV and IR finite and there are no divergences which have to be regulated. For this reason the result presented is completely scheme independent.

\paragraph{Supercurrent semi-chiral superfields}~

The computation of the ternary $\lambda$-bracket containing supercurrent semi-chiral superfields is entirely analogous to the computation for conserved currents. Technically, however, it is much more involved due to the higher conformal dimension and the appearance of free indices. In particular, the (non-supersymmetric) three-point function, $\vev{\bS^{(2)}\bS^{(1)}\bS^{(0)}}$ allows for twelve tensor structures,
\begin{equation}
    \vev{\bS_\ald^{(2)}(x_1)\bS_\bed^{(1)}(x_2)\bS_{\dot\gamma}^{(0)}(x_3)} = \f{1}{(|x_{12}|^2)^4(|x_{13}|^2)^3(|x_{23}|^2)^2}\sum_{i=1}^{12} c_i I_{i}\,,
\end{equation}
where the tensor structures are given by
\begin{equation}
    \begin{aligned}
        I_1 =& \left(T^1_{23}\right)^2\,U^{12} \,T^2_{13}\, U^{23}\, U^{32} \,,\qquad\qquad & I_2=&\, T^1_{23}\, \left(U^{13}\right)^2\, \left(T^2_{13}\right)^2 \,U^{31} \,,\\
        I_3 =&\, T^1_{23}\, \left(U^{13}\right)^2\,U^{21}\, T^2_{13}\, U^{32}\,, & I_4=&\, \left(T^1_{23}\right)^2 \,U^{12}\, \left(U^{23}\right)^2 \,U^{32} \,,\\
        I_5 =&\, T^1_{23}\, U^{12} \,U^{13}\, T^2_{13} \,U^{23}\,U^{31} \,,& I_6=&\, T^1_{23} \, U^{12}\, U^{13}\, U^{21} \,U^{23} \,U^{32} \,, \\
        I_7 =&\, U^{12}\,\left(U^{13}\right)^2 \, U^{21} \,T^2_{13} \, U^{31}\,, & I_8=&\, U^{12}\, \left(U^{13}\right)^2\, \left(U^{21}\right)^2\, U^{32} \,, \\
        I_9 =&\,T^1_{23}\, \left(U^{12}\right)^2\, \left(U^{23}\right)^2 \,U^{31} \,, & I_{10}=& \, T^1_{23} \, \left(U^{12}\right)^2\, U^{21}\,U^{23}\, T^3_{12} \,,\\
        I_{11} =&\, \left(U^{12}\right)^2\, U^{13}\, U^{21} \,U^{23}\,U^{31} \,,& I_{12} =&\,  \left(U^{12}\right)^2 \, U^{13}\,\left(U^{21}\right)^2 \,T^3_{12} \,.\\
    \end{aligned}
\end{equation}
The basic building blocks $T^i_{jk}$ and $U^{ij}$ are defined in \eqref{eq:Tconstituents}, which we repeat here for convenience,
\begin{equation}
    U^{ij} = s_{j\alpha}x_{ij}^{\alpha\ald} \bar{s}_{i\ald}\,, \qquad\qquad T^i_{jk} = \f{|x_{ij}|^2|x_{ik}|^2}{|x_{jk}|^2}\left( \f{s_{j\alpha}x_{ij}^{\alpha\ald} \bar{s}_{i\ald}}{|x_{ij}|^2} - \f{s_{k\alpha}x_{ik}^{\alpha\ald} \bar{s}_{i\ald}}{|x_{ik}|^2} \right)\,,
\end{equation}
In the expressions above we used index-free notation where all indices are contracted with commuting spinors $s_{i\alpha}$ and $\bar s_{i\ald}$. Supersymmetry fixes all the constants in this expression purely in terms of the conformal anomalies $a$ and $c$ \cite{Osborn:1998qu}. In particular, we find
\begin{equation}
\begin{aligned}
    c_1 =&\, c_2 = 2c_3 = c_4 = c_4 = c_{10} = 2c - a\,,\\
    c_6 =&\, c_7 = 2c_8 = c_9 = 2c_{11} = c_{10} = 2a \,.
\end{aligned}
\end{equation}
Expanding out the expressions of the tensor structures and contracting the appropriate indices we note that they can all be written as linear combinations of the following basic building blocks
\begin{equation}
    X_{(ij)(kl)(mn)(pq)\ald\bed\dot\gamma} = \f{\left(\bz_{12},\bz_{13}\right) \left(\bz_{ij}\right)_\ald\left(\bz_{kl}\right)_\bed\left(\bz_{mn}\right)_{\dot\gamma}}{|x_{12}|^{2\Delta_{12}}|x_{13}|^{2(9-\Delta_{12}-\Delta_{23})}|x_{23}|^{2\Delta_{23}}} \left(\bz_{pq},\dd\bz_2\right)\dd^2 \bz_1
\end{equation}
In the integrand of the ternary $\lambda$-bracket we are interested in the $\dd^2 \bz_1\dd^2 \bz_2$ component of $\bar\partial X_{(ij)(kl)(mn)(pq)\ald\bed\dot\gamma}$ and will always put $z_3 = 0$. The resulting expression for $\bar\partial X_{(ij)(kl)(mn)(pq)\ald\bed\dot\gamma}$ is given by
\begin{equation}
\begin{aligned}
    \bar\partial & X_{(ij)(kl)(mn)(pq)\ald\bed\dot\gamma} =\, \f{1}{|x_{12}|^{2\Delta_{12}}|x_{13}|^{2(9-\Delta_{12}-\Delta_{23})}|x_{23}|^{2\Delta_{23}}} \Bigg[ \left(\bz_1,\bz_{pq}\right) \left(\bz_{ij}\right)_\ald\left(\bz_{kl}\right)_\bed\left(\bz_{pq}\right)_{\dot\gamma} \\
    &\qquad+ \left(\bz_1,\bz_2\right)\left(\delta_{i}^2-\delta_{j}^2\right) \left(\bz_{pq}\right)_\ald\left(\bz_{kl}\right)_\bed\left(\bz_{mn}\right)_{\dot\gamma}
    + \left(\bz_1,\bz_2\right)\left(\delta_{k}^2-\delta_{l}^2\right) \left(\bz_{ij}\right)_\ald\left(\bz_{pq}\right)_\bed\left(\bz_{mn}\right)_{\dot\gamma}\\
    &\qquad+ \left(\bz_1,\bz_2\right)\left(\delta_{m}^2-\delta_{n}^2\right) \left(\bz_{ij}\right)_\ald\left(\bz_{kl}\right)_\bed\left(\bz_{pq}\right)_{\dot\gamma}
    + \left(\bz_1,\bz_2\right)\left(\delta_{p}^2-\delta_{q}^2\right) \left(\bz_{ij}\right)_\ald\left(\bz_{kl}\right)_\bed\left(\bz_{mn}\right)_{\dot\gamma} \\
    &\qquad+ \left(\bz_1,\bz_2\right)\left(\bz_{ij}\right)_\ald\left(\bz_{kl}\right)_\bed\left(\bz_{mn}\right)_{\dot\gamma} \left( \Delta_1 \f{(\bz_{pq},z_{12})}{|z_{12}|^2} - \Delta_2 \f{(\bz_{pq},z_{2})}{|z_{2}|^2} \right) 
    \Bigg]
    \dd^2 \bz_1\dd^2 \bz_2\,.
\end{aligned}
\end{equation}
Following the same steps as above, we can again use the heat kernel regularisation method introduced above to compute the integrated correlators. The full expression is long and not particularly illuminating, so we refrain from explicitly stating it here. Note however that in this case, the integral over the heat kernel time is not always finite. Indeed, considering a generic linear combination of basic building blocks will generically result in IR divergences of the integral and to obtain a finite result one has to put a cut-off $R \gg 1$ and carefully renormalise the result. Luckily we will not have to concern ourselves with this subtlety since as it will turn out the specific linear combination we care about produce finite integrals, both in the UV and IR, similar to the integrals presented above for the $\bJ\bJ\bJ$ $\lambda$-bracket. Computing the integral for the specific linear combination introduced above results in the final expression presented in equation \eqref{eq:3brackettoac}.

%---------------
\subsection{\texorpdfstring{$\lambda$}{lambda}-brackets from Feynman integrals}
%---------------

For the readers' convenience, we provide an explicit connection of the $\lambda$-brackets computed in this work to the universal Feynman integrals of \cite{Budzik:2022mpd}. The main object of interest we study in this article is the following $(n+1)$-ary $\lambda$-bracket,
\begin{equation}
    \left\{\bO_1 \, {}_{\lambda_1}\,  \cdots \,\bO_n \, {}_{\lambda_n}\, \bO_0 \right\}\equiv  \Qbt  \left( \prod_{k=1}^n \int_{\bbC^{2}}  \frac{d^2 z_k}{(2 \pi i)^2}e^{\lambda_k \cdot z_k} : \bO_1(z_1, \bar z_1)\cdots  \bO_n(z_n, \bar z_n) \bO_0(0,0):\right)
\end{equation}
which receives contributions from $(n-1)$-loop diagrams, as explained carefully in section 5 of \cite{Budzik:2022mpd}.

Recall that all the universal Feynman integrals $\mathcal{I}_{\Gamma}[\lambda;\delta]$ can be readily ``bootstrapped'' using quadratic identities. The relation of these integrals to the $\lambda$-brackets is as follows:
\begin{enumerate}
    \item Suppose there are no derivatives in any $\bO_i$. We formally shift the holomorphic coordinate of $\bO_i$ by $\delta_{i}$, i.e. $\bO_i(z_i,\bar{z}_i) \rightarrow \bO_i(z_i + \delta_i,\bar{z}_i)$.
    \item Each Wick contraction gives a propagator $\omega_{{\rm BM},e}$, and a set of contractions assemble a Feynman graph $\Gamma$ with the operators at the vertices and the contractions as the edges. If there are uncontracted fields left over, they will be collected at the origin. We denote those fields as $\widehat{\bO}_i(\delta_i)$. The evaluation of a bracket is then determined as the sum over all possible $(n+1)$-loop graphs.
    \item For each graph $\Gamma$, we then have the following schematic structure,
\begin{equation}
    (-1)^{\bullet}\left(\int \bar{\partial} \prod_{e\in\Gamma} \omega_e\right) : \widehat{\bO}_1(\delta_1)\cdots  \widehat{\bO}_n(\delta_n) \widehat{\bO}_0(0):
\end{equation}
where $(-1)^{\bullet}$ denotes possible signs due to the Grassmannian nature of the propagator and $\bar\partial$ arises from $\Qbt$ acting on the semi-chiral superfield. This gives us a contribution
\begin{equation}
    (-1)^{\bullet}\mathcal{I}_{\Gamma}[\partial_{\delta_{i}} + \lambda_{i};\delta_i]: \widehat{\bO}_1(\delta_1)\cdots  \widehat{\bO}_n(\delta_n) \widehat{\bO}_0(0): |_{\delta_i = 0}, \label{eq:IOOOO}
\end{equation}
where $\mathcal{I}_{\Gamma}[\lambda_i; \delta_i]$ are universal Feynman integrals in \cite{Budzik:2022mpd} with $\lambda_i$ replaced by $\partial_{\delta_i}+ \lambda_i$, acting appropriately on the corresponding operator $\widehat{\bO}_i(\delta_i)$.
\item If there are derivatives in $\bO_i$, they can be accounted for by thinking of the fields $\mathbf{f}(\delta_i)$ as a generating series
\begin{equation}
    \mathbf{f}(\delta_i) = \sum_n \frac{\delta_i^n}{n!}\partial^n \mathbf{f}
\end{equation}
and expanding \eqref{eq:IOOOO} in $\delta_i$ to the desired order before putting $\delta_i$ to zero.
\end{enumerate}
For more details on this prescription as well as details on how to compute the universal Feynman integrals, we refer the reader to \cite{Budzik:2022mpd}.

%%%%%%%%%%%%%%%%%%%%%%%%%%%%%%%%%%%%%%%%%%%%%%%%%%%%%%%%%%%%%%%%%%%%%%%%%
\section{Basic aspects of homological algebra}      %
\label{app:Linfty}                                                      %
%%%%%%%%%%%%%%%%%%%%%%%%%%%%%%%%%%%%%%%%%%%%%%%%%%%%%%%%%%%%%%%%%%%%%%%%%

$L_\infty$ algebras are a generalisation of Lie algebras where the Jacobi identities are only satisfied up to homotopy. See e.g. \cite{Kontsevich:1997vb,getzler2009lie} for an introduction. Such algebras have wide-ranging applications in physics, ranging from (closed) string field theory to QFTs and holography \cite{Zwiebach:1992ie,Hohm:2017pnh,Li:2018rnc,Chiaffrino:2023wxk}.

Let $\fg$ be a $\mathbb{Z}$-graded vector space $\fg=\bigoplus_{n\in \bbZ} \fg_n$. An $L_{\infty}$ structure on $\fg$ is a collection of multi-linear maps $[\bullet,\dots, \bullet]_k: \fg^{\otimes k} \rightarrow \fg$ of degree $2-k$, $k \geq 1$, that are graded skew-symmetric
\begin{equation}
    \left[x_{\sigma(1)}, \ldots, x_{\sigma(n)}\right]_n=(-1)^{\sgn \,\sigma} \epsilon(\sigma, x) \left[x_1, \ldots, x_n\right]_n\,.
\end{equation}
where $\epsilon(\sigma,x)$ is the Koszul sign of the permutation $\sigma$. In addition, these maps have to satisfy the following set of generalised Jacobi identities,
\begin{equation}\label{eq:GenJacobi}
    \sum_{k=1}^n(-1)^k \sum_{\sigma \in \mop{Unsh}(k, n-k)}(-1)^\sigma \epsilon(\sigma, x) \left[\left[x_{\sigma(1)}, \ldots, x_{\sigma(k)}\right]_k, x_{\sigma(k+1)}, \ldots, x_{\sigma(n)}\right]_{n-k+1}=0\,,
\end{equation}
where $\mathrm{Unsh}(k, n-k)$ is a the subset of the permutation group $S_n$ called $(k, n-k)$-unshuffle defined such that
\begin{equation}
    \sigma(1)<\sigma(2)<\cdots<\sigma(k), \quad \sigma(k+1)<\cdots<\sigma(n)\,.
\end{equation}
For small values of $n$ such identities take the form
\begin{itemize}
    \item $n=1$: It is customary to denote the 1-bracket $\left[\bullet\right]_1$ by $\dd$. The $n=1$ Jacobi identity is simply $\dd^2=0$ so that $\dd$ is a differential on $\fg$.
    \item $n=2$: In this case \eqref{eq:GenJacobi} reduces to
    \begin{equation}
        \dd\left[x_1, x_2\right]_2=\left[\dd x_1, x_2\right]+(-1)^{\left|x_1\right|} \left[x_1, \dd x_2 \right]
    \end{equation}
    which states that the differential $\dd$ satisfies a graded Leibniz rule with respect to the 2-bracket $[\bullet,\bullet]_2$. In the main text, we omit the subscript $2$ and simply denote the 2-bracket as $[\bullet,\bullet]$.
    \item  $n=3$: In this case the generalised Jacobi identity \eqref{eq:GenJacobi} yields
    \begin{equation}
        \begin{aligned}
        & \left[\left[x_1, x_2\right]_2, x_3\right]_2+(-1)^{\left(\left|x_1\right|+\left|x_2\right|\right)\left|x_3\right|} \left[\left[x_3, x_1\right]_2, x_2\right]_2+(-1)^{\left(\left|x_2\right|+\left|x_3\right|\right)\left|x_1\right|} \left[\left[x_2, x_3\right]_2, x_1\right]_2 \\
        &\qquad = \dd \left[x_1, x_2, x_3\right]_3+\left[\dd x_1, x_2, x_3\right]_3+(-1)^{\left|x_1\right|} \left[x_1, \dd x_2, x_3\right]_3+(-1)^{\left|x_1\right|+\left|x_2\right|} \left[x_1, x_2, \dd x_3\right]_3
    \end{aligned}
    \end{equation}
    which states that the binary $L_\infty$-bracket satisfies the usual Jacobi identity up to homotopy, where the deviation from the Jacobi identity for Lie algebras is given by the ternary $L_\infty$-bracket$[\bullet,\bullet,\bullet]_3$. In the main text, we omit the subscript $3$ and simply denote the ternary $L_\infty$-bracket as $[\bullet,\bullet,\bullet]$.
\end{itemize}

Heuristically, an $L_{\infty}$ algebra is a natural structure combining the notion of a Lie algebra and that of a chain complex, such that Jacobi identities are satisfied up to an infinite tower of relations. To give some more intuition into this structure let us consider some familiar special cases.

\vspace{-8pt}\paragraph{Ordinary Lie algebra}~

A Lie algebra $(\fg_0,[-,-]_2)$ is an $L_\infty$ algebra for which $\fg$ is concentrated in degree zero, i.e. $\fg = \fg_0$, and the only non-trivial $L_\infty$-bracket is the binary bracket. In this case the Jacobi identities \eqref{eq:GenJacobi} reduce to the familiar expression
\begin{equation}
    \left[\left[x_1,x_2\right]_2,x_3\right]_2 + \left[\left[x_3,x_1\right]_2,x_2\right]_2 + \left[\left[x_2,x_3\right]_2,x_1\right]_2 = 0\,,
\end{equation}
while all other generalised identities vanish identically. 

\vspace{-8pt}\paragraph{dg Lie algebra}~

A differential graded Lie algebra (dgla)  $(\fg, \dd, [\bullet,\bullet]_2)$ is an $L_\infty$ algebra for which only the 1- and 2-bracket are non-trivial.

\vspace{-8pt}\paragraph{Lie 2-algebra}~

A Lie 2-algebra is to a Lie 2=group as a Lie algebra is to a 2-group. A Lie 2-algebra is an $L_\infty$ algebra with generators concentrated in the lowest two degrees, $\fg = \fg_0 \oplus \fg_1$ with the only non-trivial $L_\infty$ brackets being the 1-, 2- and 3-bracket. The 1-bracket encodes the differential $\dd: \fg_1 \rightarrow \fg_0$, the binary bracket when restricted to degree zero is defines a Lie bracket. The ternary bracket $[\bullet,\bullet,\bullet]_3: \fg_0 \otimes \fg_0 \otimes \fg_0 \rightarrow \fg_1$ is called the Jacobiator.

%%%%%%%%%%%%%%%%%%%%%%%%%%%%%%%%%%%%%%%%%%%%%%%%%%%%%%%%%%%%%%%%%%%%%%%%%%%%%%%%%%%%
\section{(Higher) Kac Moody and Virasoro algebras}                                 %
\label{app:LieConformal}                                                           % 
%%%%%%%%%%%%%%%%%%%%%%%%%%%%%%%%%%%%%%%%%%%%%%%%%%%%%%%%%%%%%%%%%%%%%%%%%%%%%%%%%%%%

In this appendix we provide additional details on the construction and commutation relations of the infinite-dimensional symmetry algebras encountered in the main text. This appendix is largely based on the set of lecture notes \cite{williamsLectureNotes} to which we refer the reader for more details. In the main text we extensively use the $\lambda$-bracket of the higher Kac-Moody and Virasoro algebra. To complete the one-complex-dimensional story we here include a discussion of the $\lambda$-bracket in this context.

%----------
\subsection{Kac Moody algebra}
%----------

The one-dimensional affine Kac-Moody algebra associated with the Lie algebra $\fg$ is defined as a central extension of the loop algebra $L\fg\equiv \fg[z,z^{-1}]$, 
\begin{equation}
	0\rightarrow \bbC K \rightarrow \hat{\fg} \rightarrow L\fg \rightarrow 0\,.
\end{equation}
Such central extensions are classified by $H^2(L\fg)$. As vector spaces, the above sequence splits as $\hat{\fg}\simeq L\fg \oplus \bbC K$. The commutation relations are given as follows,
\begin{equation}\label{eq:KMcomrels}
\begin{aligned}
	\comm{A\otimes f(z)}{B\otimes g(z)} &= \comm{A}{B}\otimes f(z)g(z) + K \varphi_\kappa \left({A\otimes f(z)},{B\otimes g(z)} \right)\,,\\
 \comm{K}{\ \bullet\ } &= 0\,.
	\end{aligned}
\end{equation}
where $A,B\in \fg$ and $\comm{A}{B}$ is the Lie bracket in $\fg$. $\varphi_\kappa \in H^2(L\fg)$ is a 2-cocycle that takes the form
\begin{equation}
    \varphi_\kappa \left({A\otimes f(z)},{B\otimes g(z)} \right) = -\kappa(A,B) \oint\frac{dz}{2\pi i}f(z)\partial g(z)\,,
\end{equation}
where $\kappa(\bullet, \bullet)$ is a quadratic invariant polynomial on $\fg$.

The simplest example is the Heisenberg algebra $\mathbf{H}$ for which $\fg = \uu(1)$. We define a basis for $\mathbf{H}$,
\begin{equation}\label{eq:1heisenbergbasis}
    \rho_n=z^n, \quad n\in\mathbb{Z} \,,
\end{equation}
and the 2-cocycle is $\phi(z^n,z^m) = n \delta_{m+n}$. From \eqref{eq:KMcomrels} we can read off the commutation relations,
\begin{equation}
	\comm{\rho_n}{\rho_m} = n\,\delta_{m+n}\,\mathbf{1}\,.
\end{equation}
When $\fg$ is a simple Lie algebra, there is a unique quadratic invariant polynomial given by the Killing form, which in our conventions is given by $\Tr t^a t^b = \delta^{ab}$ where $\{ {t}^a\}_{a=1,\dots,\dim\,\fg}$ is a basis of $\fg$. %We then define the fields $\bJ^a(z) = \sum_{n\in \bZ}\frac{\bJ^a_n}{z^{n+1}}= \sum_{n\in \bZ} \frac{t^a\otimes\mathbf{b}_n}{z^{n+1}}$, whose modes satisfy the following commutation relations
We then define the generators ${\bJ^a_n}= {t^a\otimes\rho_n}$, satisfying the following commutation relations
\begin{equation}\label{eq:KMcommutator}
    \comm{\bJ^a_n}{\bJ^b_m} = f^{ab}{}_c \bJ^c_{m+n} + n\, \delta^{ab} \delta_{m+n}\, K\,. 
\end{equation}
As introduced in the introduction Lie conformal algebras present a useful way to rephrase the Kac-Moody algebra which easily generalises to higher dimensions. Formally, a Lie algebra is a vector space equipped with a bracket 
\begin{equation}
   [\bullet,\bullet]: \fg\otimes \fg \rightarrow \fg\,,
\end{equation}
that is bilinear, skew-symmetric and satisfies the Jacobi identity. A Lie conformal algebra in turn is a $\bbC[\partial]$-module $R$ endowed with an $\bbC$-bilinear $\lambda$-bracket 
\begin{equation}
    [\bullet\, \lambda\, \bullet]: R \otimes R \rightarrow \bbC[\lambda] \otimes R\,,
\end{equation}
which is bilinear, skew-symmetric and satisfies the Jacobi identity in a slightly generalised way. A very nice exposition can be found in \cite{2015arXiv151200821K}. We will not review all the proper definition but use the examples of Kac Moody algebra (and Virasoro algebra in the next section) to demonstrate that the $\lambda$-bracket is a useful and convenient way to repackage the singular OPE data.

A Kac Moody algebra defines a Lie conformal algebra ($\bbC[\partial] \otimes \fg+\bbC K$) as follows. Given two fields $a=a(w)$ and $b=b(w) \in \fg[\![w, w^{-1}]\!]$ their OPE is given as follows,
\begin{equation}
    a(z)b(w) = \sum_{n\in\mathbb{Z}} \frac{\left\{ a\,, b \right\}_n(w)}{ (z-w)^{n+1}}\,.
\end{equation}
where the products $\left\{ a\,, b \right\}_n$ are sometimes denoted as $a_{(n)} b$. The linear operators $a_{m} = \left\{ a\,, \bullet \right\}_{m}$ are referred to as modes of $a$. We define the $\lambda$-bracket as the generating function
\begin{equation}
    \left\{ a\,{}_\lambda \, b\right\} (w) \equiv \sum_{n\geq 0} \frac{\lambda^n}{n !} \left\{ a\,, b \right\}_n(w) \in \fg[\![w, w^{-1}]\!][\![\lambda]\!]
\end{equation}
Going back to the Kac-Moody algebra, the familiar OPE between the Kac-Moody currents reads
\begin{equation}
    \bJ^a(z)  \bJ^b(0) \sim k_G\,\frac{\delta^{ab}}{z^2} + i f^{ab}{}_{c} \,\frac{\bJ^c(0)}{z}\,,
\end{equation}
which gives rise to the $\lambda$-bracket
\begin{equation}
    \left\{ \bJ^a\,{}_\lambda \, \bJ^b \right\} =   k_G\,\delta^{ab} \lambda + f^{ab}{}_c \,\bJ^c \,.
\end{equation}

%----------
\subsection{Higher Kac-Moody algebra}
%----------

To construct the two-complex-dimensional higher Kac-Moody algebra we proceed entirely analogous to the one-dimensional case above. The definition of a higher Kac Moody algebra was given \cite{faonteHigherKacMoodyAlgebras2019}. We will not review the precise definitions but rather give an operational review of the construction of such higher algebras. 

The two-dimensional higher Kac-Moody algebra is defined as a $L_\infty$ central extension of $\fg \otimes A_2$, 
\begin{equation}
    0 \rightarrow \bbC K \rightarrow \hat\fg_{2,K} \rightarrow \fg\otimes A_2 \rightarrow 0\,,
\end{equation}
where $A_2 = \left(H^{2,\bullet}\left( \bbC^2 \backslash\{ 0 \} \right) ,\bar\partial\right)$ denotes the cochain complex modelling punctured affine space considered as an abelian dg Lie algebra. Note that in one complex dimension this simply reduces to the loop group $L\fg$. The commutation relation are given as follows,
\begin{equation}
\begin{aligned}
    \comm{A \otimes f}{B \otimes g}_2 &= \comm{A}{B} \otimes fg \,,\\
    \comm{K}{\bullet}_2 &= 0\,, \\
    \left[A\otimes f,B\otimes g,C\otimes h\right]_3 &= K \varphi_\kappa\left( A\otimes f,B\otimes g,C\otimes h \right)\,.
\end{aligned}
\end{equation}
where $A,B,C \in \fg$ and $f,g,h \in A_2$. The central extension $\varphi_\kappa \in H^2 (\fg \otimes A_2)$ is defined by a 2-cocycle,
\begin{equation}
\varphi_\kappa \left( A\otimes f,B\otimes g,C\otimes h \right) = \kappa(A,B,C)\mop{Res}_{z=0}\left( f \partial g \cdots \partial h\right)\,,  \label{eq:2KMphi}
\end{equation}
which is a tri-linear map $\hat\fg_{2,K} \times \hat\fg_{2,K} \times \hat\fg_{2,K} \rightarrow \bbC$. Therefore, in contrast to the one-dimensional case, centrally extended higher Kac-Moody algebra is not a Lie algebra. Instead, we need to generalise the notion of Lie algebra by including a ternary map resulting in an $L_\infty$ algebra, as described in Appendix \ref{app:Linfty} above.\footnote{In fact this is known on general ground \cite{Gwilliam:2018lpo,faonteHigherKacMoodyAlgebras2019}. In the spirit of radial quantisation, mode expansions are best understood as `sphere algebra' $\mathrm{Map}(S^3, \fg)$ and such centrally extended sphere algebras are naturally modelled by $L_{\infty}$ algebras.} 

The simplest example is the higher Heisenberg algebra where we take $\fg = \uu(1)$. In this case we can choose a basis of modes where the positive modes are a direct generalisation of \eqref{eq:1heisenbergbasis}, 
\begin{equation}
    \rho_{mn} = - z_1^{m}z_2^{n}\,, \qquad\qquad m,n \geq 0\,, \label{eq:2heisenbergbmn1}
\end{equation}
while the negative modes are defined as
\begin{equation}
    \rho_{mn} = - \hat\partial_{\dot +}^{-m-1}\hat\partial_{\dot -}^{-n-1} \omega_{\rm BM}\,, \qquad\qquad m,n \leq -1\,, \label{eq:2heisenbergbmn2}
\end{equation}
where $\hat\partial_\ald^{-n-1} = \f{(-1)^{-n-1}}{(-n-1)!}\partial_\ald^{-n-1}$. The remaining mixed modes are vanishing by definition. Note that this precisely corresponds to the degree zero and degree one part of the cohomology $H^{2,\bullet}(\bbC\backslash\{0\})$ respectively. In this simple example all the generators commute with each other so we immediately find
\begin{equation}
    \comm{\rho_{mn}}{\rho_{kl}}_2 = 0\,,
\end{equation}
indicating that the higher Heisenberg algebra is an abelian infinite dimensional algebra. The central extension does not appear in the binary bracket, but plugging the basis elements in \eqref{eq:2KMphi} we find the following non-trivial three-bracket
\begin{equation}
    \tbrack{\rho_{mn}}{\rho_{kl}}{\rho_{rs}} = (ks-lr)\delta_{m+k+r}\delta_{n+l+s}\,.
\end{equation}
For higher Heisenberg algebra, the 3-bracket is the only non-trivial bracket.

The higher Kac-Moody algebra is very similar to the higher Heisenberg algebra and can be obtained simply by taking the tensor product of the basis elements with a Lie algebra element. Indeed, we define the modes of the Kac-Moody algebra as 
\begin{equation}\label{eq:Jmodes}
    \bJ^a_{mn} = t^a \otimes \rho_{mn}\,.
\end{equation}
As introduced in Section \ref{sec:holotwist}, the two-bracket is non-trivial and is determined in terms of the anti-symmetric structure constants of the Lie algebra $\fg$,
\begin{equation}\label{eq:JJcommutator}
    \comm{\bJ^a_{m,n}}{\bJ^b_{k,l}}_2 = f^{ab}{}_c \bJ^c_{m+k,n+l}\,, 
\end{equation}
when both $m,n$ and $k,l$ are non-negative or when $-k-1\geq m\geq 0$ and is vanishing otherwise. It is straightforward to derive this commutator when both modes are non-negative. The brackets involving both positive and negative modes need some extra work, since the product of positive modes and negative modes
\begin{equation}
    z_{\dot +}^{m_1} z_{\dot -}^{n_2}\partial_{\dot +}^{m_2}\partial_{\dot -}^{n_2}\omega 
\end{equation}
in general, is not a linear combination of $ z_{\dot +}^{k_1} z_{\dot -}^{k_2}$ and $\partial_{\dot +}^{l_1}\partial_{\dot -}^{l_2}\omega$ unless one work in the cohomology $H^{0,\bullet}_{\bar{\partial}}(\bbC^2\backslash {0})$. More precisely, we can show that there is a nice module structure of positive modes acting on the negative modes in $\bar\partial$-cohomology
\begin{equation}\label{eq:positiveonnegative}
\begin{aligned}
    \left[ z_1^m z_2^n \times \hat{\partial}_1^{m'}  \hat{\partial}_2^{n'} \omega_{\rm BM}\right] & = \left[\hat{\partial}_1^{m'-m}  \hat{\partial}_2^{n'-n} \omega_{\rm BM}\right] \,,  \qquad & ~ &m' \geq m\geq 0\,,\quad n'\geq n\geq 0\\
    \left[z_1^m z_2^n \times \hat{\partial}^{m'}_1\hat{\partial}^{n'}_2 \omega_{\rm BM}\right] &= \left[0\right], \qquad& ~ &\text{otherwise.} 
\end{aligned}
\end{equation}
A proof of this relation can be found in Appendix \ref{subapp:proof}. Using this we can straightforwardly derive the commutation relations \eqref{eq:JJcommutator}. This restriction on the modes in the commutator is very different from the one-complex-dimensional case and is a first indication that there might be some higher structure present. Similar to the 2-Heisenberg case, the central extension appears in a ternary bracket. The central extensions are classified by $H^2(\fg\otimes A_2)$ which for simple Lie algebras $\fg$ is one-dimensional. Plugging the modes \eqref{eq:Jmodes} in \eqref{eq:2KMphi} we find the following ternary bracket,
\begin{equation}
    \left[\bJ^a_{m,n} , \bJ^a_{k,l} , \bJ^a_{r,s}\right]_3 = K\, d^{abc}\, (ks-lr)\delta_{m+k+r}\delta_{n+l+s}\,,
\end{equation}
where $d^{abc}$ is the symmetric invariant of the Lie algebra $\fg$. For a choice of $K$ we denote the two-dimensional higher Kac-Moody algebra as $\hat\fg_{2,K}$.

%----------
\subsection{Virasoro algebra}
%----------

One can think of the Virasoro algebra, $\mop{Vir} = \bbC[z,z^{-1}]\otimes \partial_z + \bbC \, c$, as a central extension of the Witt algebra, $\mop{Witt} = \bbC[z,z^{-1}]\partial_z$, the algebra of vector fields on a punctured disk, with the Lie bracket given by the Lie bracket of vector fields,
\begin{equation}
    \comm{ f(z)\partial_z}{g(z)\partial_z} = \left(f(z)\partial_z g(z)-g(z)\partial_z f(z)\right)\partial_z\,.
\end{equation}
Choosing the basis\footnote{Note the shift in order to match with the standard convention of Virasoro generators.} $\left\{ \bL_m = -z^{m+1}\partial_m \right\}_{m\in\bZ}$,  we find the commutation relations
\begin{equation}
    \left[\bL_n, \bL_m\right]=(n-m) \bL_{n+m} \,.
\end{equation}
The central extension $\varphi \in H^2(\mop{Witt})$ is defined by a 2-cocycle and can be written as the residue, 
\begin{equation}
    \varphi(\bL_n,\bL_m) = \f1{12}\oint \frac{\dd z}{2\pi \ii}\partial_z f(z)\partial^2_z g(z) \,.
\end{equation}
so that
\begin{equation}
    \varphi(\bL_n,\bL_m) = \frac{1}{12} n(n^2-1) \delta_{n+m,0} \,.
\end{equation}
Altogether, we recover the familiar form for the commutation relation of the Virasoro algebra
\begin{equation}\label{eq:commutatorVir2d}
\left[\bL_n, \bL_m\right]=(n-m) \bL_{n+m}+\frac{c}{12} n\left(n^2-1\right) \delta_{n+m, 0} \,.
\end{equation}
In some scenarios it can be useful to regard the Virasoro algebra as a Lie conformal algebra. In physical application the Virasoro algebra is obtained as the mode algebra of the holomorphic stress tensor $T(z)$. Note that in physics literature the numbering of modes is usually shifted from the maths notation so that it is customary to define $\bL_n = \left\{ T\,, \bullet \right\}_{n+1}$. We can then define the $\lambda$-bracket of the stress tensor with a primary operator as follows,
\begin{equation}
\begin{aligned}
    \left\{ {T(z)\,}_\lambda \cO(0)\right\} =& \oint_{S^1} \frac{\dd z}{2\pi \ii} \,\e^{\lambda z}\, T(z)\cO(0) \\
    =& L_{-1}\, \cO(0)+ \lambda \,L_0\, \cO(0) + L_1\, \cO(0) + \cdots \\
    =& \partial \cO(0) + h\,\lambda\, \cO(0) \,,
\end{aligned} \label{eq:TO2bracket2d}
\end{equation}
where $h$ is the conformal weight of the primary operator $\cO$. From this expression, we can straightforwardly read off the singular part of the OPE 
\begin{equation}
    T(z)\cO(0) \sim \frac{h \,\cO(0)}{z} + \partial \cO(0) \,.
\end{equation}
Similarly, the we compute the $\lambda$-bracket of two stress tensors as
\begin{equation}\label{eq:TTbracket2d}
    \left\{T \, {}_{\lambda}\, T \right\} = \lambda^3 \frac{c}{12} +\Big( 2\lambda+ \partial\Big) T\,,    
\end{equation}
which again encodes the singular part of the OPE
\begin{equation}
    T(z)T(0) \sim \frac{c/2}{z^4} +\frac{2T}{z^2} + \frac{\partial T}{z} \,.
\end{equation}
%

%------------
\subsection{Higher Virasoro algebra}
%------------

Similar to the above, we now generalise this to the higher Virasoro algebra in two complex dimensions. The higher Virasoro algebra can be obtained as a central extension of the higher $\mop{Witt}_2$ algebra which is defined as the cochain complex $\left(A_2\otimes \bbC\{ \partial_\ald \} , \bar\partial \right)$. The commutation relations are defined through the Lie bracket of the vector fields
\begin{equation}
\left[A \otimes \partial_{\ald}, B \otimes \partial_{\bed}\right]= A\, \mathcal{L}_{\partial_{\ald}}(B) \otimes \partial_{\bed}-(-1)^{|A||B|} B\, \mathcal{L}_{\partial_{\bed}}(A) \otimes \partial_{\ald}\,,
\end{equation}
where $\mathcal{L}_{\partial_z} $ is the Lie derivative. Analogous to the higher Kac-Moody algebra, we can define a basis,
\begin{equation}
    \bL_{m,n}^\ald = \rho_{m,n}\otimes \partial_\ald\,.
\end{equation}
From this expression we can straightforwardly read of the commutation relations for the non-negative modes while the commutation relations between negative modes vanish. The mixed commutation relations for positive and negative modes need some more work but can be derived using the identity in cohomology \eqref{eq:positiveonnegative}. In the end, the resulting non-vanishing commutators are given by,
\begin{equation}
    \begin{aligned}
        \comm{L_{m,n}^1}{L_{m',n'}^1} &= (m-m') \,L^1_{m+m'-1,n+n'} \,,\\
        \comm{L_{m,n}^2}{L_{m',n'}^2} &= (n-n') \,L^2_{m+m',n+n'-1} \,,\\
        \comm{L_{m,n}^1}{L_{m',n'}^2} &= n\,L^1_{m+m',n+n'-1}-m' \,L^2_{m+m'-1,n+n'} \,,  \\
        \comm{L_{m,n}^2}{L_{m',n'}^1} &= m\,L^2_{m+m'-1,n+n'}-n' \,L^1_{m+m',n+n'-1} \,.
\end{aligned}\label{eq:Lmncommutatornegative}
\end{equation}
The central extensions of the higher Witt algebra are classified by Gelfand-Fuks cohomology classes, $\varphi \in H^2(\mathrm{Witt}_2)$, which is known to be two-dimensional \cite{WilliamsThesis,Saberi:2019fkq}. Given an element $\xi=\sum \alpha^{\dot \alpha} \partial_{\dot \alpha}$ in the higher Witt algebra, its Jacobian is defined as
\begin{equation}
    [\mop{Jac}(\xi)]^{\dot \beta}_{\dot \alpha}=\partial_{\dot \alpha} \alpha^{\dot \beta}    
\end{equation}
The following expressions then define representatives for the two non-trivial 2-cocycles, 
\begin{equation}
    \begin{aligned}
        \psi_1\left(\xi_0, \xi_1, \xi_2\right) &=\,\mop{Res}_{z=0} \Tr\left(\mop{Jac}\left(\xi_0\right) \partial \mop{Jac}\left(\xi_1\right) \partial \mop{Jac}\left(\xi_2\right)\right) \,, \\
        \psi_2\left(\xi_0, \xi_1, \xi_2\right)&=\,\mop{Res}_{z=0} \Tr\left(\mop{Jac}\left(\xi_0\right)\right) \mop{Tr}\left(\partial \mop{Jac}\left(\xi_1\right) \partial \mop{Jac}\left(\xi_2\right)\right)\,.
    \end{aligned}
\end{equation}
Assuming $m,n<0$ and $k,l,r,s\geq 0$, we then find the following ternary $L_{\infty}$-bracket,
\begin{equation}
    \left[\bL_{m,n}^{\dot \alpha}, \bL_{k,l}^{\dot \beta},  \bL_{r,s}^{\dot \gamma}\right] = \tilde{A}\, \psi_1 + \tilde{C}\, \psi_2,\label{eq:[LLL]}
\end{equation}
with
\begin{align}
    \psi_1 =   & K_1 \delta_{m+k+r,3}\delta_{n+l+s,0}\delta^{\dot \alpha}_{\dot +}\delta^{\dot \beta}_{\dot +}\delta^{\dot \gamma}_{\dot +}
    + K_2 \delta_{m+k+r,0}\delta_{n+l+s,3}\delta^{\dot \alpha}_{\dot -}\delta^{\dot \beta}_{\dot -}\delta^{\dot \gamma}_{\dot -}\nn\\
    & + \Bigg[ K_3 \delta^{\dot \alpha}_{\dot +}\delta^{\dot \beta}_{\dot -}\delta^{\dot \gamma}_{\dot -}  +K_2  \delta^{\dot \alpha}_{\dot -}\delta^{\dot \beta}_{\dot -}\delta^{\dot \gamma}_{\dot +} +K_4 \delta^{\dot \alpha}_{\dot -}\delta^{\dot \beta}_{\dot +}\delta^{\dot \gamma}_{\dot -} \Bigg] \delta_{m+k+r,1}\delta_{n+l+s,2}\\
    & + \Bigg[ K_3 \delta^{\dot \alpha}_{\dot +}\delta^{\dot \beta}_{\dot -}\delta^{\dot \gamma}_{\dot +}  +K_1 \delta^{\dot \alpha}_{\dot +}\delta^{\dot \beta}_{\dot +}\delta^{\dot \gamma}_{\dot -} +K_4 \delta^{\dot \alpha}_{\dot -}\delta^{\dot \beta}_{\dot +}\delta^{\dot \gamma}_{\dot +} \Bigg] \delta_{m+k+r,2}\delta_{n+l+s,1}\,,\nn
\end{align}
and
\begin{align}
   \psi_2=   &K_1 \delta_{m+k+r,3}\delta_{n+l+s,0}\delta^{\dot \alpha}_{\dot +}\delta^{\dot \beta}_{\dot +}\delta^{\dot \gamma}_{\dot +}+ K_2 \delta_{m+k+r,0}\delta_{n+l+s,3}\delta^{\dot \alpha}_{\dot -}\delta^{\dot \beta}_{\dot -}\delta^{\dot \gamma}_{\dot -}\nn\\
    &+ \Bigg[ K_3 \delta^{\dot \alpha}_{\dot -}\delta^{\dot \beta}_{\dot -}\delta^{\dot \gamma}_{\dot +}  +K_2 \delta^{\dot \alpha}_{\dot +}\delta^{\dot \beta}_{\dot -}\delta^{\dot \gamma}_{\dot -} +K_4 \delta^{\dot \alpha}_{\dot -}\delta^{\dot \beta}_{\dot +}\delta^{\dot \gamma}_{\dot -} \Bigg] \delta_{m+k+r,1}\delta_{n+l+s,2}\\
    &+\Bigg[ K_3 \delta^{\dot \alpha}_{\dot +}\delta^{\dot \beta}_{\dot -}\delta^{\dot \gamma}_{\dot +}  + K_1 \delta^{\dot \alpha}_{\dot -}\delta^{\dot \beta}_{\dot +}\delta^{\dot \gamma}_{\dot +} +K_4 \delta^{\dot \alpha}_{\dot +}\delta^{\dot \beta}_{\dot +}\delta^{\dot \gamma}_{\dot -} \Bigg] \delta_{m+k+r,2}\delta_{n+l+s,1}\,, \nn
\end{align}
where the coefficients $K_i$ are
\begin{equation}
    \begin{aligned}
        K_1 &= kr(s-l+  lr-ks), & \quad K_2 &= ls(k-r+ lr-ks) \,,\\
        K_3 &= ks(-1+k+s+lr-ks), & \quad K_4 &= lr(1-l-r +lr-ks) \,.
    \end{aligned}
\end{equation}
In particular, the difference between the two terms is simple
\begin{equation}
\begin{aligned}
    \psi_1-\psi_2 =&\, (K_1-K_4)\delta_{m+k+r,2}\delta_{n+l+s,1}\left(\delta^{\dot \alpha}_{\dot +}\delta^{\dot \beta}_{\dot +}\delta^{\dot \gamma}_{\dot -} -\delta^{\dot \alpha}_{\dot -}\delta^{\dot \beta}_{\dot +}\delta^{\dot \gamma}_{\dot +}  \right)\\
    &+ (K_2-K_3)\delta_{m+k+r,1}\delta_{n+l+s,2}\left(\delta^{\dot \alpha}_{\dot -}\delta^{\dot \beta}_{\dot -}\delta^{\dot \gamma}_{\dot +} -\delta^{\dot \alpha}_{\dot +}\delta^{\dot \beta}_{\dot -}\delta^{\dot \gamma}_{\dot -}  \right)
    \end{aligned}
\end{equation}
The coefficient $\tilde{A}$ and $\tilde{C}$ are expected to be a linear combination of the conformal anomalies $a$ and $c$. We plan to come back to this and their relation to the $\lambda$-brackets computed in the main text in the future.

%-------------
\subsection{Proof of equation (D.23)}
\label{subapp:proof}
%-------------

In this subsection we provide a simple proof of \eqref{eq:positiveonnegative}. First of all, observe that
\begin{equation}
    \hat{\partial}_{\dot \alpha}^m \omega_{\rm BM} = (m+1) \z_{\dot \alpha}^m\omega_{\rm BM} \,,
\end{equation}
where we defined
\begin{equation}
    \z_{\ald} = \frac{\bar{z}_{\ald}}{|z|^2} \,,
\end{equation}
which satisfies the constraint $z^{\dot \alpha} \z_{\dot \alpha} = 1$. The identity \eqref{eq:positiveonnegative} only holds in cohomology and thus it is important to understand which terms are $\bar\partial$-exact. For example, noting that
\begin{equation}
 \bar{\partial}(\z_{\ald}) = -(2\pi\ii)^2 \epsilon_{\dot \alpha \dot \beta} z^{\dot\beta} \omega_{\rm BM} \,,
\end{equation}
we immediately see that expressions of the form
\begin{equation}
    \left[(z^{\dot +})^n (z^{\dot -})^m \omega_{\rm BM}\right] = [0], \qquad n+m> 0 \,,
\end{equation}
are $\bar{\partial}$-exact and thus vanish in the cohomology. Using the above we can derive a second useful cohomological identity, 
\begin{equation}
    \epsilon_{\ald\dot\gamma}\left[z^{\dot\gamma}\, \z_\bed\, \omega_{\rm BM}\right] = -\f{1}{(2\pi\ii)^2}\left[\z_\bed\, \bar{\partial} \z_\ald \right] = \f{1}{(2\pi\ii)^2}\left[\z_\ald \,\bar{\partial} \z_\bed\,\right] = -\epsilon_{\bed\dot\gamma}\left[z^{\dot\gamma}\, \z_\ald\, \omega_{\rm BM}\right] \,.
\end{equation}
From these identities we find that
\begin{equation}
    \left[\omega_{\rm BM}\right] = 2 \left[z^\dotp \z_\dotp \omega_{\rm BM}\right] = 2 \left[z^\dotm \z_\dotm \omega_{\rm BM}\right] \,.
\end{equation}
More generally we have\footnote{Note that $(z^{\dot \alpha})^n [\alpha] = [(z^{\dot \alpha})^n \alpha]$ but we cannot multiply $\z_i$, i.e. $\z_{\dot \alpha} [\alpha] \neq [\z_{\dot \alpha} \alpha]$} 
\begin{equation}
    \left[\omega_{\rm BM}\right] = (m+1) \left[(z^\dotp \z_\dotp)^m \omega_{\rm BM} \right] = (m+1) \left[(z^\dotm \z_\dotm)^m \omega_{\rm BM}\right]\,.
\end{equation}
Using this we see that
\begin{equation}
    \left[ (z^{\dot \pm})^m \hat{\partial}_{\dot \pm}^m \omega_{\rm BM} \right] =  \left[\omega_{\rm BM}\right] \,.
\end{equation}
Similarly, using the identity
\begin{equation}
    \bar\partial \left(\z_\ald\z_\bed\right) = -(2\pi\ii)^2 \epsilon_{(\bed|\dot\gamma|}\hat\partial_{\ald)} \omega_{\rm BM}\,,
\end{equation}
where the symmetrisation is only over $\ald$ and $\bed$, we find that
\begin{equation}
      \left[z^\dotm \hat{\partial}_\dotp \omega_{\rm BM} \right] = \left[z^\dotp \hat{\partial}_\dotm \omega_{\rm BM} \right] = [0]\,.
\end{equation}
Repeating the above steps in the case of multiple derivatives and/or higher powers of $z_\ald$ straightforwardly results in \eqref{eq:positiveonnegative}.

%----------
\subsection{Associativity relations of \texorpdfstring{$\lambda$}{lambda}-brackets}
%----------

Finally, we finish this appendix by listing the various associativity relations the Higher $\lambda$-brackets satisfy. These relations were instrumental in constraining the allowed expressions for the $\lambda$-brackets in Section \ref{sec:holotwist}. The whole set of associativity relations for general $n$-ary $\lambda$-brackets remains an open question. However, for the purposes of this work we only need a subset of relations derived in \cite{Budzik:2023xbr,ourpaper-higher}. The $n$-ary $\lambda$-brackets are graded symmetric under permutations of their arguments and with the $\lambda$ parameters. The last entry of the $\lambda$-brackets behaves in a slightly different manner from the others but if we define $\lambda_n \equiv -\sum_{i=1}^{n-1} \lambda_i - \partial$, with the latter symbol acting outside the bracket they are graded symmetric under permutation of all arguments. 

The binary $\lambda$-bracket $\{\bullet \,_{\lambda} \bullet \}$ is best under control. In this case the bracket has to satisfy the following set of axioms:
\begin{enumerate}
	\item $\{\partial A\,{}_\lambda\,B\} = - \lambda \{A\,{}_\lambda\,B\}$\,,
	\item $\partial \{A\,{}_\lambda\,B\} = -\lambda \{  A\,{}_\lambda\,B\}+\{ A\,{}_\lambda\,\partial B\}$\,, which implies 
	$\{ A\,{}_\lambda\,\partial B\} = (\partial + \lambda) \{A\,{}_\lambda\,B\}$\,,
	\item $\{A\,{}_\lambda\,B\} = (-1)^{|A||B|} \{B\,{}_{-\lambda- \partial }\,A\}$\,,
	\item $\{A  \, {}_{\lambda_1} \{ B\, {}_{\lambda_2} \,C \}\} - (-1)^{(|A|+1)(|B|+1)}\{B  \, {}_{\lambda_2} \{ A\, {}_{\lambda_1} \,C \}\} +(-1)^{|A|} \{\{ A\, {}_{\lambda_1} \,B \} \, {}_{\lambda_1+\lambda_2} C   \}=0$\,,
\end{enumerate}
where in the equations above $|A|$ and $|B|$ denote the fermion parity of the operators.

%%%%%%%%%%%%%%%%
\newpage
\bibliographystyle{JHEP}
\bibliography{twistbib}		
\end{document}